\documentclass[conference]{IEEEtran}
\IEEEoverridecommandlockouts
\usepackage{cite}
\usepackage{amsmath,amssymb,amsfonts}
\usepackage{algorithmic}
\usepackage{graphicx}
\usepackage{textcomp}
\usepackage{amsthm}
\usepackage{tabularray}
\usepackage{xcolor}
\usepackage{bbding}
\usepackage{makecell}
\usepackage{algorithm}  
\usepackage{color}
\usepackage{multirow}
\usepackage{bm}
\usepackage{array}
\usepackage{booktabs}
\usepackage{tabularx,booktabs}
\usepackage{multicol}
\usepackage{cleveref}
\usepackage{balance}
\usepackage{lastpage}
\usepackage{tabulary}
\usepackage{subfigure}
\usepackage{etoolbox}
\usepackage{bbm}
\usepackage{enumitem}
\usepackage{mathrsfs}
\usepackage{multicol}
\usepackage{amsfonts,amssymb}
\usepackage{ulem}
\usepackage{cancel}
\usepackage{fancyhdr}
\usepackage{setspace}
\usepackage{stackengine}
\usepackage{threeparttable}
\usepackage{url}
\usepackage{pifont}

\usepackage{amsthm}
\theoremstyle{definition}

\usepackage{tcolorbox}
\tcbuselibrary{skins, breakable, theorems}

\usepackage{nomencl}
\makenomenclature

\def\BibTeX{{\rm B\kern-.05em{\sc i\kern-.025em b}\kern-.08em
    T\kern-.1667em\lower.7ex\hbox{E}\kern-.125emX}}

\newcommand{\blue}[1]{\textcolor{black}{#1}}

\begin{document}

\bstctlcite{IEEEexample:BSTcontrol}

\title{Large Language Model (LLM) for Telecommunications: A Comprehensive Survey on Principles, Key Techniques, and Opportunities
\thanks{
Hao Zhou, Chengming Hu, Ye Yuan, Yufei Cui, Can Chen, Yili Jin, Haolun Wu, Dun Yuan, Li Jiang, and Xue Liu are with the School of Computer Science, McGill University, Montreal, QC H3A 0E9, Canada. (emails:\{hao.zhou4, chengming.hu, ye.yuan3, yufei.cui, can.chen, yili.jin, haolun.wu, dun.yuan, li.jiang3\}@mail.mcgill.ca, xueliu@cs.mcgill.ca);

Di Wu is with the School of Electrical and Computer Engineering, McGill University, Montreal, QC H3A 0E9, Canada. (email: di.wu5@mcgill.ca);

Charlie Zhang is with Samsung Research America, Plano, Texas, TX 75023, USA. (email: jianzhong.z@samsung.com);

Xiangbin Wang is with the Department of Electrical and
Computer Engineering, Western University, London, ON N6A 3K7, Canada.
(e-mail: xianbin.wang@uwo.ca);

Jiangchuan Liu is with the School of Computing Science, Simon Fraser
University, Burnaby, BC V5A 1S6, Canada. (e-mail: jcliu@sfu.ca).
}}

\author{\IEEEauthorblockN{Hao Zhou, Chengming Hu, Ye Yuan, Yufei Cui, Yili Jin, \\ Can Chen, Haolun Wu, Dun Yuan, Li Jiang, Di Wu, Xue Liu,  \IEEEmembership{Fellow, IEEE},\\ Charlie Zhang, \IEEEmembership{Fellow, IEEE}, Xianbin Wang, \IEEEmembership{Fellow, IEEE}, Jiangchuan Liu, \IEEEmembership{Fellow, IEEE}. }
}

\maketitle

\thispagestyle{fancy}            
\chead{This paper has been accepted by IEEE Communications Surveys and Tutorials. } 
\renewcommand{\headrulewidth}{1pt}      
\pagestyle{plain}

\begin{abstract}
Large language models (LLMs) have received considerable attention recently due to their outstanding comprehension and reasoning capabilities, leading to great progress in many fields.   
The advancement of LLM techniques also offers promising opportunities to automate many tasks in the telecommunication (telecom) field.
%
After pre-training and fine-tuning, LLMs can perform diverse downstream tasks based on human instructions, paving the way to artificial general intelligence (AGI)-enabled 6G.
Given the great potential of LLM technologies, this work aims to provide a comprehensive overview of LLM-enabled telecom networks. 
In particular, we first present LLM fundamentals, including model architecture, pre-training, fine-tuning, inference and utilization, model evaluation, and telecom deployment. 
Then, we introduce LLM-enabled key techniques and telecom applications in terms of generation, classification, optimization, and prediction problems.
Specifically, the LLM-enabled generation applications include telecom domain knowledge, code, and network configuration generation. After that, the LLM-based classification applications involve network security, text, image, and traffic classification problems.
Moreover, multiple LLM-enabled optimization techniques are introduced, such as automated reward function design for reinforcement learning and verbal reinforcement learning.
Furthermore, for LLM-aided prediction problems, we discussed time-series prediction models and multi-modality prediction problems for telecom.
Finally, we highlight the challenges and identify the future directions of LLM-enabled telecom networks. 
\end{abstract}


\begin{figure*}[!t]
\centering
\includegraphics[width=0.96\linewidth]{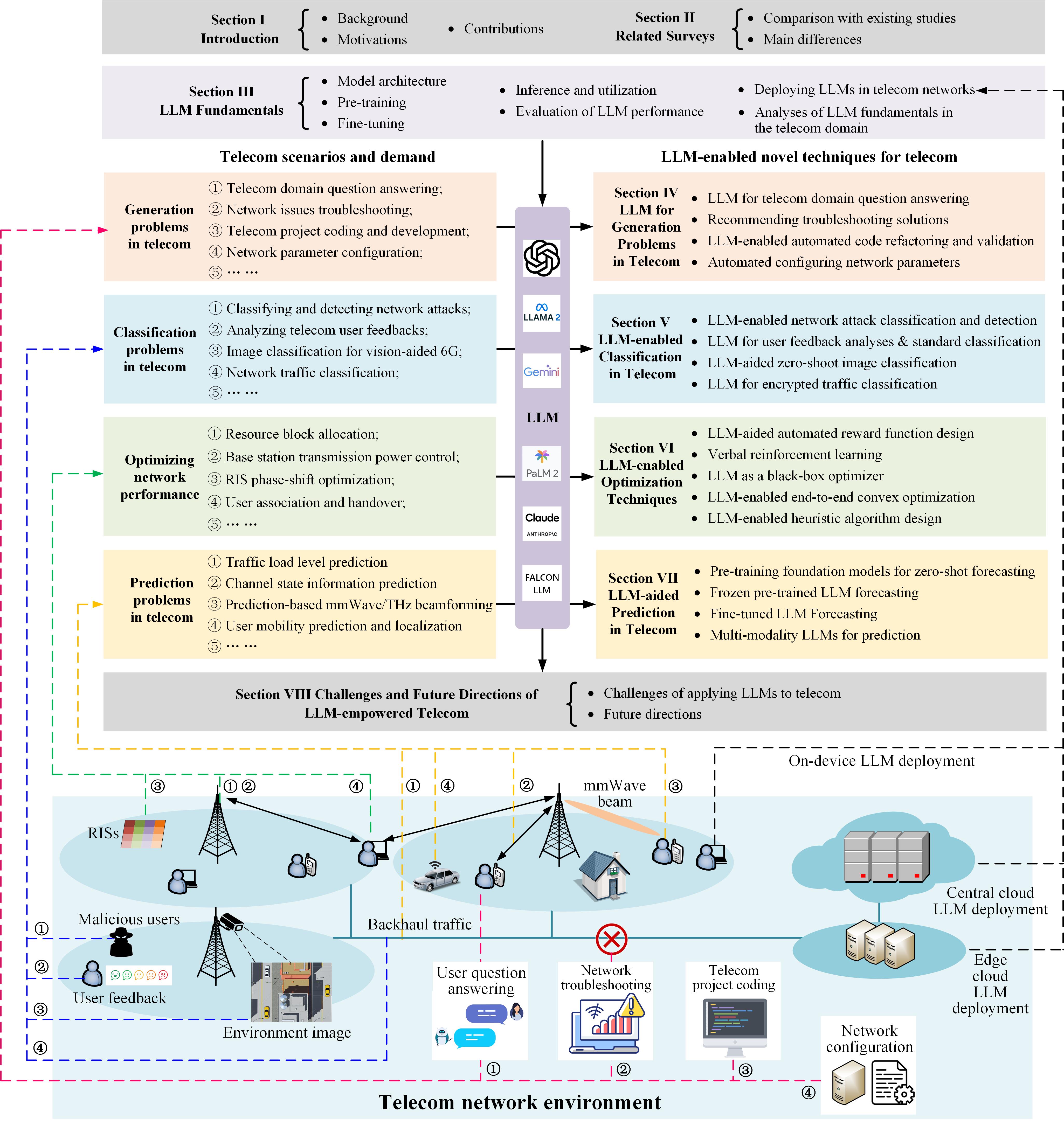}
\caption{Organization and key topics covered in this work.}
\label{fig1}
\end{figure*}

\section{Introduction}

While 5G networks have entered the commercial deployment stage, the academic community has started the exploration of envisioned 6G networks.
In particular, 6G networks are expected to achieve terabits per second (Tbps) level data rates, $10^7/km^2$ connection densities, and lower than 0.1 $ms$ latency\cite{zhang20196g}. 
\blue{To achieve these goals, the International Telecommunication Union (ITU) has defined six key use cases for envisioned 6G networks\cite{itu-2030}. Specifically, three cases are extensions of IMT-2020 (5G), namely immersive communication, hyper-reliable and low-latency communication, and massive communication, and the other three novel usage cases are ubiquitous connectivity, integrated sensing and communication, and AI and communication.} 
These novel techniques have shown satisfactory performance towards 6G requirements, but the complexity of network management also significantly increased.
From 3G, 4G LTE to 5G and envisioned 6G networks, telecommunication (telecom) networks have become a complicated large-scale system, including core networks, transport networks, network edge, and radio access networks\cite{zhou2022knowledge}. 
\blue{Moreover, 6G ubiquitous connectivity aims to address presently
uncovered areas, e.g., rural and sparsely populated areas, by integrating other access systems such as satellite communications.}
\blue{In addition, 6G integrated sensing and communication is designed to improve applications requiring sensing capabilities, i.e., assisted navigation, activity detection, and environmental monitoring.}
\blue{Despite the potential benefits, such highly integrated network architecture and functions may lead to a huge burden on} 6G network management, including network configuration and troubleshooting, product design and coding, standard specification development, performance optimization and prediction, etc.

To handle such complexity, machine learning (ML) has become one of the most promising solutions, and there have been a large number of studies on artificial intelligence (AI)/ML-enabled wireless networks, e.g., reinforcement learning-based network management\cite{zhou2021ran}, deep neural network-enabled channel state information (CSI) prediction\cite{luo2018channel}, and federated learning for distributed model training in wireless environments\cite{zhang2022federated}. 
For example, convex optimization has been applied to optimize network performance, but it requires problem-specific transformation for convexity. By contrast, reinforcement learning will transform the problem into a unified Markov decision process (MDP), and then interact with the environment to explore optimal policies. Compared with conventional optimization algorithms\cite{zhou2023survey}, reinforcement learning overcomes the complexity of dedicated problem reformulation, and can better handle environmental uncertainties, e.g., the growing diversity of user preferences, and more distributed and heterogeneous resources in future telecom networks.
These studies have demonstrated the importance of incorporating ML to improve the efficiency, reliability, and quality of telecom services.

Recently, large language model (LLM) techniques have attracted considerable interest from both academia and industry. 
Unlike previous ML algorithms, these large-scale models with a huge amount of parameters have shown versatile comprehension and reasoning capabilities in various fields such as health care\cite{singhal2023towards}, law\cite{colombo2024saullm}, finance\cite{wu2023bloomberggpt}, education, and so on \cite{zhao2023survey}. 
For instance, Wu \textit{et al.} introduced a BloombergGPT model that is trained on a wide range of financial data with 50 billion parameters, and the Med-PaLM2 developed by Google achieves 86.5\% correct rate on the medical question answering dataset \cite{singhal2023towards}.
\blue{LLM technologies have many promising features such as in-context learning (ICL), step-by-step learning, and instruction following\cite{Wei-2023-arXiv-symbol}}. 
\blue{Existing studies have shown that LLMs can answer telecom-domain questions, generate troubleshooting reports, develop project code, and configure networks, which will significantly lower the difficulty of 6G ubiquitous connectivity management.} 
\blue{Meanwhile, for 6G integrated sensing and communication, LLMs can understand and process multi-modal data, e.g., text, satellite or street camera images, 3D LiDAR maps and videos. It provides a promising approach to simulate and understand the 3D wireless signal transmission environment.}

Despite the great potential, LLM's real-world application is still at a very early stage, especially for domain-specific scenarios.
For instance, telecom is a broad field that includes various knowledge domains, e.g., signal transmissions, protocols, network architectures, devices, and different standards. 
LLM is expected to properly understand and generate content that aligns with real-world details and specific requirements of telecom applications\cite{bariah2023understanding}. However, such specific telecom-related requirements are rare in the existing knowledge base of general-domain LLMs.
Therefore, applying a general-domain LLM directly to telecom tasks may lead to poor performance. 
Meanwhile, fine-tuning \blue{LLMs} on telecom datasets may improve LLM's performance of domain-specific tasks, but the telecom-specific dataset collection and filtering still require careful design and evaluation.
In addition, many telecom tasks require multi-step planning and thinking, e.g., a simple coding task can include multiple steps, indicating dedicated prompting and analyses from the telecom perspective\cite{du2023power}.

Given the above opportunities and challenges, this work presents a comprehensive survey of LLM-enabled telecom networks. 
Different from existing studies that focus on one specific aspect such as edge intelligence \cite{shen2024large,lin2023pushing}, grounding and alignment\cite{xu2024large}, this work provides a comprehensive survey on fundamentals, key techniques, and applications of LLM-enabled telecom. 
\blue{To be specific, this work focuses on generative models that were originally developed for language tasks, i.e., language models, and it also involves more diverse techniques and broad application scenarios such as optimization and prediction problems.}
\blue{In this survey, the term “foundation models” refers to models specifically developed from scratch for applications that are beyond pure language-related tasks, such as the prediction foundation models in Section VII-B, while “LLM-enabled” or “LLM-aided” approaches denote methods that repurpose existing pre-trained language models for telecom tasks. Moreover, when referring to LLMs, it means that the inputs to the model are purely text, and the model generates purely text as outputs, even if the model can accept inputs in other modalities, such as GPT-4V and GPT-4o. When discussing the multi-modal inputs, we explicitly describe them as multi-modal large language models or multi-modal LLMs.}

Although LLM development is originally motivated by natural language tasks, it is worth noting that there have been diverse state-of-the-art explorations that are beyond the conventional language processing tasks, e.g., coding and debugging \cite{xiang2023toward}, recommendation \cite{li2023prompt}, LLM-enabled agents \cite{xi2023rise}, instruction-based optimization \cite{yang2023large}, network time-series prediction and decision making \cite{wu2024large}, etc. 
These LLM-inspired techniques have become crucial pillars of LLM studies, and exploring these techniques is crucial to take full advantage of LLM capabilities.  
Fig.\ref{fig1} presents the organization of this work, in which the left side indicates the telecom scenarios and demand, and the right side shows the LLM-enabled techniques.
To better present the detailed application scenarios, the bottom of Fig.\ref{fig1} shows telecom environments that include radio access networks, network edge, central cloud, and other network elements such as regular users, malicious users, mmWave beam, environment image sensing, RISs, backhaul traffic, etc. 
\blue{Meanwhile, we categorize key telecom applications into generation, classification, optimization, and prediction problems to better distinguish different scenarios and customized designs\footnote{\blue{Note that although the classification, optimization, and prediction capabilities are all based on the LLM's inference and generation capabilities, this organization can significantly reduce the reader's difficulty in understanding the LLM's potential for telecom applications.}}}

In particular, we focus on the following topics:

1) \textbf{LLM fundamentals}: 
Understanding LLM fundamentals is the prerequisite for developing advanced applications in telecom networks\cite{zhao2023survey}. 
Compared with existing studies\cite{shen2024large,lin2023pushing,xu2024large}, this work presents a more comprehensive overview of the model architecture, pre-training, fine-tuning, inference and utilization, and evaluation.  
Additionally, it presents different approaches to deploy LLMs in telecom networks, such as central cloud, network edge, and mobile LLM \cite{lin2023pushing}. It further analyzes LLM fundamentals from the telecom application perspective, e.g., training or fine-tuning telecom-specific \blue{LLMs}, and the importance of prompting and multi-step planning techniques for telecom tasks.

2) \textbf{LLM for generation problems in telecom}:
Generating desired content is the most common usage of LLM, and here we investigate the applications to specific telecom scenarios. 
In particular, it involves answering telecom-domain questions, generating troubleshooting reports, project coding, and network configuration.  
It shows that LLM's generation capabilities are particularly useful in text and language-related telecom tasks to save human effort, e.g., automated code refactoring and design\cite{du2023power}, recommending troubleshooting solutions\cite{bosch2022integrating}, and generating network configurations\cite{DBLP:journals/corr/abs-2401-06786}.

3) \textbf{LLM-based classification for telecom}: Classification is a common task in the telecom field, and we present LLM-enabled network attack classification and detection, telecom text, image, and traffic classification problems. 
For instance, there have been many studies on visioned-aided blockage prediction and beamforming for 6G networks \cite{charan2021vision}, and some LLM can provide \blue{zero-shot} image classification capabilities, overcoming the training difficulties of conventional algorithms in complicated signal transmission environments \cite{matsuuravisual}.

4) \textbf{LLM-enabled optimization techniques}: Optimization techniques are of great importance to telecom networks, e.g., resource allocation and load balancing\cite{zhou2023survey}, and LLM offers new opportunities \cite{zhou2023survey}. 
In particular, we introduce LLM-aided automated reward function design for reinforcement learning, verbal reinforcement learning, LLM-enabled black-box optimizer, end-to-end convex optimization, and LLM-aided heuristic algorithm design.
For example, reinforcement learning has been widely used for network optimization, but the reward functions are usually manually designed with a trial-and-error approach\cite{booth2023perils}. LLM can provide automated reward function designs, and such an improvement can significantly promote reinforcement learning applications in the telecom field.

\begin{table*}[!t]
\caption{\blue{Comparison of this work with existing surveys}}
\centering
\small
\setstretch{1.3}
\begin{threeparttable} 
\resizebox{1\textwidth}{!}{%
\begin{tabular}{|m{0.65cm}<{\centering}|m{0.8cm}<{\centering}|m{0.8cm}<{\centering}|m{0.8cm}<{\centering}|m{0.7cm}<{\centering}|m{0.8cm}<{\centering}|m{0.8cm}<{\centering}||m{1.35cm}<{\centering}|m{1.1cm}<{\centering}|m{0.8cm}<{\centering}|m{1.1cm}<{\centering}||m{1.1cm}<{\centering}|m{0.5cm}<{\centering}|m{0.7cm}<{\centering}|m{1cm}<{\centering}||m{1.3cm}<{\centering}|m{0.7cm}<{\centering}|m{0.84cm}<{\centering}|m{1.05cm}<{\centering}||m{1.5cm}<{\centering}|m{1.2cm}<{\centering}|}
\hline 
\multirow{4}*{Ref.} & \multicolumn{6}{c||}{ \textbf{LLM fundamental techniques}}&\multicolumn{4}{c||}{ \textbf{Generation applications}} & \multicolumn{4}{c||}{ \textbf{Classification applications }}  & \multicolumn{4}{c||}{ \textbf{Optimization techniques}}  & \multicolumn{2}{c|}{ \textbf{\makecell{Prediction \\ techniques}}} \\
\cline{2-21} 
& Archit- ecture & Pre-training & Fine-tuning  & Infe- rence  & Evalu- ation & Deploy- ment  & Question answering & Troubles- hooting  &  Coding  & Network config.  & Network attacks &  Text  &  Image    &  Network traffic  &  \makecell{LLM \\-aided RL}  & Black- box    & Convex  & Heuristic  & Time series LLM  &  Multi- modality  \\
\hline
\cite{shen2024large} &  & \checkmark &  &  &  & \checkmark &  &  &  &  &  &   &  &  &  &   &  &   &  & \\
\hline
\cite{lin2023pushing} &  & \checkmark &  &  &  & \checkmark &  &  &  &  &  &   &  &  &  &   &  &   &  & \\
\hline
\cite{xu2024large} &  &  &  &  &  & \checkmark &  &  &  &  &  &   &  &  &  &   &  &   &  & \checkmark\\
\hline
\cite{tarkoma2023ai}   &  & \checkmark & \checkmark  &    &    &  \checkmark  &  &  & \checkmark & \checkmark &  &   &  & \checkmark &  &   &  &   &  & \\
\hline
\cite{chen2023big}  &    &  \checkmark  &  &  &  & \checkmark &  &  &  &  &  &   &  &  &  &   &  &   &  & \checkmark\\
\hline
\cite{bariah2023large} &  &  &  &  &  &  &  &  &  &  &  &   & \checkmark &  &  &   &  &   &  & \checkmark \\
\hline
\cite{xu2024large2}   &  & \checkmark & \checkmark &  &  &  &  &  &  &  &  &   &  &  &  &   &  &   &  & \checkmark \\
\hline
\blue{\cite{javaid2024leveraging}}   &  & \blue{\checkmark} &  &  &  & \blue{\checkmark} &  &  &   & \checkmark &  &   &  &  & \blue{\checkmark}  &   &  &   &  &  \blue{\checkmark} \\
\hline
\blue{\cite{maatouk2024large}}   &  &  \blue{\checkmark}  & \blue{\checkmark}  &  &  &  &  & \blue{\checkmark}  &  & \blue{\checkmark}  &  &   &  &  &  &   &  &   &  &  \\
\hline
\blue{\cite{shao2024wirelessllm}}   &  \blue{\checkmark}  & \blue{\checkmark}  &  \blue{\checkmark} & \blue{\checkmark} &  & \blue{\checkmark} &  &  &  &  &  &   &  &  &  &   &  &   &  &  \\
\hline
\blue{\cite{huang2023large}}  &  & \blue{\checkmark} & \blue{\checkmark} & \blue{\checkmark} &  &  &  &  &  & \blue{\checkmark} &  &   &  &  &  &   &  &   &  &  \\
\hline
Our work &  \checkmark  &  \checkmark  &  \checkmark  &  \checkmark  &  \checkmark  &  \checkmark    & \checkmark   &  \checkmark  &  \checkmark   &  \checkmark   &   \checkmark  &  \checkmark  &  \checkmark  & \checkmark &  \checkmark   &  \checkmark   &  \checkmark  &  \checkmark   &  \checkmark & \checkmark\\
\hline
\end{tabular}}   
\begin{tablenotes}    
        \footnotesize       
        \item[1] Multi-modality is discussed in several existing studies but not from the prediction perspective. Table \ref{tab1} divides key topics from LLM fundamentals to \\ optimization and prediction to better align with the organization of our work.   
\end{tablenotes} 
\end{threeparttable}  
\label{tab1}
\end{table*}

5) \textbf{LLM-aided prediction in telecom}:
Prediction techniques are crucial for telecom networks, such as CSI prediction\cite{luo2018channel}, prediction-based beamforming\cite{charan2021vision}, and traffic load prediction\cite{wang2020cellular}. Existing studies have started exploring one-model-for-all time-series models. After pre-training on a large corpus of diverse time-series data, such a model can learn the hidden temporal patterns, and then generalize well across different prediction tasks without extra training. 
We will first introduce how to pre-train foundation models, and then present frozen pre-trained and fine-tune-based \blue{LLMs}. In addition, the potential of multi-modal LLM is discussed for telecom prediction tasks.

6) \textbf{Challenges and future directions}:
Finally, we identify the challenges and future directions of LLM-empowered telecom. The challenges focus on telecom-domain \blue{LLM} training, practical LLM deployment, and prompt engineering for telecom applications. The future directions include LLM-enabled planning, model compression and fast inference, overcoming hallucination problems, retrieval augmented-LLM, and economic and affordable \blue{LLMs}.

In summary, the main contribution of this work is that we provide a comprehensive survey of the principles, key techniques, and applications for LLM-enabled telecom networks, ranging from LLM fundamentals to novel LLM-inspired generation, classification, optimization and prediction techniques along with telecom applications.
This work covers nearly 20 telecom application scenarios and LLM-inspired novel techniques, aiming to be a roadmap for researchers to use LLMs to solve various telecom tasks. 
The rest of this paper is organized as Fig. \ref{fig1}. Section \ref{sec-related} discusses related surveys, and Section \ref{sec-fundamental} presents LLM fundamentals. \blue{Sections} \ref{sec-gene}, \ref{sec-class}, \ref{sec-optimize}, and \ref{sec-predict} focus on generation, classification, optimization, and prediction problems and telecom applications, respectively.
Finally, Section \ref{sec-challenge} identifies the challenges and future directions, and Section \ref{sec-conc} concludes this work.

\section{Related Surveys}
\label{sec-related}

\blue{Table \ref{tab1} compares this work with existing studies \cite{shen2024large,lin2023pushing,xu2024large,tarkoma2023ai,chen2023big,bariah2023large,xu2024large2,javaid2024leveraging,maatouk2024large,shao2024wirelessllm,huang2023large},} including LLM fundamental techniques such as pre-training and fine-tuning, and other key topics ranging from question answering to multi-modality. 
Firstly, Table \ref{tab1} shows that most existing studies focus on the fundamental techniques of \blue{LLMs}, e.g., pre-training \blue{LLMs} for telecom tasks in general \blue{\cite{shen2024large,lin2023pushing,javaid2024leveraging,maatouk2024large,shao2024wirelessllm,huang2023large}}.
LLM deployment is discussed in many existing studies, including central cloud \cite{shen2024large,xu2024large2}, network edge \cite{lin2023pushing}, and mobile execution\cite{xu2024large}. 
Due to the storage and computational resources constraint at the network edge, Lin \textit{et al.} also summarized various techniques in  \cite{lin2023pushing} that can be used to improve the LLM training efficiency at the network edge, such as parameter-efficient fine-tuning, split edge learning, and quantized training.

\blue{Meanwhile, researchers have investigated various network application scenarios for LLM and generative AI (GAI), such as integrated satellite-aerial-terrestrial networks~\cite{javaid2024leveraging}, secure physical layer communication~\cite{zhao2024generative}, semantic communication~\cite{liang2023generative}, and vehicular networks~\cite{zhang2024generative}. For instance, Javaid \textit{et al.} studied the application of LLMs to integrated satellite-aerial-terrestrial networks, including resource allocation, traffic routing, network optimization, etc \cite{javaid2024leveraging}. Huang \textit{et al.} presents a general overview of LLM for networking, involving network design, configuration, and security \cite{huang2023large}.
Du \textit{et al.} present a novel concept named “AI-generated everything”, discussing the interactions between (GAI) and different network layers~\cite{du2024age}.}
In addition, sensing has become an important part of future 6G networks, and the multi-modal LLM are discussed in several existing studies, e.g., integrated sensing and communication with LLM\cite{xu2024large, chen2023big}, multi-modal input to \blue{LLMs} for intelligent sensing and communication\cite{bariah2023large}, and multi-modal sensing\cite{xu2024large2}.
These studies are very valuable explorations of LLM-enabled telecom networks by focusing on model training and deployment. 
However, LLM techniques are rapidly progressing and many LLM-inspired novel techniques and applications have been recently proposed. 
This work is different from existing studies in the following aspects:

1) In terms of LLM fundamentals, we provide comprehensive overviews and analyses, ranging from model architecture and pre-training to LLM evaluation and deployment. 
For instance, prompt engineering is of great importance for using LLM technology, but some crucial techniques such as \blue{chain-of-thought (CoT)}\cite{wei2022chain} and step-by-step planning are not discussed in many existing studies\cite{shen2024large,lin2023pushing,xu2024large,tarkoma2023ai,chen2023big,bariah2023large,xu2024large2}. Understanding these prompt design skills is the prerequisite for advanced telecom applications. 
By contrast, this work provides detailed analyses of chain-of-thought along with telecom applications, e.g., LLM-aided automated wireless project coding with multi-step prompting and thinking\cite{du2023power}.
Meanwhile, we also systemically analyzed the features of different LLM deployment strategies in telecom, while existing studies usually involve one single deployment\cite{shen2024large,xu2024large2, lin2023pushing, xu2024large}.

2) In terms of LLM-inspired techniques, this work presents the most state-of-the-art novel algorithms and designs. For instance, reinforcement learning has been widely applied to telecom optimization problems, but the reward function design requires considerable human effort\cite{booth2023perils}. 
Existing studies have shown that LLM can be used for automated reward function design, achieving a comparable performance as human manual designs\cite{song2023self,kwon2023reward,ma2023eureka}. 
Such a technique may bring revolutionary changes to reinforcement learning techniques, which have great potential for telecom applications.  
In addition, time-series LLM is also a promising technique for telecom, enabling one-model-for-all prediction\cite{gruver2023large}. 
However, these novel techniques are not mentioned in most existing studies.

\definecolor{Abbey}{rgb}{0.301,0.317,0.337}
\begin{table}
\centering
\caption{Summary of existing general and domain-specific \blue{LLMs}.}
\label{tab:resource_model}
\setstretch{0.9}
\resizebox{0.45\textwidth}{!}{%
\begin{tblr}{
  column{1} = {c},
  column{2} = {c},
  column{3} = {c},
  column{4} = {c},
  column{5} = {c},
  cell{2}{1} = {r=8}{},
  cell{10}{1} = {r=3}{},
  cell{13}{1} = {r=3}{},
  cell{16}{1} = {r=3}{},
  cell{19}{1} = {r=3}{},
  cell{22}{1} = {r=3}{},
  cell{25}{1} = {r=3}{},
  vlines,
  hline{1-2,10,13,16,19,22,25, 28} = {-}{},
}
\textbf{Domain}         & \textbf{Model}    & \textbf{Size} & \textbf{Pre-train} & \textbf{Latest Update} \\
General~                & GPT-4-Turbo       & -                 & -                  & Mar 2024             \\
                        & Claude-3 Opus~    & ~-                & ~-                 & Mar 2024             \\
                        & Gemini-1 Ultra    & -                 & -                  & Dec 2023             \\
                        & Mistral-Large     & -                 & -                  & Feb 2024             \\
                        & Llama-2           & 70B               & 10T tokens         & Jul 2023             \\
                        & Qwen-1.5          & 72B               & -                  & Feb 2024             \\
                        & DeepSeek          & 67B               & 2T tokens          & Jan 2024             \\
                        & Baichuan-2 Turbo  & 13B               & -                  & Sep 2023             \\
Healthcare              & MedGPT            & -                 & -                  & Jul 2021             \\
                        & ChatDoctor        & 7B                & 100K               & Jun 2023             \\
                        & Med-PaLM         & 540B               & 760B                  & Dec 2022             \\
Finance                 & finBERT           & 110B              & 1.8M               & Aug 2019             \\
                        & FinMA             & 30B               & 1T tokens          & Jun 2023             \\
                        & BloombergGPT      & 50B                & 569-770B tokens    & Dec 2023             \\
Time Series             & TabLLM            & 3B                & ~50,000 rows       & Mar 2023             \\
                        & LLMTime           & 70B               & -                  & Oct 2023             \\
                        & TIME-LLM          & 7B                & -                  & Jan 2024             \\
{Autonomous \\driving } & Driving with LLMs & 7B                & 110k               & Oct 2023             \\
                        & Dilu              & -                 & -                  & Feb 2024             \\
                        & DriveGPT4         & 70B               & 112k               & Mar 2024             \\
{Law }                  & LexiLaw           & 6B                & -                  & May 2023             \\
                        & JurisLMs          & 13B               & -                  & July 2023             \\
                        & ChatLaw           & 13B               & 980k               & July 2023             \\
{Recommen- \\dation }   & M6-Rec            & 300M              & 1G                 & May 2022             \\
                        & TallRec           & 7B                & ~100 samples       & Oct 2023             \\
                        & AgentCF           & 175B              & 20k samples        & Oct 2023             
\end{tblr}}
\label{tab-llmsummary}
\end{table}

3) In terms of telecom applications, we systematically summarize various LLM application scenarios, including question answering, network troubleshooting, coding, network configuration, network attack classification and security, text and image classification, etc.
Compared with existing studies, we presented more comprehensive overviews and analyses of using LLM techniques to solve various problems in the telecom domain. 
For each application, this work provides technical details such as framework, pre-training steps, and prompt designs, which are more informative than existing studies that focus on general system-level designs.

\blue{Moreover, Table \ref{tab-llmsummary} summarizes various general and domain-specific LLMs, demonstrating that LLMs have received considerable attention across many fields.
Researchers have trained various domain-specific LLMs for their application scenarios, including healthcare\cite{li2023chatdoctor}, finance\cite{wu2023bloomberggpt}, time series\cite{jin2024timellm}, autonomous driving\cite{xu2023drivegpt4}, and recommendation systems\cite{zhang2023agentcf}, etc. 
For instance, DriveGPT4 is designed to provide interpretable end-to-end
autonomous driving\cite{xu2023drivegpt4}.
LLM is also used for the automated design of reward functions in robot control\cite{ma2023eureka}, achieving better performance than human manual designs.
%
%
Thus, given the rapid progress and great potential of LLMs, a comprehensive survey is expected to summarize the latest and potential applications of LLMs in the telecom field.} 
To this end, this work answers such a question: What are the most state-of-the-art techniques inspired by \blue{LLMs}, and how can these techniques be used to solve telecom domain problems?
The answer to this question is crucial for building intelligent next-generation telecom networks.

\section{LLM Fundamentals}
\label{sec-fundamental}
\begin{figure}[!t]
\centering
\includegraphics[width=0.95\linewidth]{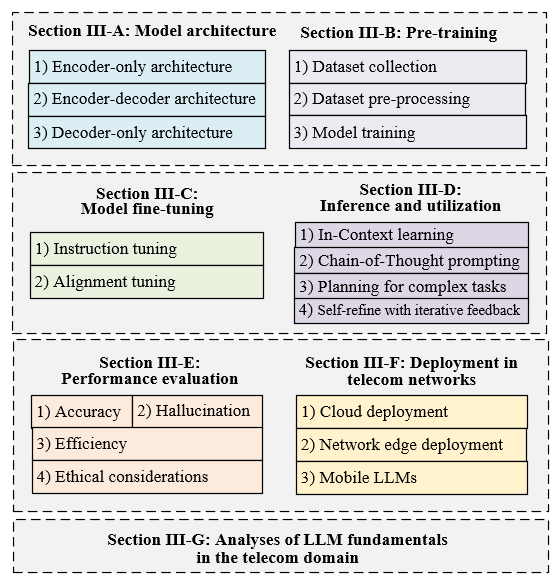}
\caption{\blue{Organization and key topics of Section \ref{sec-fundamental}}.}
\label{fig-fundamental}
\end{figure}

This Section will introduce LLM fundamentals, and the overall organization is shown in Fig.\ref{fig-fundamental}. 
It presents a thorough overview of LLM fundamentals, including the model architecture, pre-training, fine-tuning, inference and utilization, and model evaluation. We further discuss how \blue{LLMs} can be deployed in telecom networks such as central cloud, network edge, and mobile devices. Finally, we analyze LLM fundamentals from the telecom application perspective, e.g., training or fine-tuning \blue{LLMs} for the telecom domain.

\subsection{Model Architecture}
\label{sec-modelarchi}
The fundamental component of contemporary \blue{LLMs} is the transformer scheme~\cite{vaswani2017attention}, which leverages an attention mechanism to capture global dependencies between inputs and outputs.
%
%
%
%
%
Transformers process raw inputs by tokenizing them and applying embeddings and positional encodings. 
The vanilla transformer architecture comprises two main components: the encoder and the decoder. 
The encoder's role is to extract features and understand the relationships among all input tokens. 
It employs self-attention, also referred to as bidirectional attention, allowing each token to attend to every other token in both directions. 
Conversely, the decoder is responsible for producing the output sequence while taking into account the input sequence and previously generated tokens. 
It initially applies a masked attention mechanism, known as causal attention, ensuring that the current token only attends to previously generated tokens.
Additionally, the decoder employs cross-attention, where the query comes from the decoder, and the key and value are from the encoder, enabling the decoder to integrate information from both the input sequence and the already generated tokens.
Beyond the basic version of the attention mechanism, various variants are developed to capture the different relationships among tokens, such as multi-head attention~\cite{vaswani2017attention}, multi-query attention~\cite{shazeer2019fast}, and grouped-query attention~\cite{ainslie2023gqa}.
Current architectures can be classified into three distinct categories: encoder-only architecture, encoder-decoder architecture, and decoder-only architecture.

\subsubsection{\rm \textbf{Encoder-only architecture}}

Models with an encoder-only structure solely comprise an encoder. These models are tailored for language understanding tasks, where they extract language features for downstream applications such as classification. A prominent example is bidirectional encoder representations from transformers (BERT)~\cite{devlin2018bert}. BERT is pre-trained with two main objectives: the masked language model objective, which aims to reconstruct randomly masked tokens, and the next sentence prediction objective, designed to ascertain if one sentence logically follows another.
There have been many variants of this model, such as RoBERTa~\cite{liu2019roberta}, which enhances the performance on downstream tasks\footnote{Here downstream tasks refer to a series of target tasks that can be solved by the pre-trained model, e.g., text classification, natural language inference.}, 
and ALBERT~\cite{lan2019albert} that introduces two parameter-reduction techniques to accelerate BERT's training process.

\subsubsection{\rm \textbf{Encoder-decoder architecture}}
The foundational transformer block employs an encoder-decoder architecture, wherein the encoder relays keys and values generated by its self-attention module to the decoder for cross-attention processing. For example, the study in \cite{raffel2020exploring} introduces the text-to-text transfer transformer, a unified framework that reformulates all text-based language tasks into a text-to-text format, thereby facilitating the exploration of transfer learning within natural language processing (NLP).
BART is another well-known model with standard transformer architecture\cite{lewis2019bart}, which employs a denoising autoencoder approach for pre-training sequence-to-sequence models. It introduces arbitrary noise into text and is trained to reconstruct the original content, effectively combining elements of BERT's bidirectional encoding and GPT's causal decoding methodologies.

\subsubsection{\rm \textbf{Decoder-only architecture}}
Decoder-only architectures specialize in unidirectional attention, allowing each output token to attend only to its past tokens and itself. Both prefix and output tokens undergo identical processing within the decoder. Decoders are further distinguished based on their attention mechanisms into causal and non-causal decoders. In causal decoders, every token is restricted to attending to its past tokens and itself; in non-causal decoders, prefix tokens can attend to all tokens within the prefix.
Causal decoders are predominantly adopted in popular \blue{LLMs}, such as the GPT series~\cite{brown2020language}, PaLM~\cite{chowdhery2023palm}, and  LLaMA~\cite{touvron2023llama}. 
Non-causal decoders~\cite{zeng2022glm} resemble encoder-decoder frameworks in their ability to bidirectionally process the prefix sequence and autoregressively generate output tokens sequentially.

\blue{Note that LLM is a complicated system, and there are multiple approaches to apply LLMs to the telecom field, ranging from pre-training and fine-tuning, to prompting. For instance, pre-training an LLM from scratch by using telecom-domain datasets, fine-tuning a general domain LLM for specific telecom tasks, or using general domain LLMs by prompting directly. The following will introduce the key procedures and features of each approach, in which Sections \ref{sec-pretrain}, \ref{sec-finetune}, and \ref{sec-prompt} introduce the procedures of pre-training, fine-tuning, and prompting, respectively.}

\subsection{LLM Pre-training}
\label{sec-pretrain}
The aim of pre-training language models is to predict the next word within a sentence. After being trained on extensive datasets, \blue{LLMs} exhibit emergent capabilities in comprehension and reasoning.
This subsection will introduce dataset collection, preprocessing, and model training techniques.

\subsubsection{\rm \textbf{Dataset collection and preprocessing}}
Datasets for training language models fall into two primary categories: general and specialized.
General datasets comprise a diverse range of sources, such as web pages, literature, and conversational corpora. For instance, web pages like Wikipedia\cite{guo2020wiki} can contribute to a language model's broad linguistic understanding.
Meanwhile, literary works also serve as a rich reservoir of formal and lengthy texts\cite{beltagy2019scibert}. These materials are crucial for teaching \blue{LLMs} complex linguistic constructs, facilitating the modelling of long-range dependencies.
Specialized data involves scientific texts and programming-related data. For example, scientific literature comprises a wealth of formal writing imbued with domain-specific knowledge, encompassing academic papers and textbooks. On the other hand, programming data drawn from online question-answering platforms like Stack Exchange \cite{xu2022systematic}, along with public software repositories such as GitHub, provide raw material rich with code snippets, comments, and documentation.
Incorporating these specialized texts into the training of \blue{LLMs} can significantly improve LLM's performance in reasoning and domain-specific knowledge applications.
However, before pre-training, it is critical to preprocess the collected datasets, which often contain noisy, redundant, irrelevant, and potentially harmful data. The preprocessing procedure may include quality filtering, de-duplication\cite{chowdhery2023palm}, privacy redaction\cite{laurenccon2022bigscience}, and tokenization \cite{kudo2018subword}.

\subsubsection{\rm \textbf{Model training}}
In the model training process, two pivotal hyperparameters are the batch size and the learning rate. For the pre-training of \blue{LLMs}, a substantial batch size is required, and recent studies suggest incrementally enlarging the batch size to bolster the stability of the training process~\cite{chowdhery2023palm}.
In terms of learning rate adjustments, a widely used strategy is to start with a warm-up phase and then succeed with a cosine decay pattern. This approach helps in achieving a more controllable learning rate schedule.
To enhance the training scalability, several key techniques are proposed. For instance, 3D parallelism encompasses data parallelism, pipeline parallelism, and tensor parallelism. Data parallelism involves the replication of the model's parameters and optimizer states across multiple GPUs\cite{li2021terapipe}, allocating to each GPU a subset of data to process and subsequently aggregate the computed gradients. Pipeline parallelism, as detailed in \cite{huang2019gpipe}, assigns distinct layers of an LLM to various GPUs, allowing the accommodation of larger models within the confines of GPU memory. Tensor parallelism operates on a similar premise by decomposing the tensors\cite{shoeybi2019megatron}, especially for the parameter matrices of LLM, facilitating the distribution and computation across multiple GPUs.   
Meanwhile, ZeRO is also a useful technique \cite{rajbhandari2020zero}, which conserves memory by retaining only a portion of the model's data on each GPU. The remainder of the data is accessible across the GPU network, as needed, effectively addressing memory redundancy concerns.

\subsection{LLM Fine-tuning}
\label{sec-finetune}
Fine-tuning refers to the process of updating the parameters of pre-trained \blue{LLMs} to adapt to domain-specific tasks. Although the pre-trained LLM already has vast language knowledge, they lack specialization in specific areas. Fine-tuning overcomes this limitation by allowing the model to learn from domain-specific datasets, making the LLM more effective on specific applications. This subsection will introduce two fine-tuning strategies: instruction and alignment tuning.

\subsubsection{\rm \textbf{Instruction tuning}}
\label{sec-instruction}
Instruction tuning is a method for fine-tuning pre-trained \blue{LLMs} using a collection of natural language-formatted instances. This technique aligns closely with supervised fine-tuning and multi-task prompted training, enhancing the LLM's ability to generalize to unseen tasks, even in multilingual contexts~\cite{Wei-ICLR-2022-Finetuned}. The process involves collecting or constructing instruction-formatted instances and employing these to fine-tune \blue{LLMs} in a supervised manner, typically using sequence-to-sequence loss for training. Models like InstructGPT and GPT-4 have demonstrated the effectiveness of instruction tuning in meeting real user needs and improving task generalization~\cite{Ouyang-arxiv-2022-Training,OpenAI-OpenAI-2023-GPT-4v}.
Instruction-formatted instances usually consist of a task description, an optional input, a corresponding output, and possibly a few examples as demonstrations. These instances can originate from various sources, such as traditional NLP task datasets, daily chat data, and synthetic data. Existing research has reformatted traditional NLP datasets with natural language task descriptions to aid \blue{LLMs} in understanding tasks, proving particularly effective in enhancing task generalization capabilities~\cite{Wei-ICLR-2022-Finetuned}. 
The design and quality of instruction instances significantly will impact the model's performance. Scaling the instructions, for instance, tends to improve generalization ability up to a certain point, beyond which additional tasks may not yield further gains~\cite{chung2022scaling}. Diversity in task descriptions and the number of instances per task are also critical, with a smaller number of high-quality instances often sufficing for significant performance improvements~\cite{Wei-ICLR-2022-Finetuned}.

\subsubsection{\rm \textbf{Alignment tuning}}
\label{sec-alignment}
Alignment tuning aims to ensure \blue{LLMs} adhere to human values, preventing outputs that could be harmful, biased, or misleading. This concept emerges from the realization that while the LLM excels in various NLP tasks, they may inadvertently generate content that deviates from ethical norms or human expectations~\cite{brown2020language}. 
Collecting human feedback is central to the alignment-tuning process. In particular, it involves curating responses from diverse human labellers to guide the LLM toward generating outputs that align with the predefined criteria. Approaches to collecting this feedback include ranking-based methods, where labellers evaluate the quality of model-generated outputs, and question-based methods, where labellers provide insights on specific aspects of the outputs, such as their ethical implications~\cite{Ziegler-arxiv-2019-Fine-Tuning}.

A prominent technique in alignment tuning is reinforcement learning from human feedback (RLHF), where the model is fine-tuned using reinforcement learning algorithms based on human feedback. This process typically starts with supervised fine-tuning using human-annotated data, followed by training a reward model that reflects human preferences, and finally, fine-tuning the LLM using this reward model. Despite its effectiveness, RLHF can be computationally intensive and complex, necessitating alternative approaches for practical applications~\cite{Christiano-NeurIPS-2017-Deep}.
An alternative method for RLHF is direct optimization through supervised learning, which bypasses the complexities of reinforcement learning. This method relies on constructing a high-quality alignment dataset and directly fine-tuning \blue{LLMs} to adhere to alignment criteria. Although less resource-intensive than RLHF, this approach requires careful dataset construction and may not capture the full range of human values and preferences as effectively as RLHF~\cite{zhou-arxiv-2023-lima}.
\blue{Additionally, researchers have introduced the direct preference optimization (DPO) technique~\cite{rafailov2023directpreferenceoptimizationlanguage}, which eliminates the need for a reward model and allows the model to align directly with preference data. Some advanced LLMs, like Llama 3, utilize both RLHF with proximal policy optimization (PPO) and DPO. However, both RLHF and DPO depend on high-quality human preference data, which is limited and costly to acquire. To address this challenge, methods such as Constitutional AI~\cite{bai2022constitutionalaiharmlessnessai} and RL from AI feedback (RLAIF) ~\cite{lee2023rlaifscalingreinforcementlearning} have been developed to generate preference data using LLMs, enabling models to learn from AI feedbacks and facilitating knowledge transfer between models.}

\subsection{LLM Inference and Utilization by Prompting}
\label{sec-prompt}
\begin{figure*}[t]
\centering
\setlength{\abovecaptionskip}{0pt} 
\includegraphics[width=0.98\linewidth]{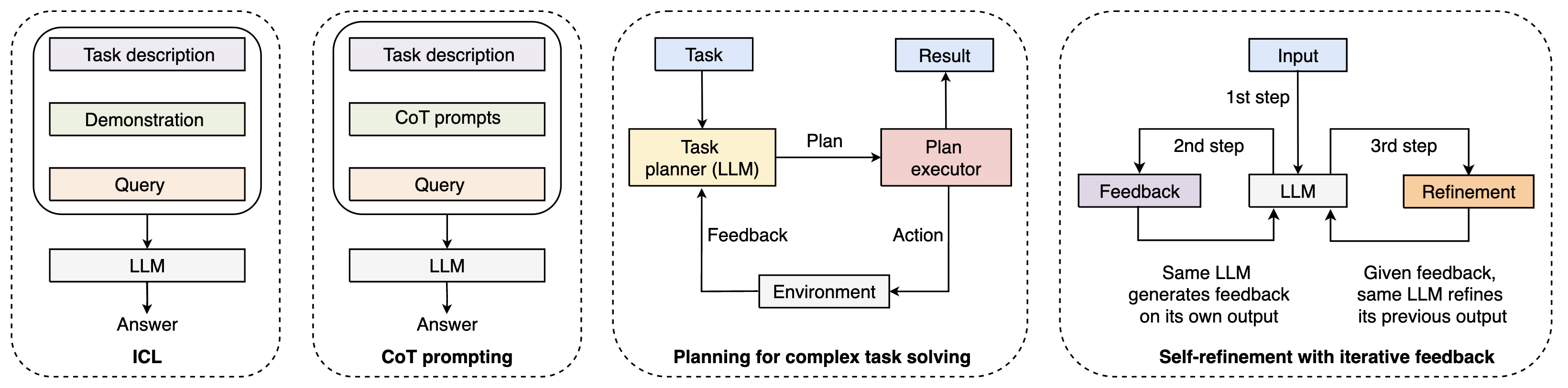}
\caption{\blue{Comparison of various prompt engineering techniques~\cite{zhao2023survey, madaan2024self}. Different from task-specific examples in ICL demonstrations, CoT prompting additionally incorporates intermediate reasoning steps in demonstrations. Prompt-based planning breaks down intricate tasks into manageable sub-tasks and outlines action sequences for their resolution. Self-refine enhances LLM's outputs through iterative feedback and refinement.}}
\label{fig-prompting}
\end{figure*}

Prompt engineering is the process in which users design various inputs for AI models to generate desired outputs.
Compared with fine-tuning, prompting has no requirements for extra training, producing output instantly based on user inputs. 
It indicates a straightforward approach to using \blue{LLMs}, and the rapid response and training-free features make it a promising method for telecom applications. 
\blue{This subsection will introduce key techniques in prompt engineering, including ICL, CoT prompting, LLM for complex planning, and self-refinement with iterative feedback.
The comparisons among different prompt engineering techniques are shown in Fig.~\ref{fig-prompting}.}

\subsubsection{\rm \textbf{In-context learning (ICL)}}
\label{subsec-icl}
ICL, first introduced with GPT-3 \cite{brown2020language}, utilizes formatted natural language prompts and integrates task descriptions and examples to guide \blue{LLMs} in task execution. This approach allows the LLM to recognize and perform new tasks by leveraging contextual information. 
The design of demonstrations is critical for ICL, encompassing the selection, format, and order of examples. 
The format of demonstrations involves converting selected examples into a structured prompt, integrating task-specific information and possibly incorporating reasoning enhancements like CoT \cite{chung2022scaling, Wei-arxiv-2022-chain}. The ordering addresses LLM biases, arranging demonstrations based on similarity to the query or employing information-theoretic methods to optimize information conveyance \cite{Liu-ACL-2022-What, Lu-ACL-2022-Fantasically}.
ICL's underlying mechanisms include task recognition and task learning, and then \blue{LLMs} can use pre-trained knowledge and structured prompts to infer and solve new tasks. Task recognition involves \blue{LLMs} identifying the task type from the provided examples, leveraging pre-existing knowledge from pre-training data \cite{Wies-2023-arXiv-the}. Task learning, on the other hand, refers to LLMs acquiring new task-solving strategies through the given demonstrations, a capability that becomes more pronounced with increasing model size \cite{Oswald-arxiv-2022-Transformers}. Recent studies suggest that larger \blue{LLMs} exhibit an enhanced ability to surpass prior knowledge and learn from the demonstrations provided in ICL settings \cite{Wei-arxiv-2023-Larger}.

\subsubsection{\rm \textbf{Chain-of-thought (CoT) prompting}}
\label{subsec-cot}

CoT prompting is an advanced strategy to enhance LLM's performance on complex reasoning tasks, such as arithmetic, commonsense, and symbolic reasoning, by incorporating intermediate reasoning steps into prompts \cite{Wei-arxiv-2022-chain}. Differing from ICL's input-output pairing, CoT prompting enriches prompts with sequences of reasoning steps, guiding \blue{LLMs} to bridge between questions and answers more effectively.
Initially proposed as an ICL extension, CoT augments demonstrations from mere input-output pairs to sequences comprising inputs, intermediate reasoning steps, and outputs \cite{Wei-arxiv-2022-chain}. These steps help \blue{LLMs} navigate complex problem-solving more transparently and logically, though they typically require manual annotation. However, creative phrasings such as “\textit{Let's think step by step}” can trigger \blue{LLMs} to generate CoTs autonomously, which significantly simplifies the CoT implementation.

Despite improvements, CoT prompting faces challenges such as incorrect reasoning and instability. The enhancement strategies include better prompt design by utilizing diverse and complex reasoning paths, advanced generation strategies, and verification-based methods. These methods address generation issues by exploring multiple paths or validating reasoning steps, thus improving result accuracy and stability.
Furthermore, extending beyond linear reasoning chains, recent studies propose tree- and graph-structured reasoning to accommodate more complex problem-solving processes\cite{Yao-arxiv-2023-Tree}. 
%
In addition, CoT prompting significantly benefits large-scale \blue{LLMs} (over 10B parameters) and tasks requiring detailed step-by-step solutions. However, it may underperform in simpler tasks or when traditional prompting is already effective \cite{Wei-arxiv-2022-chain}.

\subsubsection{\rm \textbf{Planning for complex task solving}}
\label{subsec-planning}

While ICL and CoT prompting provide a straightforward approach for task solving, they often fall short in complex scenarios like mathematical reasoning and multi-hop question answering \cite{Qian-2022-arXiv-limitations}. To this end, prompt-based planning has emerged, breaking down intricate tasks into smaller and manageable sub-tasks and outlining action sequences for their resolution.
The planning framework for \blue{LLMs} encompasses three main components: the task planner, the plan executor, and the environment. The task planner devises a comprehensive plan to address the target task, which could be represented as a sequence of actions or an executable program \cite{Zhou-arxiv-2022-Least}. This plan is then carried out by the plan executor, which can range from text-based models to code interpreters, within an environment that provides feedback on the execution results \cite{Wang-arXiv-2023-Plan, Yao-arxiv-2023-Tree}.
In plan generation, \blue{LLMs} can utilize text-based approaches to produce natural language sequences or code-based methods for generating executable programs, enhancing the verifiability and precision of the planned actions \cite{Wang-arXiv-2023-Plan}. Feedback acquisition follows, where the LLM evaluates the plan's efficacy through internal assessments or external signals, refining the strategy based on outcomes from different environments \cite{Yao-arxiv-2023-Tree}.
In addition, the refinement process is crucial for optimizing the plan based on received feedback, and the corresponding methods include reasoning, backtracking, memorization, etc \cite{Yao-arxiv-2023-Tree}.

\subsubsection{\rm \blue{\textbf{Self-refinement with iterative feedback}}}

\blue{Considering that LLMs may not generate correct answers initially, self-refine has recently emerged to improve their outputs through iterative feedback and refinement. 
Among these studies, Madaan \textit{et al.}~\cite{madaan2024self} first use an LLM to generate initial outputs, and then employ the same LLM to provide specific feedback on these outputs. 
Note that this feedback is actionable, containing concrete steps to further improve the initial outputs. With such specific and actionable feedback, the same LLM can iteratively refine its outputs until performance converges.
Furthermore, Hu \textit{et al.}~\cite{hu2024TrafficLLM} propose a self-refined LLM (named TrafficLLM) specifically designed for communication traffic prediction, which leverages in-context learning to enhance predictions through a three-step process: traffic prediction, feedback generation, and prediction refinement. 
Following the comprehensive feedback, refinement demonstration prompts enable the same LLM to refine its performance on target tasks.}



\subsection{Evaluation metrics of LLM}
\label{sec-evaluation}
Evaluating the performance of \blue{LLMs} is a multifaceted task and receives increasing attention.
This subsection focuses on the evaluation metrics that encompass various dimensions, including accuracy, hallucination, efficiency, and human alignment etc. 
Each of these aspects plays a crucial role in determining the overall  applicability of \blue{LLMs} in real-world scenarios such as telecom networks.
Firstly, accuracy is paramount in evaluating LLM technologies as it directly impacts the model's reliability and trustworthiness. It measures how well an LLM can understand and process natural language queries, generate relevant and correct responses, and perform specific tasks like translation, summarization, or question-answering. Benchmarks and standardized datasets are often used to quantitatively evaluate the model's accuracy. 


Secondly, hallucination refers to instances where the LLM generates incorrect or factual inconsistent information, often presenting it with a high degree of confidence. This phenomenon can significantly undermine the credibility of LLM-generated content. Evaluating an LLM's tendency to hallucinate involves analyzing the model's responses for factual accuracy, consistency, and relevance to the input prompt. 
Recent studies show that traditional automatic metrics for summarization such as ROUGE~\cite{lin2004rouge} and BERTScore~\cite{zhang2019bertscore} show sub-optimal performance on factual consistency measurement~\cite{gao2022dialsummeval}. 
Thereafter, some novel metrics have been proposed to detect hallucination errors, such as AlignScore in \cite{ zha2023alignscore}. 


Then, the efficiency of LLMs indicates the computational resources required for training and inference, as well as the speed at which these models can generate responses. As the size of LLM expands, this expansion leads to significant issues regarding environmental sustainability and the ease of access to these technologies~\cite{samsi2023words, faiz2023llmcarbon}.
Evaluating an LLM's efficiency involves a detailed assessment of its performance relative to the consumed resources. Key metrics for this assessment include the energy usage during operations, the time it takes to process information, and the financial burden associated with acquiring and maintaining the necessary hardware infrastructure. Additionally, it's important to consider the efficiency of data usage during training, as optimizing data can reduce computational requirements~\cite{mcdonald2022great}.


The last metric is human alignment. Manual evaluation for LLM alignment to human values generally offers a more holistic and precise assessment compared to automated evaluation~\cite{chang2023survey}. 
This is supported by numerous studies, such as~\cite{ziems2024can, liang2022holistic}, which incorporate human alignment evaluation to provide a more in-depth analysis of their methods' performance. 
Human alignment assesses the degree to which the language model’s output aligns with human values, preferences, and expectations. 
It also considers the ethical implications of the generated content, ensuring that the language model produces text that respects societal norms and user expectations, promoting a positive interaction with human users.

\begin{table*}[!t]
\caption{Summary of LLM deployment strategies}
\centering
\setstretch{1.05}
\small
\resizebox{1\textwidth}{!}{
\begin{tabular}{|m{1.8cm}<{\centering}|m{7cm}<{\centering}|m{9.5cm}<{\centering}|}
\hline 
LLM deployment strategies & Main features \& Advantages    & Potential issues \& Difficulties   \\
\hline
Cloud deployment&  Cloud deployment is the most straightforward method for LLM deployment. \blue{LLMs} are usually computationally demanding, and cloud servers can provide abundant computational and storage resources for model training, fine-tuning, and inference. & Cloud deployment indicates higher end-to-end latency for implementing user requests, since the inquiries have to be first uploaded and then processed and downloaded. It may prevent the application of some latency-critical applications such as robot control, vehicle-to-vehicle communications, unmanned aerial vehicle (UAV) control, etc.\\
\hline
Network edge deployment &  Network edge deployment can be an appealing approach to shorten the response time and save backhaul bandwidth to the central cloud. It enables rapid user request processing at edge servers or cloud, achieving shorter end-to-end delay than the central cloud-based approach.   &  Network edge servers are usually resources-constrained, indicating limited computational and storage resources for \blue{LLM} fine-tuning and inference. Therefore, some techniques may be exploited, e.g.,  efficient parameter-efficient fine-tuning\cite{ding2023parameter}, split edge learning\cite{lin2024efficient}, and quantized training\cite{dettmers2024qlora}. In addition, model compression is also a promising direction for edge LLM deployment.  \\
\hline
On-device deployment &  On-device LLM is considered a very promising direction to deploy \blue{LLMs} directly at user devices. It enabled customized \blue{LLMs} based on specific user requests. Meanwhile, on-device LLMs have the lowest service latency by processing tasks locally. Therefore, it has great potential for implementing real-time tasks.  &  Despite the great advantages, on-device LLMs are still in the very early stages, and the main challenge is to overcome the very limited computational and storage resources at user devices. Apple has proposed a technique to store \blue{LLM} parameters on flash memory\cite{alizadeh2023llm} and achieve a 20 times faster inference speed. Qualcomm also announced a new mobile platform to support popular small-scale LLMs\cite{qualcomm}. Therefore, how to utilize limited computation resources to achieve faster inference is the key to on-device LLM deployment. \\
\hline
Cache-based deployment & Cached-based approach is proposed by \cite{lin2023pushing} based on mobile edge computing architecture.
Specifically, the authors propose to store the full-precision parameters in the central cloud, quantized parameters in the edge cloud, and the frozen parameters at the user devices, enabling more flexible model training and migration.   & Such a distributed deployment approach is promising to save the model store and migration cost. However, compared with on-device deployment, the cache-based method also requires complicated coordination strategies for model update and synchronization, e.g., model update and synchronization frequency and the quantization bit version selection.  \\
\hline
Cooperative deployment  &  Cooperative deployment is proposed in \cite{yang2023edgefm}, which involves the interactions between local small models and cloud-based large models. In particular, it assumes that the local model can collect and submit sensor data selectively to the large model, and the large model will update the small-scale local models based on its domain-specific knowledge.   &  The cooperative deployment is a feasible solution to connect small-scale local \blue{LLMs} to large cloud models. However, the local model updating frequency should also be carefully determined to reduce the burden on cloud \blue{LLMs}. In addition, note that the inference is still updated locally, and therefore the required computational resources are still challenges. To this end, it may be combined with on-device LLMs to address the resource issues.   \\
\hline
\end{tabular}}
\label{tab-deploy}
\end{table*}

\subsection{LLM Deployment in Telecom Networks} 
\label{sec-deploy}
Practical deployment is the prerequisite for advanced applications of LLM technologies in telecom networks. In particular, it indicates how \blue{LLMs} can be deployed within the current telecom network architecture, e.g., central cloud, network edge or even user devices. The LLM has great demands for computational and storage resources. For instance, GPT-4 has 1.76 trillion parameters and the model size is 45 GB\cite{OpenAI-OpenAI-2023-GPT-4v}, posing a heavy burden on network storage capacities. Fine-tuning an \blue{LLM} with 7 billion parameters, such as GPT-4-LLM\cite{peng2023instruction}, could take nearly 3 hours on an 8$\times$80GB A100 machine, which is extremely time-consuming \cite{zhang2023instruction}. 
In addition, the inference time of \blue{LLMs} will also contribute to overall network latency, which is related to hardware support, batch size, parallelism, model pruning, etc\cite{wang2023tabi}.    
Therefore, it is of great importance to deploy \blue{LLMs} appropriately to better serve the telecom network demand.
We summarize the existing deployment schemes in Table \ref{tab-deploy}, including cloud, network edge, on-device, cache-based, and cooperative deployment. We present the details of each strategy as the following:

\subsubsection{\rm \textbf{Cloud deployment}}: Considering LLM's high demand for computational and storage resources, deploying \blue{LLMs} in the central cloud is a straightforward solution, which can provide substantial computational resources to support the fine-tuning and inference of \blue{LLMs}\cite{shen2024large}. Shen \textit{et al.} investigate LLM-enabled autonomous edge AI\cite{shen2024large}, in which the network edge devices can send the user request and datasets feature to \blue{LLMs} in the cloud, and then the LLM can send back the task planning and AI model configuration to network edge devices through the backhaul. After that, the network edge and user devices can collaborate to make edge inferences. Cloud deployment can easily adapt to existing telecom network architecture, and a few pieces of extra hardware are needed since the LLM is deployed in the virtual cloud. However, cloud deployment suffers from long response time and high bandwidth costs since all data has to be transmitted to the cloud, and then LLM will process the request and finally download the LLM's output\cite{lin2023pushing}. The long response time may prevent the applications on latency-critical tasks, e.g., vehicle-to-vehicle networks and unmanned aerial vehicle control. In addition, the frequent multimodal information exchange, such as images and videos, between end users and cloud LLM will lead to extra bandwidth costs.

\begin{figure*}[!t]
\centering
\includegraphics[width=1\linewidth]{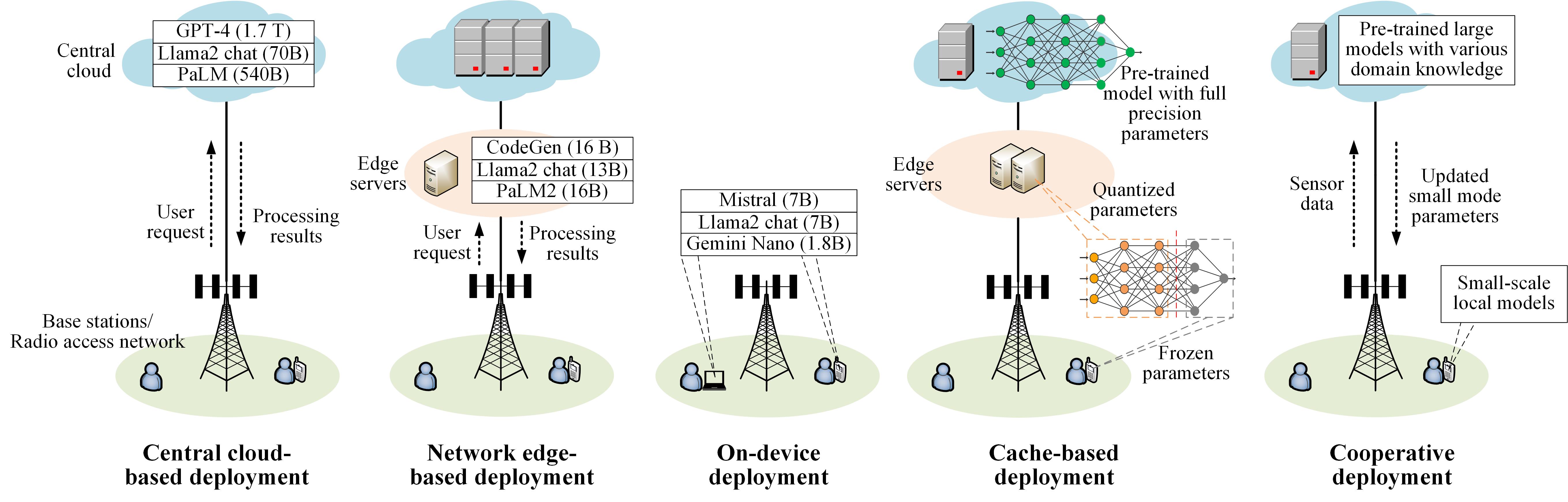}
\caption{Illustration of different LLM deployment strategies.}
\label{fig-deploy}
\end{figure*}

\subsubsection{\rm \textbf{Network edge deployment:}} 
Here network edge refers to edge cloud or BSs that are closer to users than central cloud. Network edge deployment can be an appealing approach to shorten the response time and save bandwidth.  
However, compared with the central cloud, network edge devices usually have limited computational and storage capacities. To this end, multiple techniques can be exploited.    
For the storage capacity challenge, parameter sharing and model compression may be applied. In particular, \blue{LLMs} for different downstream tasks may share the same parameters, which can be exploited to save the storage capacity. On the other hand, other technologies may be applied to reduce the computational resources demand in fine-tuning and inference, including parameter-efficient fine-tuning\cite{ding2023parameter}, split edge learning\cite{lin2024efficient}, and quantized training\cite{dettmers2024qlora}. With these techniques, deploying \blue{LLMs} at the network edge becomes a practical strategy.

\subsubsection{\rm \textbf{On-device deployment:}} There are multiple benefits of deploying \blue{LLMs} on user-side mobile devices, e.g., fast responses and local customization based on the user's specific requirements. However, such a deployment is also challenging since \blue{LLMs} are usually storage- and computation-intensive. Xu \textit{et al.} introduced a split learning approach based on collaborative end-edge-cloud computing, aiming to deploy LLM agents at mobile devices and network edge \cite{xu2024large}. Specifically, the authors assume that \blue{LLMs} with less than 10B parameters such as LLAMA-7B can operate on mobile devices, providing real-time inference services. 
Meanwhile, \blue{LLMs} with more than 10B parameters such as GPT-4 are deployed on network edge servers, using global information and historical memory to assist the mobile LLM in processing complex tasks. Such a collaboration enables higher flexibility by exploiting mobile \blue{LLMs}. However, the study of on-device LLM is still in a very early stage, and it requires considerable efforts to prove the feasibility of such a design\cite{xu2023llmcad}.   
For instance, Apple has proposed a technique to store \blue{LLM} parameters on flash memory\cite{alizadeh2023llm}, achieving a 20 times faster inference speed than using GPU with limited dynamic random-access memory (DRAM) capacity. 
Similarly, Qualcomm has recently announced the Snapdragon 8s Gen 3 mobile platform, which supports popular small-scale LLM such as Llama 2 and Gemini Nano\cite{qualcomm}.  
These studies may pave the way to effective inference of on-device LLMs.

\subsubsection{\rm \textbf{Cache-based deployment}}
Lin \textit{et al.} proposed a cache-based method in \cite{lin2023pushing}, which utilizes the mobile edge computing architecture to store, cache, and migrate models in edge networks.
Specifically, they propose to store the full-precision parameters in the central cloud, quantized parameters in the edge cloud, and finally the frozen parameters at the user devices.  
Such a separate model caching enables more flexible model training and migration. 
For instance, edge clouds or servers can apply low-precision computation by using quantized training, improving the edge training speed with limited computational resources. 
In addition, storing the frozen parameters on user devices can save the storage capacities of the edge cloud, reducing the latency caused by full model migration. 
However, the cache-based method may require complicated coordination strategies for model update and synchronization, e.g., model update and synchronization frequency and the quantization bit version selection.

\subsubsection{\rm \textbf{Cooperative deployment}}
Lin \textit{et al.} proposed a novel EdgeFM approach in \cite{yang2023edgefm}. In particular, the edge devices will collect the sensor data from the environment, and then the local model can evaluate the uncertainty features of the collected data and the real-time network dynamics. After that, the local EdgeFM model will selectively upload the unseen data classes to query large models in the cloud, and the large models can periodically update a customized small-scale model at the network edge. 
Therefore, when the network environment changes, at the early stage,  the local model can frequently query large models in the cloud, and then it can execute customized small models on edge devices at the
late stage. 
Such a cooperative deployment can reduce the system overhead, and enable dynamic customization of local small models for edge devices.
The experiment in \cite{yang2023edgefm} shows 3.2x lower 
end-to-end latency and achieve 34.3\% accuracy improvement than the baseline.

Finally, Fig. \ref{fig-deploy} illustrates different LLM deployment strategies. 
\blue{Note that LLM’s requirements for storage and computational resources are the main motivations for developing various deployment strategies.
For instance, the model size of Llama3-8b is around 5 GB, and therefore it is possible to be implemented at the network edge or even user devices, i.e., Snapdragon 8s Gen 3 mobile platform recently developed by Qualcomm. Similarly, Gemini Nano is less than 2 GB, and such a small size allows on-device deployment, e.g., Google plans to load Gemini Nano to its Pixel 8 smartphones.  
By contrast, large-scale LLMs require much more computation resources. For example, inference with Llama3-70b consumes at least 140 GB of GPU RAM. Using 2-bit quantization, the Llama3-70b can be implemented on a 24 GB consumer GPU, but such a low-precision quantization will significantly degrade the model accuracy. 
To this end, hybrid deployment methods such as cache-based and cooperative deployment are proposed. The key objective is to take advantage of large-scale LLM's high accuracy, while reducing the dependency on computational resources.} 
\blue{On the other hand}, these approaches may be combined, e.g., deploying small-scale on-device LLMs and then using larger cloud models to update the local models periodically. 
Given these deployment methods, many critical problems can then be investigated, e.g., service delay evaluation and task offloading, which still require more research efforts. For example, Chen \textit{et al.} proposed a NETGPT scheme in \cite{chen2024netgpt}, involving offload architecture, splitting architecture, and synergy architecture for cloud-edge collaboration.

\subsection{Analyses of LLM Fundamentals in the Telecom Domain}
Previous Sections \ref{sec-modelarchi} to \ref{sec-deploy} have covered the key techniques of LLM fundamentals, ranging from model architecture and pre-training to evaluation and deployment in telecom networks. This subsection will analyze how these fundamental techniques can be applied to the telecom domain.

For telecom applications, pre-training an \blue{LLM} from scratch can be time-consuming. 
It first requires extensive dataset collection, and the dataset preprocessing has to consider the format of complicated telecom equations and theories.  
Meanwhile, it also requires considerable computational resources to pre-train \blue{LLMs}, leading to heavy burdens for telecom networks.
By contrast, a more efficient approach is to fine-tune a general-domain \blue{LLM} for specific telecom-domain tasks. 
Applying LLM technologies to the telecom domain requires an in-depth understanding of these fine-tuning techniques, such as instruction and alignment tuning methods.     
In particular, instruction tuning involves carefully constructing and selecting instruction datasets, employing strategic tuning methodologies, and considering practical implementation aspects. 
These strategies will significantly improve the performance, generalization, and user alignment of LLM technologies in the telecom domain.
On the other hand, alignment tuning is a multifaceted process involving the setting of ethical guidelines, collection of human feedback, and application of advanced fine-tuning techniques such as RLHF. 
However, adapting these state-of-the-art fine-tuning techniques to telecom environments is still an open question. The fine-tuning process is usually task-specific, which requires professional knowledge of various telecom domain tasks. Instruction tuning can be a promising method for building a telecom-LLM by using existing telecom knowledge, but the dataset collection can be difficult due to  privacy issues.  

Prompting techniques are especially useful for solving real-time telecom tasks with stringent delay requirements, e.g., resource allocation and user association. It means that \blue{LLMs} can directly learn from the inputs and generate desired outputs without extra training, avoiding the tedious model training process in conventional ML algorithms.
%
For instance, ICL provides a framework for leveraging the LLM in new task domains without explicit retraining, with its effectiveness heavily influenced by the design and structure of demonstrations. 
%
Meanwhile, CoT prompting has emerged as a potent method for eliciting deeper reasoning capabilities in \blue{LLMs}, applicable to a range of complex reasoning tasks. While still evolving, this approach opens new avenues for LLM application across diverse problem domains such as the telecom field.
In addition, prompt-based planning represents a sophisticated approach to navigating complex tasks, enhancing LLM's problem-solving capabilities through structured action sequences, feedback integration, and continuous plan refinement. 
Such planning capabilities are very important for telecom applications since many telecom tasks involve multi-step thinking with complicated procedures. For instance, the resource allocation may include multi-layer controllers\cite{zhou2021ran}, and optimization problems can involve several agents and elements\cite{zhou2023cooperative}. Therefore, multi-step planning and thinking should be carefully designed for LLM-enabled telecom applications.

Evaluation metrics are critical to assess the \blue{LLM}'s performance in telecom environments. For instance, efficiency is one of the most important metrics that should be considered in telecom applications since many tasks require rapid or even real-time responses. Therefore, \blue{LLMs} with long inference times may be inappropriate for these mission-critical applications, e.g., Ultra-Reliable Low Latency Communications (URLLC).    
In addition, evaluating the performance of \blue{LLMs} should also include their proneness to hallucination and ethical standards, e.g., LLM may make misleading or even wrong decisions in network management. 
As LLM design and models continue to evolve and integrate more deeply into various aspects of society, the criteria for their evaluation will likely expand and become more sophisticated. 
%
Ensuring that \blue{LLMs} are accurate, reliable, efficient, and ethically responsible is essential for their sustainable and beneficial integration into human-centric applications.

Finally, practical deployment is the prerequisite for applying LLM to telecom networks. Compared with other domains such as education or healthcare, many telecom tasks have stringent requirements for delay and reliability, which require more efficient and reliable model output. Meanwhile, telecom devices usually have limited computational and storage resources. Therefore, efficient model training, fine-tuning, inference and storage techniques should be explored\cite{lin2023pushing}. 
With previous knowledge and analyses, we will present detailed LLM-inspired techniques and applications in telecom tasks in terms of generation, classification, optimization, and prediction problems in the following sections.

\section{LLM for Generation Problems in Wireless Networks}
\label{sec-gene}

The outstanding generation capability is one of the most attractive features of LLMs. 
This section first introduces the motivations for applying the LLM technique to telecom-related generation tasks, and then it presents detailed application scenarios, including telecom domain knowledge generation, code generation, and network configuration generation.

\begin{table*}[!t]
\caption{Summary of LLM-aided generation-related studies in the network field.}
\centering
\small
\setstretch{1.07}
\resizebox{1\textwidth}{!}{%
\begin{tabular}{|m{1.5cm}<{\centering}|m{0.8cm}<{\centering}|m{6cm}<{\centering}|m{6.5cm}<{\centering}|m{5cm}<{\centering}|}
\hline
Topics & Refer- ences & Proposed LLM-aided generation schemes & Key findings \& Conclusion & Telecom application opportunities\\
\hline
\multirow{20}*{\makecell{Domain \\ knowledge \\ generation}} & \cite{holm2021bidirectional}  &   Adapting a BERT-like model to the telecom domain and testing the model performance by question answering downstream task in the target domain.  &  The proposed technique achieved $F1$ score of 61.20 and $EM$ score of 36.48 on question answering in a small-scale telecom question answering dataset.   &   \multirow{20}*{\makecell{Techniques such as customizing \\\blue{LLMs} to understand \\ and apply telecom-specific \\ language, evaluating their \\genuine understanding of \\domain knowledge generation\\ can contribute to more \\efficient, reliable, and secure \\telecom service applications. \\ Existing studies have \\demonstrated the capability \\of LLM techniques to be applied in \\telecom, including question\\ answering, literature review,\\ generating troubleshooting report.\\ It shows great promises to build\\ next generation communication \\ networks. }   }     \\
\cline{2-4}
&  \cite{marzo2021natural}      &   It proposed a multi-stage BERT-based approach to understand the textual data of telecom trouble reports, and then generate a ranked solution list for new troubles based on previously solved troubleshooting tickets.    &  1) Presenting more information in the query can produce a better list of recommended solutions; 2) Creating a small candidate list is the key to reducing the model latency.   &        \\
\cline{2-4}
&  \cite{bosch2022integrating}     &   It combines a BERT-like method with transfer learning for trouble report retrieval, leveraging non-task-specific telecom data and generalizing the model to unseen scenarios.     &  The experiment includes nearly 18500 trouble reports, showing that combining pre-trained telecom-specific language models with fine-tuning strategies outperforms pure domain adaptation fine-tuning.    &          \\
\cline{2-4}
&  \cite{soman2023observations}    &  Question answering test on various \blue{LLMs}, e.g., GPT 3.5, GPT 4, and Bard, including telecom knowledge and product questions.  &  Bard and GPT4 show promise with respect to accuracy and could be useful for telecom domain question and answering. LLM's summarization requires reliability tests.  &        \\
\cline{2-4}
&  \cite{wang2024grammar}  & Integrating domain-specific grammars into \blue{LLMs} to guide the generation of structured language outputs, enhancing performance in domain-specific tasks.  &  Demonstrating the efficacy of integrating domain-specific grammars with \blue{LLMs} in enhancing their ability to generate structured language outputs tailored to specific domains. It emphasizes the potential of this approach to significantly improve LLM performance in domain-specific tasks. &  \\
\cline{2-4}
\hline
 \multirow{19}*{\makecell{Code \\ generation}} & \cite{du2023power}  &  Using \blue{LLMs} to generate Verilog code for wireless communication system development in FPGA. The experiment was implemented in the OpenWiFi project. &  The LLM is capable of refactoring, reusing, and validating existing code. With proper design and prompting, \blue{LLMs} can generate more complicated projects with multi-step scheduling. LLM greatly reduced the coding time of undergraduate and graduate students by 65.16\% and 68.44\%, respectively.  & \multirow{5}*{\makecell{ Code is the cornerstone of modern \\ communication networks, and the LLM \\ provide promising opportunities to \\ improve the efficiency and reliability\\ of codes, and meanwhile greatly save \\ human effort. \\
 a) The LLM can refactor and \\ validate existing code.\\ This is very useful in\\  telecom filed, since \\ the network architecture is \\ constantly evolving and updated;\\
 b) With proper prompting, the LLM \\can generate complicated projects\\ with multi-step scheduling \\ requirements, which is very \\ common in telecom filed due to\\ complicated network elements with\\ diverse functions. }   }\\
\cline{2-4}
&  \cite{mani2023enhancing}  &  It proposed a framework to use LLM to generate task-specific code for traffic analyses and network life-cycle management.   &   Combining the LLM with proper libraries, such as GPT-4 and NetworkX, can achieve 88\% and 78\% coding accuracy for traffic analysis and network lifecycle management tasks, respectively.   &   \\
\cline{2-4}
&  \cite{xiang2023toward}  &  Employing four students to reproduce the results of existing network studies with the assistance of \blue{LLMs}. & The students successfully reproduced networking systems by prompting engineering ChatGPT. They also achieve much lower lines of code by using ChatGPT, e.g., one of them is only 20\% of the open-source existing version.  &        \\
\cline{2-4}
&   \cite{zhang2022repairing}  &   Using LLM techniques for automated program repair of introductory level Python projects.  &  The proposed scheme successfully repaired a larger fraction of programs (86.71\%) compared to the baseline (67.13\%), and adding few-shot examples will raise the ratio to 96.50\%. &        \\
\cline{2-4}
& \cite{thakur2023benchmarking}  &  Fine-tuning pre-trained \blue{LLMs} on Verilog datasets collected from GitHub and Verilog textbooks and then generating Verilog projects.  &  Fine-tuning \blue{LLMs} over a specific language can improve the coding correct rate by 26\%.     &        \\
\hline
\multirow{12}*{\makecell{ Network \\ configuration \\ generation}} & \cite{DBLP:conf/cnsm/DzeparoskaLTL23} &   It proposed a three-stage LLM-aided progressive policy generation pipeline for intent decomposition. & Through evaluating a service chain use case, the paper found \blue{LLMs} could generalize to new intents through few-shot learning and concluded leveraging \blue{LLMs} for policy generation is promising for automatic intent-based application management. & \multirow{3}*{\makecell{Telecom network operators\\
can leverage the LLM for network \\
configuration generation in \\
various ways. This includes automatic \\
network provisioning, optimization \\
and performance tuning, security \\
and compliance configuration, \\
fault diagnosis and troubleshooting, \\
and network virtualization. The LLM \\
enables efficient, reliable, and \\
secure generation of network \\
configurations, reducing manual \\
effort and improving network \\
management in telecom \\
environments.
}   }\\
\cline{2-4}
& \cite{DBLP:journals/corr/abs-2309-06342} &  It proposed a multi-stage framework that utilizes \blue{LLMs} to automate network configuration by taking in natural language requirements and translating them into formal specification, high-level configurations, and low-level device configurations.  & The results showed that state-of-the-art LLM technologies like GPT-4 are capable of generating fully working configurations from natural language requirements without any fine-tuning.  &        \\
\cline{2-4}
& \cite{DBLP:conf/hotnets/MondalTBMV23}  &  It proposed a framework that combines \blue{LLMs} with verifiers, using localized feedback from verifiers to automatically correct errors in configurations generated by the LLM.  & The proposed scheme is able to synthesize reasonable though imperfect configurations with significantly reduced human effort, and coupling \blue{LLMs} with verifiers providing localized feedback is necessary for real-world use configurations despite requiring more testing.  &        \\
 \hline
\end{tabular}}
\label{tab-llm-generation}
\end{table*}

\subsection{Motivations of Using LLM-based Generation for Telecom}
This subsection will introduce the key motivations of using LLM-enabled generation for telecom applications.
Firstly, LLM can make telecom knowledge more accessible.
\blue{LLMs} have been pre-trained on many real-world datasets and equipped with considerable knowledge from various fields. Therefore, question-answering has become the most well-known application of LLMs. 
With domain-specific datasets from websites and textbooks, the LLM can extract professional knowledge from existing publications and then generate appropriate answers based on users' requests. For instance,  Maatouk \textit{et al.} build a telecom knowledge dataset in \cite{maatouk2023teleqna}, including 25,000 pages from research publications, overview, and standards specifications. With proper training and fine-tuning, such a dataset can greatly contribute to a Telecom-GPT, providing a systematic overview of hundreds of publications and standards. With reasoning and comprehension capabilities, professional telecom knowledge will become much more accessible to all researchers and even benefit the whole society.

Meanwhile, LLM's generation capabilities can also automate many tasks that are usually time-consuming.
For instance, developing new standard specifications usually requires considerable writing, discussions, and reviews. 
By contrast, given enough historical reports and proper prompts, the LLM can produce a draft standard instantly, and then the experts can review it accordingly. 
Moreover, the experts' comments can be fed directly to \blue{LLMs}, and then the LLM can produce a new version efficiently, significantly saving human efforts on writing and revising paper works.
Similarly, LLM technologies have been used to generate code in many existing studies, which is one of the most time-consuming tasks of modern industry\cite{xiang2023toward,zhang2022repairing}. \blue{LLMs} can refactor and improve existing codes, contributing to developing telecom projects.


In addition, \blue{LLMs} can easily learn from the provided existing examples, which is known as ICL. 
This capability is particularly useful in generation tasks, and \blue{LLMs} can quickly generalize the given examples to related unseen scenarios.
Meanwhile, if the initial generated output can not satisfy the requirements, users can also send the feedback directly to the LLM input, and then the LLM agent will revise the generation accordingly. 
This user-friendly generation approach will lower the difficulty of applying LLM techniques to generation tasks in telecom, which usually requires considerable professional knowledge and experience.

Given the above motivations and advantages, it is crucial to exploit LLM's generation capabilities and apply them to telecom networks. 
Table \ref{tab-llm-generation} summarized LLM-aided generation-related studies and telecom application opportunities. 
In the following, we will introduce domain knowledge generation, code generation, and network configuration generation.

\subsection{Domain Knowledge Generation}
\label{sec-knowledge}

Generating domain-specific knowledge is an important application of LLM technologies in telecom. In particular, it refers to creating comprehensive summaries, overviews, and interpretations of telecom standards, technologies, and research findings. By leveraging vast datasets of technical documents, research papers, and standards specifications, LLM agents can produce detailed explanations and summaries that are tailored to the user's level of expertise and interest. This not only democratizes access to telecom knowledge but also serves as a bridge to fill the gap between experts and non-expert users in the telecom field.

\subsubsection{ \rm \textbf{Understanding telecom domain knowledge}}
Telecom is a broad field, and there are various domains of knowledge such as signal transmissions, network architectures, communication protocols, and industry standards. For instance, signal transmission is fundamental telecom knowledge, involving the differences between amplitude, frequency, and phase modulation, as well as the distinctions between digital and analog signals. 
Meanwhile, communication protocols refer to sets of rules that ensure standardized data transmission, allowing for interoperability among diverse systems. Knowledge of these protocols is fundamental for the development and maintenance of robust communication networks. Additionally, telecom standards are equally important. Standards such as 3G, 4G, and the emerging 5G for mobile communications, as well as IEEE 802.11 for Wi-Fi, play a critical role in global telecom networks\cite{ibarrola2023evolution}. They facilitate the seamless operation of devices and services across different networks.

A thorough understanding of the above telecom knowledge is not only vital for the development of new technologies and services, but also for ensuring that systems are interoperable and secure. 
The depth of understanding in telecom knowledge directly impacts the ability to innovate, secure, and solve problems within the telecom field. The integration of \blue{LLMs}, trained with domain-specific datasets, offers promising avenues for automating knowledge generation and facilitating access to complex telecom content, thus bridging the gap between experts and general users.

\begin{figure*}[!t]
\centering
\includegraphics[width=0.95\linewidth]{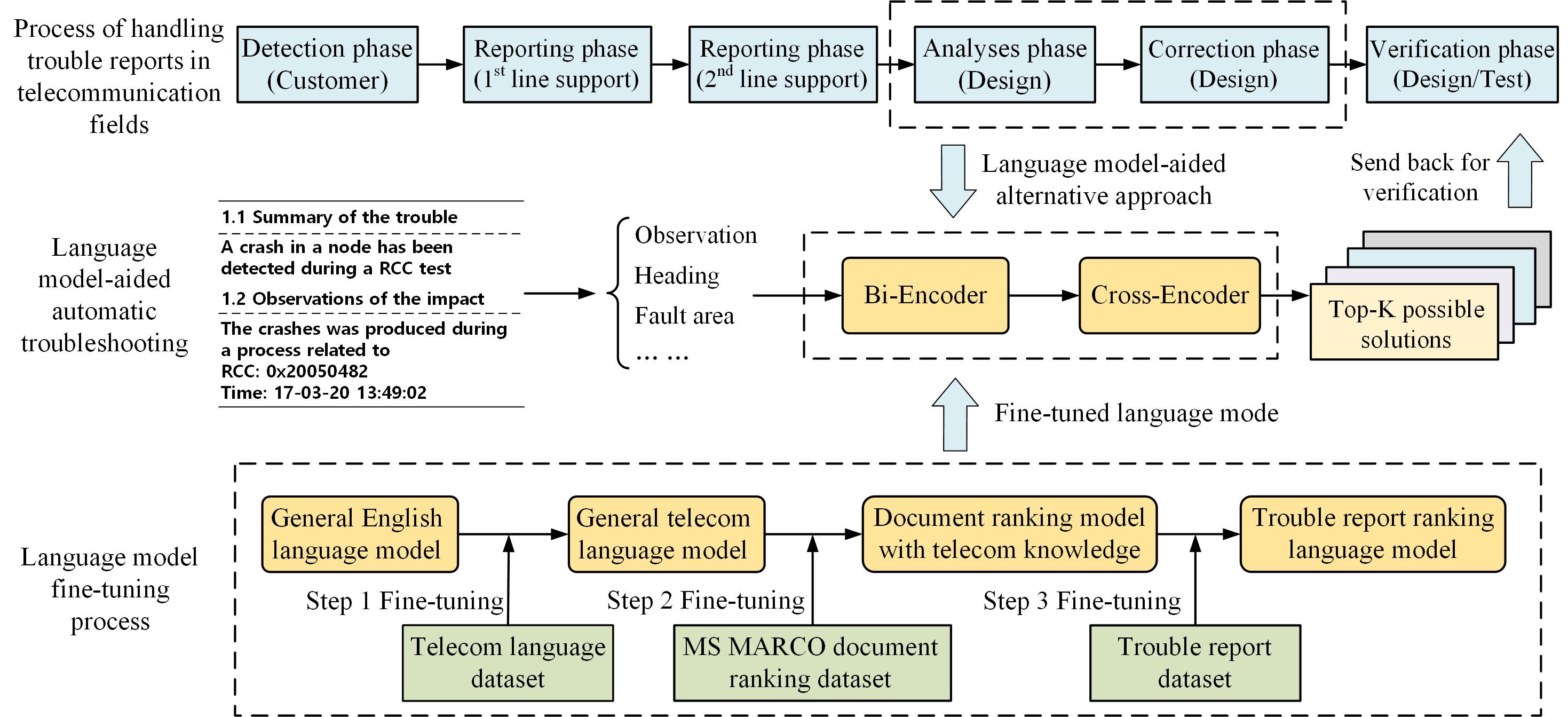}
\caption{Using language models for automated troubleshooting in telecom fields\cite{bosch2022integrating}.}
\label{fig-gene-answer}
\end{figure*}

\subsubsection{\rm  \textbf{Training \blue{LLMs} with telecom-specific data}}

Training \blue{LLMs} with telecom-specific data involves curating and preprocessing vast amounts of domain-specific information to fine-tune the models, aiming to generate accurate and relevant content within the telecom field. This process is crucial as it tailors the LLM's capabilities to understand and generate content that aligns with specific telecom requirements.
It can be summarized by following steps:
\begin{itemize}
    \item The first step in training the LLM with telecom-specific data is the collection of datasets. These datasets may include technical documents, research papers, standards specifications, and other forms of professional literature prevalent in the telecom sector. For example, Holm \textit{et al.}~\cite{holm2021bidirectional} created a small-scale TeleQuAD to train the question-answering capabilities of the build Bert-based model. Similarly, 185,000 trouble reports\cite{bosch2022integrating} are included to train a Bert-like model to generate automated troubleshooting tickets. However, these datasets are usually inaccessible to the public. By contrast, Maatouk \textit{et al.}~\cite{maatouk2023teleqna} introduced a large dataset of telecom knowledge to provide systematic overviews and detailed explanations of standards and research findings. 
    
    \item Following dataset collection, the preprocessing stage involves cleaning and organizing the data to make it suitable for training. This step may include removing irrelevant information, correcting errors, and converting the data into a format that is compatible with the ML model. The study \cite{gu2021domain} shows that preprocessing large-scale datasets for LLM training can improve the model's learning efficiency and output quality.
    
    \item Finally, it is worth noting that there are two main approaches to train \blue{LLMs}, which are training the model from scratch or fine-tuning a general-domain LLM. 
    In particular, training the model from scratch may produce better performance since the model can specialize in telecom language, but it is also time-consuming. On the other hand, 
    the fine-tuning process adapts the pre-trained LLM to the telecom domain. This step involves training the model on the collected telecom-specific dataset, allowing it to adjust its parameters to better understand and generate telecom content. Fine-tuning enables the model to grasp the unique terminologies, concepts, and contexts of the telecom field, significantly enhancing its generation capabilities. Although fine-tuning a pre-trained LLM is much more efficient than training from scratch, the experiment in \cite{holm2021bidirectional} proves that training the model on telecom-domain text from scratch can achieve better performance than fine-tuning a general-domain model.   
\end{itemize}

The integration of telecom-specific data into LLM training is not just about enhancing the model's knowledge base; it's about equipping the LLM with the ability to understand the nuances and complexities of the telecom field. This tailored training approach ensures that the LLM can generate content that is not only informative but also practical and applicable to real-world telecom challenges.


\subsubsection{\rm  \textbf{Using LLM to telecom knowledge-related generation tasks}}

After proper training or fine-tuning, using \blue{LLMs} to generate telecom domain knowledge is a transformative approach that leverages the model's ability to process and synthesize vast amounts of information into coherent, accessible content tailored to the needs of various stakeholders in the telecom field. This capability extends from generating summaries of complex technical documents to answering specific queries with detailed explanations, thereby facilitating a deeper understanding of telecom technologies, standards, and practices. In the following, we present some existing applications of telecom knowledge-related generation tasks.

\textbf{ Telecom-domain question answering:} 
Question answering is one of the most well-known applications of LLM technologies.
Using \blue{LLMs} to answer domain-specific questions is grounded in the model's ability to interpret and articulate complex information in a manner that is both comprehensive and understandable.
For example,  Soman \textit{et al.} evaluated the capabilities and limitations of existing pre-trained general domain LLMs in \cite{soman2023observations}, including GPT-3.5, GPT-4, Bard, and LLaMA. For instance, one telecom-domain question is "\textit{What are the different 5G spectrum layers?}"
GPT-4 identifies the bands as below 1 GHz, 1-6 GHz and above 6 GHz, while LLaMA identifies the frequency bands as below 600 MHz, 600 MHz-24 GHz and above 24 GHz. These differences could be caused by different data sources of GPT-4 and LLaMA in the pre-training period. However, this could easily confuse or even mislead users without professional knowledge, which shows the importance of training a telecom-domain LLM specifically.  
Holm \textit{et al.}~\cite{holm2021bidirectional} further investigate how various training methods can affect the model performance, e.g., pre-training a model using telecom knowledge from scratch or fine-tuning an existing general-domain model.  
In summary, LLM-enabled question answering democratizes access to advanced telecom knowledge, making it accessible to a broader audience, including researchers, practitioners, and the general public.
In addition, LLM agents can also tailor the generated content based on the user's level of expertise and specific interests. 
There is an increasing number of commercial LLM products for generative question answering over business documents, e.g., nexocode and Caryon.  
By leveraging the comprehensive understanding and generation capabilities of LLM technologies, the telecom industry can enhance the accessibility of complex information, support educational endeavours, and streamline development processes.

\begin{figure*}[!t]
\centering
\setlength{\abovecaptionskip}{0pt} 
\includegraphics[width=0.95\linewidth]{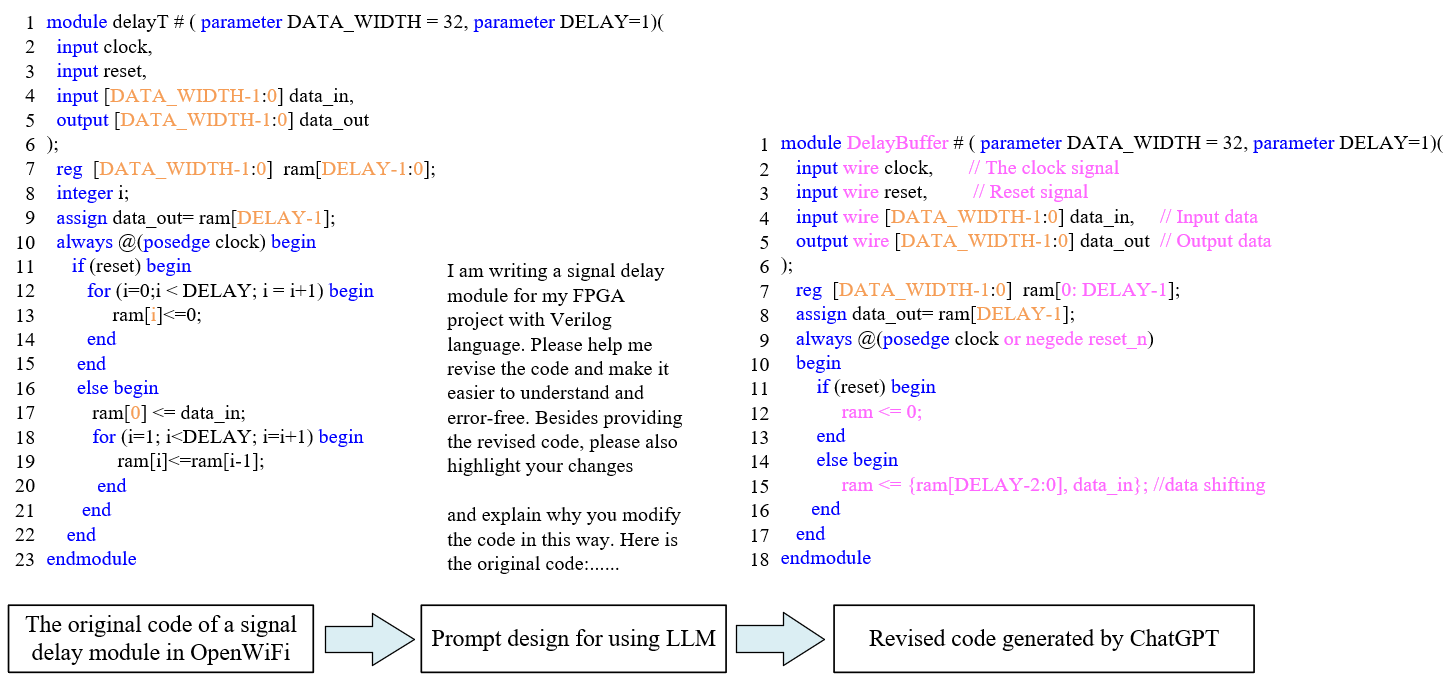}
\caption{Using ChatGPT to improve the code quality of OpenWiFi project\cite{du2023power} (The pink fonts show the main changes).}
\label{fig-code}
\end{figure*}

\textbf{ Generating troubleshooting solutions for telecom trouble reports:}
Telecom networks are complicated large-scale systems, and it is critical to identify, analyze and then resolve both software and hardware
faults, which are known as trouble reports.
The authors in \cite{marzo2021natural} and \cite{bosch2022integrating} investigated using language models to understand previous trouble reports and then generate recommended solutions.  
Grimalt applied a BERT-based model to generate and rank multiple possible solutions for a given system fault in \cite{marzo2021natural}, which archives a nearly 55\% correct rate. 
Then, Bosch \cite{bosch2022integrating} improved the model in \cite{marzo2021natural} by including transfer learning and non-task-specific telecom data to improve the generalization capabilities on handling unseen trouble reports.  
Fig.\ref{fig-gene-answer} summarizes the proposed scheme in \cite{marzo2021natural} and \cite{bosch2022integrating}. One can observe that the analysis and correction phases can be time- and effort-consuming, which usually requires professional knowledge of telecom networks and devices. 
To this end, a language model-enabled method is proposed. It considers trouble report observation, headings, and fault areas as input and generates the top-K possible solutions. Then, the generated candidate solutions are sent back for verification. 
In particular, the fine-tuning process of the language model consists of three main steps, including the telecom language dataset, MS MARCO document ranking dataset\cite{bajaj2016ms}, and trouble report dataset. 
Here, the MS MARCO dataset is included to train question-answering and ranking models, in which a large number of question-answer pairs are collected from search engines\cite{bajaj2016ms}. 
Fig.\ref{fig-gene-answer} proves that using language models to generate solutions for automated troubleshooting can significantly improve overall efficiency, enabling faster response and repair for telecom.


%
%
%


Finally, it is worth noting that these models may generate misleading or even wrong solutions, which can be caused by different data sources, training strategies, and so on\cite{soman2023observations}. 
For instance, the best correct rate in \cite{bosch2022integrating} is around 60\%, and therefore, verification is crucial before real-world implementation.

\subsection{Code Generation}
\label{sec-code}
Efficient and reliable code is of paramount importance to intelligent communication networks. 
Recent studies have demonstrated the strong coding capability of LLMs, including commonly-used languages (e.g., Python\cite{nijkamp2022codegen, zhang2022repairing}) and hardware description languages (e.g., Verilog~\cite{du2023power, thakur2023benchmarking}). 
For instance, Zhang \textit{et al.}~\cite{zhang2022repairing} apply the LLM to build an automatic program repair system for introductory Python programming
assignments, and the experiment on 286 real student programs achieves a repair rate of 86.71\%. For hardware description languages like Verilog for FPGA development, Du \textit{et al.}~\cite{du2023power} show that LLM can reduce nearly 50\% of the coding time for undergraduate and postgraduate students and improve the quality by 44.22\% for undergraduates and 28.38\% for postgraduates. 
Existing studies\cite{nijkamp2022codegen, zhang2022repairing, du2023power, thakur2023benchmarking} have shown that LLM can refactor and improve existing codes. 
In addition, well-crafted prompts and designs can tackle complex, multi-step coding challenges encompassing multiple sub-tasks.
Given these potentials, introducing LLM-aided coding into telecom can greatly save human effort in coding, validating, and debugging while providing more efficient and reliable codes for telecom network scheduling and management projects.

\subsubsection{\rm  \textbf{LLM for code refactoring}}
Code refactoring is a common task that is frequently involved when developing wireless communication systems. 
Code refactoring aims to improve the readability, efficiency, and reliability of existing code~\cite{lacerda2020code}.   
For instance, good readabilities can lower the difficulty of long-term maintenance and reuse of existing code modules. 
Readability is also a critical requirement for wireless networks since the network architectures and protocols are constantly evolving and updated, e.g., from WiFi 6 to 6E and WiFi 7, and from RAN to cloud RAN and Open RAN.    
However, real-world projects usually include multiple contributors with different coding styles and mixed qualities. Such an issue could be very common in telecom, which are considered as complicated large-scale systems that include multiple modules with diverse functions. 
Therefore, improving code readability, efficiency, and reliability becomes more important for the telecom field.  

Fig. \ref{fig-code} shows an example from \cite{du2023power}, which applies ChatGPT to revise the original code of an open-source FPGA-based project OpenWiFi\cite{jiao2020openwifi}. The pink fonts indicate the changes made by ChatGPT. In particular, ChatGPT suggests using meaningful names for modules and variables, e.g., replacing the name “\textit{DelayT}” with “\textit{DelayBuffer}”. Meanwhile, four comments are added to improve the readability of the input and output. 
The input and output data type specification ``\textit{wire}" is added from line 2 to line 5, providing more explicit definitions and higher reliability. ChatGPT also recommends adding the negative edge of active-low reset signals in the “\textit{always}” block in line 9 of the revised code. Du \textit{et al.} \cite{du2023power} explained that such an asynchronous reset is more reliable and the system can make instant responses when detecting errors, without waiting for the rising edge of the clock signal.

In addition, code validation is also an important task for telecom project development. 
Du \textit{et al.}~\cite{du2023power}  utilized ChatGPT to generate an error-free testbench for effective OpenWiFi project validation. 
However, the fine-tuning process is not investigated, which can be a prerequisite for effectively generating hardware description languages. Different from the aforementioned studies, Thakur \textit{et al.}~\cite{thakur2023benchmarking} fine-tuned a pre-trained LLM on Verilog datasets collected from GitHub and textbooks, demonstrating that fine-tuned LLMs can improve the coding correct rate by 26\% on a specific language.

\subsubsection{\rm \textbf{LLM-aided code generation with multi-step scheduling}}
\label{sec-code-multi}
Previous sections have shown that LLM can be used for fundamental coding tasks. However, real-world telecom project development is usually much more complicated by including multi-step scheduling and several sub-tasks. Xiang \textit{et al.} applied LLM to regenerate the code of existing studies in \cite{xiang2023toward}, and the authors suggested that ChatGPT does not respond well to monolithic prompts like "\textit{implement this technique in the following steps}". Instead, a more practical method is to send a detailed modular prompt each time.   
Such a step-by-step approach is also investigated in \cite{mani2023enhancing} and \cite{du2023power}. Specifically, Mani \textit{et al.}~\cite{mani2023enhancing} applied \blue{LLMs} to network graph manipulation, and the prompt design is decoupled into the application prompt and code generation prompt. Specifically, the application prompt can provide task-specific prompts based on templates and user queries, and then the code generation prompt can use plugins and libraries to instruct \blue{LLMs}.   
The experiment shows that combining the LLM with proper libraries, such as GPT-4 and NetworkX, can achieve 88\% and 78\% coding accuracy for traffic analysis and network lifecycle management tasks, respectively.   
Du \textit{et al.} investigated a more complicated coding task in \cite{du2023power} by using Verilog to build a Fast Fourier Transform (FFT) module. A failure is first observed by using the following prompt:
 \begin{itemize}
  \item[] 
  \begin{tcolorbox}
  [title = {A failed prompt in \cite{du2023power} to generate FFT module. }]
  \textit{Help me write an FFT module for my FPGA system in Verilog language. Here are details of my specifications}: ...\\
  \textit{I also provide you with the instantiation template}: ...
  \end{tcolorbox}
\end{itemize}
The generated code failed because: a) FFT computation is a complicated task with several sequential or parallel subtasks; b) The LLM lacks the capabilities of multi-step scheduling. 
To this end, the authors decouple the problem into four steps:
\begin{itemize}
    \item Step 1: Asking ChatGPT to generate two simple IP cores that are frequently used in the following FFT design:
    \begin{tcolorbox}
    \textit{I am working on an FPGA project in Verilog. Please write two IP cores for me. The first IP core is for butterfly computation for FFT. Here is its template}:....\\
    \textit{The second IP core is for complex multiplication in FFT. I will use it to multiply the output of a butterfly computation with the twiddle factor provided... Here is a template of the IP core}...
    \end{tcolorbox}
    
    \item Step 2: Showing ChatGPT a simple 2-point FFT example with templates and suggestions and then asking ChatGPT to produce a 4-point FFT IP core:
    \begin{tcolorbox}
    \textit{"I am writing a four-point DIF-FFT on FPGA. You can use the following IP cores to build the target four-point FT IP core. Here is the template of butterfly computation IP Core..."}\\
    \textit{"And here is the template of the two-point FFT IP Core.."}\\
    \textit{"Further, I also have some suggestions for you..."}
    \end{tcolorbox}
    
    \item Step 3: Asking ChatGPT to develop an eight-point FFT module based on the generated 4-point FFT in Step 3:
    \begin{tcolorbox}
    \textit{"I am writing an eight-point DIF-FFT on FPGA. Apart from IP cores Given in Question One,..., you can also use the fft\_4\_point IP core generated in Answer one. You need to look back to Question-1 and Answer-1 for detailed input/output information on the four IP cores. Once again, I want to emphasize that:..."}
    \end{tcolorbox}
    \item Step 4: Finally, asking ChatGPT to generate a 16-point FFT using the 8-point FFT that has been generated in Step 3. This step is repeated in \cite{du2023power} by asking for a $2N$-point FFT module based on previously generated $N$-point FFT modules. 
\end{itemize}
Steps 1-4 is an obvious step-by-step CoT approach. Instead of asking for an 8-point FFT module directly, it starts from two simple IP cores and then provides examples of 2-point FFT modules with detailed suggestions. This is a very useful technique for LLM-aided project design in telecom networks, decoupling the objective into several steps with detailed examples and suggestions.

Finally, we summarize some key lessons from existing studies on the use of LLM for code generation. Firstly, step-by-step prompt design is an important lesson that has been demonstrated in several existing studies\cite{du2023power, xiang2023toward, mani2023enhancing}. 
Decoupling the complicated multi-step scheduling problem into several stages will lower the difficulty for LLM's understanding. 
For instance, in 5G cloud RAN simulation, we can divide the network into cloud, edge, and users, and then use LLM to generate the code for each part sequentially.
Secondly, examples and pseudo-code are important for code generation. The LLM has excellent ICL capabilities, quickly learning from examples and generalizing to other scenarios. 
Xiang \textit{et al.}~\cite{xiang2023toward} also reveal that implementation with pseudocode first can produce stabilized data types and structures, avoiding other changes when implementing the following components.  
There have been many codes for the telecom field in GitHub and textbooks, taking advantage of these existing examples is crucial to use LLM techniques.
Then, a significant amount of human effort can be saved in code generation by using \blue{LLMs} for debugging and testing. Xiang \textit{et al.} \cite{xiang2023toward} also shows that most errors can be solved by sending the error message to the LLM. Many of these errors are related to data types, which can be avoided by specifying key variables’ data types. This lesson is also proved in \cite{du2023power}, in which the LLM specified the data types of inputs and outputs to improve the reliability of existing code.
Finally, LLM-aided coding can lower the requirement for professional knowledge \cite{du2023power, xiang2023toward}. In particular, Du \textit{et al.} \cite{du2023power} show that both undergraduate and graduate students can benefit from the assistance of \blue{LLMs}, achieving comparable coding qualities. Xiang \textit{et al.} \cite{xiang2023toward} prove that undergraduate students can reproduce the results of some existing network studies by using the LLM. 

\begin{figure*}[!t]
\centering
\setlength{\abovecaptionskip}{0pt} 
\includegraphics[width=1\linewidth]{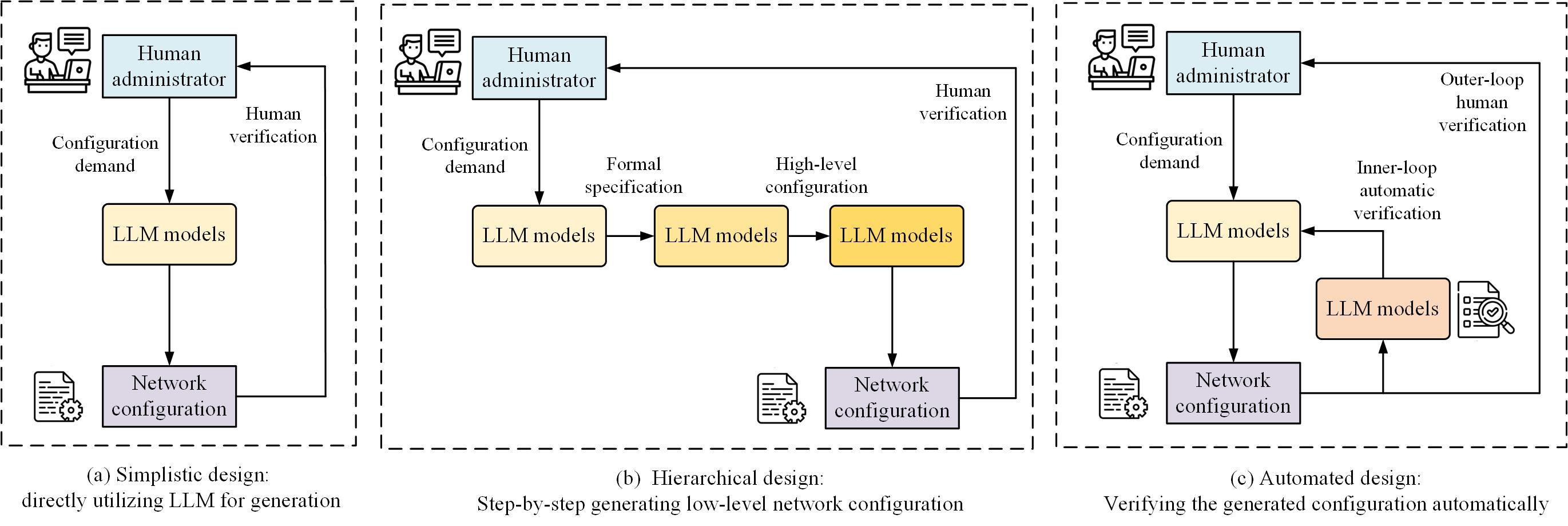}
\caption{Frameworks for LLM-based network configuration generation.}
\label{fig-configgen}
\end{figure*}

\subsection{Network Configuration Generation}
\label{sec-configuration}
Network administrators orchestrate the flow of information within a network. They can guide data from source to destination by configuring a complex set of parameters for network elements. 
These configurations impact a wide range of devices and services, such as switches, routers, servers, and network interfaces. To ensure a reliable data stream, these settings require precise calibration across all network functionalities.
Over the past ten years, both academic institutions and the commercial sector have embraced the concept of Software-Defined Networking (SDN)~\cite{DBLP:journals/comsur/NunesMNOT14} as a means to streamline network management, marking a shift away from the older, more rigid networking models. SDN offers numerous advantages; nonetheless, adjusting network settings remains a task that often requires manual input. Such manual adjustments can be expensive, as they demand the skills of specialized developers familiar with various network protocols, and meanwhile such manual configurations are also intricate and prone to errors.
Numerous initiatives have been launched with the aim of streamlining the translation of overarching network guidelines into individual settings for each network component. Such efforts focus on reducing human errors by creating verifiable and reliable configuration outputs through rigorous checks~\cite{DBLP:conf/nsdi/El-HassanyTVV18,DBLP:conf/apnoms/ChenJWLYFX23}. Nonetheless, setting up network configurations is still considered as a labour-intensive, intricate, and costly endeavour for network operators.

Recent advancements have demonstrated that the LLM possesses the ability to generate cohesive and contextually relevant content. They can answer questions and sustain in-depth conversations with users. Applications like GitHub Copilot and Amazon CodeWhisperer exemplify these advancements, assisting with a variety of programming-related tasks. These developments inspire confidence that the LLM can also be utilized to generate network configurations~\cite{DBLP:journals/corr/abs-2401-06786,DBLP:conf/cnsm/DzeparoskaLTL23}.

One notable development of LLM-aided network configuration is CloudEval-YAML~\cite{DBLP:journals/corr/abs-2401-06786}, a benchmark that provides a realistic and scalable assessment framework specifically for YAML configurations in cloud-native applications. This benchmark utilizes a hand-crafted dataset and an efficient evaluation platform to thoroughly examine the performance of \blue{LLMs} within this context. 
Dzeparoska \textit{et al.}~\cite{DBLP:conf/cnsm/DzeparoskaLTL23} have introduced a pioneering method that employs the few-shot learning capabilities of the LLM to automate the translation of high-level user intents into executable policies. This approach facilitates dynamic, automated management of applications without the necessity for predefined procedural steps. In a related vein, Wang \textit{et al.}~\cite{DBLP:journals/corr/abs-2309-06342} have developed NETBUDDY, a multi-stage pipeline that leverages \blue{LLMs} to translate high-level network policies specified in natural language into low-level device configurations. NETBUDDY first uses an LLM to convert the input into a formal specification, such as a data structure to express reachability. 
It then generates forwarding information and configuration scripts from the formal specification. Finally, NETBUDDY interacts with an LLM multiple times to sequentially provide topology, addressing details and prototype programs to automatically generate vendor-agnostic configurations for the switches and routers. The evaluation of the network emulator demonstrates NETBUDDY's ability to enforce path policies and dynamically modify existing deployments. 
In addition, Mondal \textit{et al.}~\cite{DBLP:conf/hotnets/MondalTBMV23} presented Verified Prompt Programming (VPP) to improve GPT-4's ability to synthesize router configurations. VPP combines GPT-4 with verifiers like Batfish~\cite{DBLP:conf/nsdi/FogelFPWGMM15}, which check configurations for syntax errors and semantic differences. Experiments showed that VPP presented $10\times$ leverage performance for translating a Cisco configuration to Juniper format by identifying and fixing syntax errors, structural mismatches, attribute differences, and policy behaviours through 20 automated prompts. Implementing no-transit policies across 6 routers achieved $6\times$ leverage performance with 12 automated prompts guiding GPT-4 to resolve syntax, topology, and semantic policy errors.

Fig.~\ref{fig-configgen} summarizes three frameworks for LLM-based network configuration generation. In particular, the first framework employs a simplistic design, directly utilizing LLM to generate network configurations from natural language. However, the generated configurations may be inaccurate and require human inspection and improvement. In the second framework~\cite{DBLP:journals/corr/abs-2309-06342}, a hierarchical design is employed, where multiple LLMs collaborate to generate low-level network configurations step-by-step, aiming to enhance the final output. 
The verification scheme is crucial to evaluate the quality of the produced configuration, which may be placed in the second design as in~\cite{DBLP:conf/hotnets/MondalTBMV23} and \cite{DBLP:journals/corr/abs-2309-06342} to check the syntax, compilability, and correctness of the generated output. 
The third framework~\cite{DBLP:conf/hotnets/MondalTBMV23} is an automated design, incorporating an automatic verifier once the configuration is generated. This verifier validates the configuration and allows the LLM to automatically refine the output. 
While human inspection is still necessary, this approach significantly reduces the extent of manual intervention required. It is worth noting that these frameworks are not mutually exclusive and can be combined. For instance, in the hierarchical design, an automatic verifier can be added after each LLM iteration.

These existing studies have demonstrated the potential of using the LLM to configure networks automatically, which can be very useful in configuring telecom network settings.
The LLM offers promising opportunities for the automation of tedious tasks, reduction of human error and cost, and rapid prototyping and deployment of network infrastructure. 
However, telecom networks are complex systems with numerous interdependent components, and there are still many challenges to applying LLM technologies to telecom network configuration, e.g., contextual understanding, error handling and verification, security concerns, and interoperability between vendors and devices. 
For example, networks often comprise devices from various vendors, each with its own configuration language and parameters. The LLM must be capable of understanding and generating configurations that are compatible across these diverse environments. In addition, network configurations must adhere to security best practices. The LLM must be equipped to understand and apply these practices consistently to avoid creating security vulnerabilities.

\begin{table*}[!t]
\caption{Summary of LLM-based Generation for Telecom. }
\centering
\small
\setstretch{1.1}
\resizebox{1\textwidth}{!}{%
\begin{tabular}{|m{2cm}<{\centering}|m{2cm}<{\centering}|m{4cm}<{\centering}|m{3.5cm}<{\centering}|m{4cm}<{\centering}|m{4cm}<{\centering}|}
\hline
LLM-based generation applications & Specific scenarios & Main features & Prompt and fine-tunning requirement & Advantages compared with conventional approaches & Applications for telecom fileds\\
\hline
\multirow{8}*{\makecell{Domain \\ knowledge \\ generation}}  & Telecom-domain question answering  & Question answering is the most well-known application of LLMs. It represents a significant step forward in the ongoing effort to bridge the knowledge gap in telecom and empower individuals and organizations within the field, including telecom question answering, literature summary and review, etc.  &  General domain \blue{LLMs} can also answer telecom-related questions, but fine-tuning a telecom-specific LLM can provide more reliable and professional answers. CoT prompting may improve the answer quality. & The use of LLM techniques signifies a shift towards more efficient and accessible knowledge dissemination methods than any existing textbooks, websites, and tutorials for their comprehension and reasoning capabilities.   &\multirow{8}*[+5ex]{\makecell{1) Building a telecom-domain \\ LLM is a promising direction \\ to make telecom-knowledge \\ more accessible for both \\ professional researchers and \\ the public.\\ 2) Automated troubleshooting \\ is another promising \\ application to automate the\\ problem-solving process\\ in telecom fields.\\ 3) The LLM also have the \\ potential of generating \\ other language-related tasks, \\ e.g., specifications\\  and protocols. }}\\
\cline{2-5}
&  Generating solutions based on trouble reports   &  Using language models to generate troubleshooting solutions automatically. It considers trouble observations and fault information as input, and produces recommended solutions.   & The language model must be fine-tuned on telecom-domain language and trouble reports datasets. An extra document ranking fine-tuning is required to realize recommendation functions. &     Automated troubleshooting is a very promising technique to greatly save human time and effort, since the conventional approach relies on expert knowledge and trial-and-error tests.  &          \\
\hline
\multirow{6}*{\makecell{Code \\ generation}}  & Code refactoring &  Using the LLM for fundamental code refactoring and design validations, improving the code quality automatically without human intervention.   &  The prompt input is easier since no multi-step scheduling is involved. Fine-tuning the LLM based on existing codes can improve the quality of the generated code.  & \multirow{3}*[+4.5ex]{\makecell{ 1) Improving the readability, \\ efficiency, and reliability\\ of the project. \\ 
2) Considerably saving human \\ effort on coding, debugging,\\ and testing the project.\\ 
3) Lowering the requirement \\ for professional knowledge \\ when developing a system.  }}  & 
\multirow{3}*[+4.5ex]{\makecell{ Coding is one of the most\\ time-consuming part in \\ wireless system development.\\ Incorporating LLM-aided \\ coding can greatly save human \\ effort and improve the code \\ quality. However, datasets may \\ be required for fine-tuning,\\ which can be collected from\\ GitHub or wireless textbooks.}}\\
\cline{2-4}
& Coding tasks with multi-step scheduling & The LLM can also be used to generate complicated projects with multi-step scheduling and sub-tasks.   &  The input prompts have to be carefully designed in a CoT approach with appropriate examples and templates.  &    &  \\
\hline
Network configuration generation  & Automatic network configuration generation by using \blue{LLMs}.  &   Using the LLM to generate network configurations automatically, and then verify by \blue{LLMs} or humans.  & The prompt must be carefully designed due to the complexity of network configurations, e.g., dividing the prompts into the task-specific part and code generation part.   & The LLM enables efficient generation of network configurations, reducing manual effort and cost in the telecom industry.  & Applications include automatic network provisioning, performance tuning, security and compliance configuration, etc.\\
\hline
\end{tabular}}
\label{tab-genesummary}
\end{table*}

\subsection{Discussions and Analyses}

LLM techniques have promising generation capabilities for telecom applications, and Sections \ref{sec-knowledge} to \ref{sec-configuration} have introduced various scenarios for generating telecom knowledge, troubleshooting reports, code, and network configuration. 
Table \ref{tab-genesummary} summarized the main features, input and fine-tuning requirements, advantages, and telecom applications. 
In the following, we summarize the key findings and analyses.

\blue{Firstly, multi-step planning capabilities are crucial for telecom-related generation tasks.
Telecom networks are large-scale complicated systems, and many tasks require dedicated planning and scheduling.  
For example, the study in~\cite{du2023power} demonstrated that using a one-step prompt to generate a complicated 64-point FFT module is impractical, while step-by-step planning can achieve a satisfactory result. 
Similarly, step-by-step reasoning and planning are also useful to reproduce the results of existing publications for code generation problems~\cite{xiang2023toward}. 
Therefore, multi-step planning, i.e., step-by-step prompt design, is critical to obtain the desired output.} For instance, the prompt design~\cite{mani2023enhancing} is decoupled into the application prompt and code generation prompt, in which the application prompt focuses on task-specific requirements, and the code generation prompt uses plugins and libraries to instruct \blue{LLMs}.

Meanwhile, LLM-enabled generation can significantly save humane efforts.
Existing studies have shown that LLM has excellent capabilities for code refactoring and validation, which are usually solved manually with considerable human effort. 
Applying such a technique to the telecom field will significantly save human labour on projecting coding, validating, and debugging. 
For instance, Zhang \textit{et al.}~\cite{zhang2022repairing} introduce that the LLM can successfully repair 86.71\% programs for introductory level Python projects, and adding few-shot examples will raise the ratio to 96.50\%.
In addition, \blue{LLMs} can also abstract fundamental knowledge in the network field from textbooks, journals, and specification standards, which avoids the time-consuming literature review process.

\blue{LLMs} have been trained on many real-world datasets from web pages like Wikipedia, and they can be further fine-tuned on domain-specific datasets, e.g., telecom\cite{maatouk2023teleqna} and cybersecurity\cite{aghaei2022securebert}. 
Despite the satisfactory performance, there is no guarantee for the correctness of the generated output. 
Such risks are avoided when the generated code or network configuration can be verified by pre-designed test cases.    
However, when using the LLM to summarize or extract knowledge from existing literature, the quality of generated knowledge is hard to validate.   
For example, \blue{LLMs} may produce wrong numbers or units, and these mistakes can easily mislead users without professional knowledge.
To this end, efficient validation schemes are crucial to evaluate the performance of generated solutions, especially for coding and network configuration problems.
Human verification is a simplistic and straightforward approach, but it requires considerable human labour and can be inefficient. 
Therefore, automatic validation is the key to improve the overall efficiency of the whole pipeline, e.g., sending the code implementation error message to a LLM for automatic debugging\cite{xiang2023toward}, and using LLMs to validate the network configuration files.

\section{LLM-enabled Classification Problems}
\label{sec-class}

Classification problems are extensively studied within telecom networks.
Accurate and robust classification is crucial for improving network service quality and performance. 
This section will introduce the motivations and capabilities of LLM technologies in addressing a range of classification problems, including attack classification and detection, text classification, image classification, and encrypted traffic classification.

\begin{table*}[!t]
\caption{Summary of LLM-aided classification-related studies and telecom applications.}
\centering
\small
\setstretch{1.05}
\resizebox{1\textwidth}{!}{%
\begin{tabular}{|m{1.5cm}<{\centering}|m{0.8cm}<{\centering}|m{5cm}<{\centering}|m{6cm}<{\centering}|m{6cm}<{\centering}|}
\hline
Topics & Refer-ences & Proposed LLM-aided generation schemes & Key findings \& Conclusions & Telecom application opportunities\\
\hline
\multirow{25}*{\makecell{ Security \\ related \\ classification }} & \cite{aghaei2022securebert} &  Building a specialized cybersecurity language model (named SecureBERT) through fine-tuning RoBERTa~\cite{liu2019roberta} on a cybersecurity corpus.   & SecureBERT excels at understanding text within a cybersecurity context, which enables a strong generalization capability across various telecom security tasks.  &  \multirow{25}*{\makecell{By fine-tuning pre-trained general \\ LLM  models \cite{aghaei2022securebert, ameri2021cybert, yin2020apply} or \\ building security-specific \\ models from scratch~\cite{ferrag2024revolutionizing, 9804700}, LLM \\ models exhibit the advantage in \\ understanding security context, \\ enabling the application \\ of LLM techniques across a \\ range of telecom security tasks.\\ 
Existing studies show that \\ 
LLM-based method can outperform\\
existing ML and DL models in\\
terms of classification and\\
detection accuracy. LLM techniques \\ 
can also provide incident recovery\\
suggestions. However,\\
it is essential to initially create relevant \\ training and testing datasets extracted from \\ security-related telecom language corpora.} }     \\
\cline{2-4}
& \cite{ferrag2024revolutionizing} & Building a security-specific LLM from scratch designed for detecting network cyber threats, involving several steps: data preparation, data tokenization, model training, and model deployment.  & SecurirtBERT showcases the powerful predictive capabilities of security-specific \blue{LLMs} in identifying various types of attacks, significantly outperforming the traditional ML and DL models.  &          \\
\cline{2-4}
& \cite{ameri2021cybert} & Building a novel classifier of cybersecurity feature claims (named CyBERT) by fine-tuning a pre-trained BERT language model~\cite{devlin2018bert} on industrial control device documents. A large repository is created to gather industrial device information encompassing 41073376 words.  & CyBERT enables the effective identification of cybersecurity claim-related sequences, with an accuracy improvement of 19\% in comparison to the general BERT text classifier~\cite{devlin2018bert}.   & \\
\cline{2-4}
& \cite{yin2020apply} & Applying transfer learning to a BERT model~\cite{devlin2018bert} to extract changeable token embeddings from vulnerability descriptions. A pooling layer is positioned at the top to extract sentence-level semantic features. & The exploitability prediction framework (named ExBERT) not only accurately predicts software vulnerabilities but also learns sentence-level semantic features and captures long dependencies within descriptions.   & \\ 
\cline{2-4}  
& \cite{9804700}  & Applying a BERT model~\cite{devlin2018bert} to tokenize URLs within HTTP requests and then passing these tokens to a multilayer perceptron model to distinguish normal and anomalous HTTP requests. &  By integrating NLP with web attack detection, BERT~\cite{devlin2018bert} demonstrates strong capabilities in understanding web requests and SQL language, achieving remarkable detection performance that significantly surpasses that of traditional ML detection methods.   &  \\
\hline
\multirow{5}*{\makecell{Text \\ classification}} & \cite{aftan2023using}   &  It applied an AraBERT model to classify telecom customer satisfaction in Saudi Arabia by using the Twitter dataset.  &  BERT-based model obtained more accurate and stable results than conventional CNN and RNN algorithms.   & \multirow{5}*[+2ex]{\makecell{ LLM techniques have inherent \\ advantages in processing \\ text-related tasks. Existing studies have \\ shown that the LLM can achieve a\\ comparable performance as existing  CNN \\ or  RNNs. It is promising for text-related \\ telecom tasks such as standard developing \\ and user feedback processing\cite{terra2021q}. }   }\\
\cline{2-4}
&  \cite{bariah2023understanding}   &  Fine-tuning several \blue{LLMs}, e.g., BERT, distilled BERT, RoBERTa and GPT-2, to the telecom domain languages, and using them for 3GPP standard classification problems. &    With proper pre-processing and fine-tuning, the experiment in \cite{bariah2023understanding} can achieve an 80\% accuracy even if only 20\% of the text segments are used.  &          \\
\hline
\multirow{5}*{\makecell{ Image \\ classification}} &  \cite{matsuuravisual}
& It investigates the zero-shot image classification capabilities of LLaVA model, which means using the model directly without any extra training.  &  The performance can be significantly improved with a combination of carefully crafted prompts, hierarchical classification strategies, and adjusted model temperatures. 
& \multirow{5}*{\makecell{  Images are important information \\ for telecom sensing. Enabling efficient image \\ classification can be very useful for\\ many telecom applications, including \\ vision-aided sensing, mmWave beamforming\\ \cite{charan2021vision}, user localization\cite{yao2022joint}, and so on.
}   }\\
\cline{2-4}
& \cite{pratt2023does}   &  Using LLM's inherent knowledge to generate descriptive sentences with crucial discriminating characteristics of the image categories. &  This simple approach can effectively improve the zero-shot image classification accuracy on a range of benchmarks. & \\
\hline
\multirow{10}*{\makecell{ Network \\ traffic \\ classification}} & \cite{shi2023bfcn} & Capturing long-distance contextual relations within traffic sequence through BERT, and then integrating packet-level token semantic features at the forward and backward positions of BiLSTM, which enhances the BiLSTM attention to packet-level features.   & BiLSTM can capture relevant features of front and rear token sequences after BERT extracts general features of encrypted traffic, learning the long-distance relations within token sequences. 
& \multirow{5}*[+5ex]{\makecell{The LLM facilitate effective encrypted traffic \\ classification, a critical technique in telecom \\ network management while protecting data \\ and user privacy. Note that the assumption of \\ clean pre-training data presents challenges in \\ secure traffic classification. This vulnerability \\ is exposed particularly when attackers craft \\ a poisoned model with backdoors by inserting \\ low-frequency words as toxic embeddings. \\ Such manipulation allows attackers to deceive \\ the normally fine-tuned model during specific \\ classification tasks. }   }\\
\cline{2-4} & \cite{lin2022bert}   &  Building an Encrypted Traffic BERT (named ET-BERT), which aims to learn generic traffic representations from large-scale unlabeled encrypted traffic. & ET-BERT showcases strong effectiveness and generalization across five encrypted traffic classification tasks, e.g., General Encrypted Application Classification~\cite{van2020flowprint}, Encrypted Malware Classification, Encrypted Traffic Classification on VPN~\cite{draper2016characterization}, etc.  &          \\
 \hline
\end{tabular}}
\label{tab-llm-classification}
\end{table*}

\subsection{Motivations and Classification Capabilities of \blue{LLMs}}

Conventional classification techniques heavily rely on statistical methods. 
However, with the recent advancements in telecom networks, there has been a surge in multi-modal and heterogeneous network data, e.g., numerical traffic data, textual security logs, and environmental images, which presents significant challenges for traditional classification techniques, indicating a need for more advanced and adaptable approaches.
Recently, LLM techniques have shown their capability to effectively process multi-modal and heterogeneous data across both natural language and computer vision fields. 
These capabilities position them as a promising research direction for addressing classification problems within telecom networks. 

Firstly, LLM technologies can contribute to telecom security by automated security language understanding and classification.
Inspired by numerous advancements in NLP, LLM excels at explaining textual contents and transforming them into informative representations, such as GPT~\cite{radford2018improving} and BERT~\cite{devlin2018bert}. 
With their strong capabilities, LLMs have recently shown exceptional superiority in attack detection, aiming to enhance the security of telecom networks\cite{ferrag2024revolutionizing}.   
Through fine-tuning pre-trained models or developing LLMs from scratch, LLMs retain the functional capabilities in general English while gaining a thorough understanding of the specialized security language, allowing \blue{LLMs} to effectively identify and respond to security threats in telecom networks.


Secondly, LLM techniques have inherent advantages in text-related classification tasks, which are very useful for text processing and classification in the telecom field, e.g., customer textual feedback, telecom standard specifications, technique reports and publications. 
For example, enhancing the quality-of-experience (QoE) of telecom services hinges on a comprehensive understanding of customer feedback, which may include various real-world topics, ranging from signal strength to sending messages and calls \cite{mitra2013context}.
Given the superiority across various text-related tasks, LLMs have strong capabilities to classify customer comments and extract useful feedback, allowing telecom operators to enhance service quality by properly understanding user satisfaction levels.

In addition, \blue{LLMs} can extract visual features from the dynamic and complex telecom environment.
The integration of computer vision and image processing into the telecom field, such as equipping BSs with cameras to pinpoint user locations, can boost network efficiency in the dynamically changing wireless environment. 
Although primarily focusing on processing and understanding textual information, some \blue{LLMs} also have remarkable image processing capabilities in vision-related tasks\cite{alayrac2022flamingo}, including image-to-text generation~\cite{tewel2022zerocap} and object detection~\cite{du2022learning}, etc. 
Consequently, this integration enables the LLM to analyze both visual and network data, which can effectively bridge the gap between textual and visual data analysis, leading to a more comprehensive approach for network management.


Finally, LLM's zero-shot classification capabilities have been demonstrated in multiple existing studies, such as text and image classification tasks\cite{matsuuravisual,pratt2023does}. 
In particular, it means that the LLM can be used to classify and detect objects by using the real-world knowledge learned in the pre-training phase, and no extra training is required for target tasks.  
Such zero-shot classification capabilities can be appealing for telecom networks since many telecom classification tasks need rapid responses, e.g., network attack detection\cite{ferrag2024revolutionizing}, image processing and classification\cite{charan2021vision}.    
With the above potential and motivations, in the following, we will introduce LLM-enabled attack classification and detection, text classification, image classification, and encrypted traffic classification problems.

\subsection{LLM for Telecom Security and Attack Detection}
\label{sec-security}

\begin{figure*}[t]
\centering
\setlength{\abovecaptionskip}{0pt} 
\includegraphics[width=0.95\linewidth]{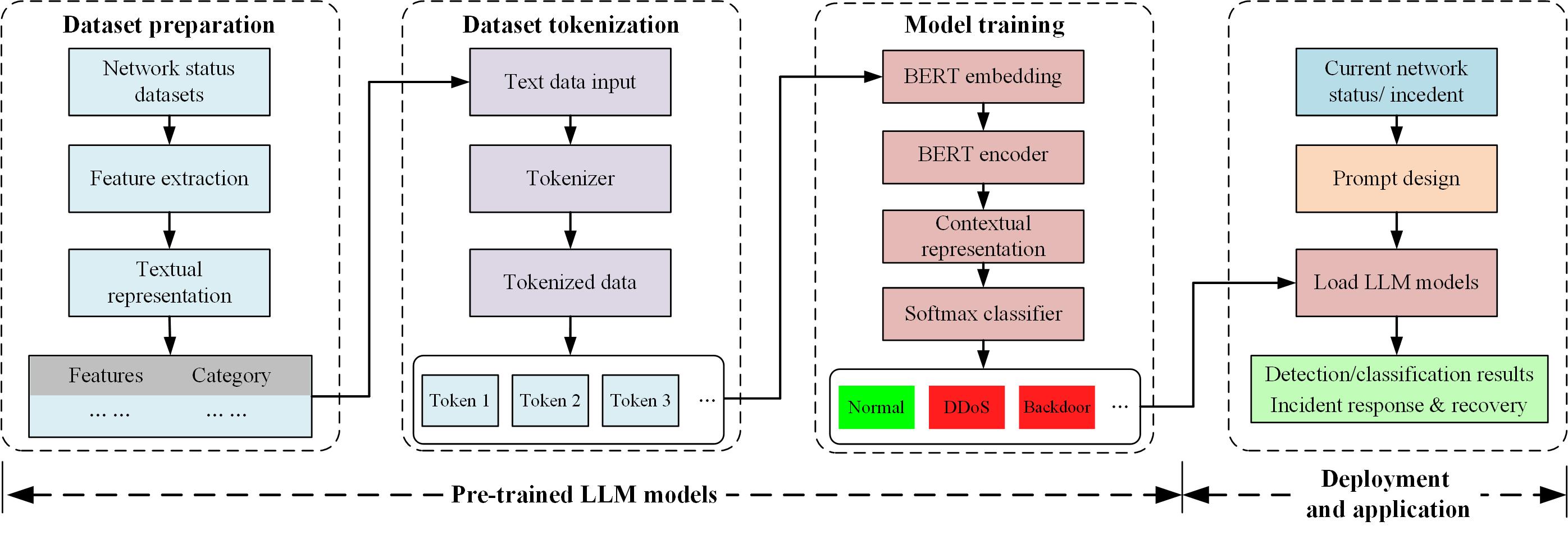}
\caption{Framework of LLM-based attack detection\cite{ferrag2024revolutionizing}.}
\label{fig-attackdetection}
\end{figure*}

The numerous advancements in telecom have led to more complex and interconnected infrastructures with a wide range of technologies, protocols, and services, which can pose significant challenges in controlling and monitoring telecom security.
The growing threats and incidences of hostile attacks have exposed severe vulnerabilities in telecom.
For instance, Denial of Service (DoS) can decrease network availability by overwhelming systems, and Man-in-the-Middle (MITM) attacks can violate network integrity by secretly modifying communications between two parties.
This underscores the requirement for robust attack detection mechanisms to monitor the network system against malicious activities.
However, with the evolution of current telecom networks, a surge of multi-modal network data can be captured, including numerical measurements such as traffic loads and CSI, and descriptive textual contents with device logs and network configurations.   
The data contains a substantial amount of redundant and correlated information, potentially obscuring critical patterns in attack detection, which poses significant challenges to achieving accurate attack detection. 

Recently, NLP has achieved numerous successes in capturing informative features from multi-modal and heterogeneous data across various application scenarios, including sentiment analysis, speech recognition, and machine translation, among others.  
Specifically, LLM techniques have emerged as a promising direction across various NLP applications, which are beneficial to explaining textual inputs and transforming them into quantitative representations. 
The common method to apply \blue{LLMs} across various domains involves employing general-domain models as baselines, followed by fine-tuning them for specific domain tasks.
%
To enhance the security of telecom networks through LLM techniques, it is important to note that the security language, such as ransomware, API, OAuth, and keylogger, significantly differs in structure and semantics from the general linguistic language. 
This suggests that a conventional LLM may find it challenging to fully understand the specific vocabulary inherent to security-related texts, potentially leading to limited generalization ability in security applications.
To this end, existing studies that employ LLM techniques for attack detection can be categorized into two primary directions as follows:

\subsubsection{\textbf{Fine-tuning pre-trained \blue{LLMs}}}

Existing studies have leveraged pre-trained LLMs and adapted them to achieve specific security objectives through model fine-tuning \cite{aghaei2022securebert}.
For instance, Aghaei \textit{et al.}~\cite{aghaei2022securebert} introduce a specialized cybersecurity language model named SecureBERT, which is capable of processing texts with cybersecurity implications and effectively applied across a broad range of cybersecurity tasks, including phishing detection, intrusion detection, code and malware analysis, etc. 
In particular, SecureBERT applies a cybersecurity corpus comprising 1.1 billion words, divided into 2.2 million documents, with each document averaging 512 words through the Spacy text analytic tool~\cite{spacy}. 
To build the security-customized tokenizer, a byte pair encoding method is employed to extract 50265 tokens from the cybersecurity corpus to generate the initial token vocabulary.
Among all the extracted tokens, SecureBERT and RoBERTa~\cite{liu2019roberta} share 32592 mutual tokens, while SecureBERT identifies 17673 tokens specific to the cybersecurity corpus, including firewall, breach, crack, ransomware, malware, phishing, and vulnerability, among others. 
Each token is represented by an embedding vector with dimensions identical to those in pre-trained RoBERT, augmented with random Gaussian noise added to the embedding factor of each token.
SecureBERT emulates the architecture framework of RoBERT~\cite{liu2019roberta}, encompassing twelve transformer and attention layers, which are trained on the specifically collected corpus through the customized tokenizer tailored to the unique task requirements.

%
SecureBERT is evaluated to predict masked security-related words within a sentence, which is the task known as masked language models.
The testing dataset is generated by extracting sentences from cyber-security reports with 17341 records.
The experiment shows that SecureBERT outperforms RoBERTa, powerful model on general language, in predicting masked words within a sentence with a security context, as illustrated in the following examples\cite{aghaei2022securebert}: 

 \begin{tcolorbox}
  [title = {Comparisons between SecureBERT and RoBERTa in masked tasks\cite{aghaei2022securebert}}]
\textit{\textbf{Task 1:} “Information from these scans may reveal opportunities for other forms $<$\textbf{mask}$>$ establishing operational resources, or initial access.”}

\textit{\textbf{SecureBERT:} reconnaissance. \\
\quad \textbf{RoBERTa:} early.} 
\vspace{5pt}

\textit{\textbf{Task 2:} “Search order $<$\textbf{mask}$>$ occurs when an adversary abuses the order in which Windows searches for programs that are not given a path.”}

\textit{\textbf{SecureBERT:} hijacking. \\
\textbf{RoBERTa:} abuse.} 
\vspace{5pt}

\textit{\textbf{Task 3:} “Botnets are commonly used to conduct $<$\textbf{mask}$>$ attacks against networks and services.”}

\textit{\textbf{SecureBERT:} DDoS. \\
\textbf{RoBERTa:} automated.} 
  \end{tcolorbox}

The three predicted terms \textit{reconnaissance, hijacking, and DDoS} are prevalent in cybersecurity corpora. SecureBERT accurately understands the security context to predict these masked words, whereas RoBERTa exhibits incorrect prediction, underscoring the advantages of SecureBERT in security-related language tasks.

\subsubsection{\textbf{Building security-specific \blue{LLMs} from scratch}}

In addition to fine-tuning, another strategy is to build an LLM from scratch specifically designed for network-based attack detection. 
For example, Ferrag \textit{et al.}~\cite{ferrag2024revolutionizing} designed SecurityBERT for detecting the ever-evolving cyber threat landscape, which involves several steps: data preparation, data tokenization, model training, and model deployment, as shown in Fig.~\ref{fig-attackdetection}. 
In particular, the authors utilize a publically available dataset EdgeIIoTset~\cite{9751703} related to the Internet of Things (IoT) and Industrial IoT (IIoT) connectivity protocols, categorized into five types of threats: DoS/D-DoS attacks, information gathering, MITM attacks, injection attacks, and malware attacks.
Then, to leverage the power of \blue{LLMs}, null features are eliminated during the feature extraction in \cite{ferrag2024revolutionizing}, and both numerical and categorical features are converted into textual representations. 
Specifically, each feature is combined with its column name and value and then subjected to hashing. The hashed values from the same instance are merged into a sequence, which generates a fixed-length textual representation of the network traffic data while maintaining privacy. 
After that, ByteLevelBPETokenizer~\cite{wolf2019huggingface} is subsequently applied to segment the textual representations of the network traffic data. This segmentation process breaks down the text into smaller subwords, expected to be found in the tokenizer’s vocabulary.
After the pre-training phase, the model is fine-tuned on a labelled dataset~\cite{9751703} by adding a Softmax activation function at the output layer, which allows SecurityBERT to enable the learned contextual representations in the specific task of attack detection. 
Finally, in the deployment phase of Fig.~\ref{fig-attackdetection}, once attacks are identified through SecurityBERT, FalconLLM is further employed to determine the severity and negative impact of identified attacks, leading to the formulation of potential mitigation strategies and recovery procedures.

SecurityBERT is employed to identify normal events and 14 distinct attacks in \cite{ferrag2024revolutionizing}, such as DDoS\_UDP, DDoS\_ICMP, SQL\_Injection, Vulnerability\_Scanner, etc. 
The experiment shows that SecurityBERT achieves the average accuracy, recall, and F1-score of 0.98, 0.84, and 0.84, respectively, demonstrating the strong classification capabilities of security-specific \blue{LLMs} in identifying various types of attacks.
In addition, SecurityBERT significantly surpasses the performance of traditional ML and deep learning models such as decision tree, convolutional neural networks (CNN), recurrent neural network (RNN), and long short-term memory (LSTM).

Finally, to develop security-specific LLM technologies for telecom networks, it is essential to initially create relevant training and testing datasets extracted from security-related telecom language corpora. 
Following model fine-tuning with security-customized tokenizers, these language models can significantly boost performance across various telecom security tasks, including cyber threat intelligence, vulnerability analysis, and threat action extraction~\cite{aghaei2019threatzoom, aghaei2020threatzoom}.


\subsection{Text Classification}
Text classification and processing is a useful technique for the telecom industry, and the applications include user enquiries and intent classification and analyses\cite{terra2021q}, automated trouble report classification\cite{yayah2021automated}, standard specification classification\cite{bariah2023large}, and so on.
In the following, we introduce two applications in telecom customer feedback analyses and specification classification.

\subsubsection{ \rm \textbf{Using LLMs for telecom user feedback classification and analyses}}
Understanding user feedback is crucial for telecom operators to improve the QoE and maintain customer satisfaction and loyalty. 
For instance, Vieira \textit{et al.} applied CNN and LSTM networks in \cite{terra2021q} for sentiment analysis and topic classification, and the analysis proved that 78.3\% of the complaints are related to weak signal coverage, and 92\% of these regions have coverage problems considering a specific cellular operator.
These analyses can be particularly useful for telecom operators to improve service quality such as signal coverage and strength.  
However, user feedback can be complicated by involving service experiences, suggestions, recommendations, and complaints.  
In addition, the feedback can be collected from various sources, e.g., social media, websites, phone calls, and company collection. 
These challenges require more advanced ML techniques to better classify and capture user's intentions. 
The LLM has shown superb performance in a range of text-related tasks, e.g., question answering, summarization, dialogue, and sentiment analysis, outperforming many existing techniques even in zero-shot settings. 
For instance, Aftan \textit{et. al} applied AraBERT model to classify telecom customer satisfaction in Saudi Arabia by using the Twitter dataset \cite{aftan2023using}, and the BERT-based model obtained more accurate and stable results than conventional CNN and RNN algorithms.
In addition, using \blue{LLMs} to analyze customers' experience and intent has attracted considerable interest from both industry and academia, e.g., Microsoft has proposed to use LLMs to generate, validate, and apply user intent taxonomies\cite{shah2023using}. 
Therefore, it shows great promise in integrating LLM technologies into the telecom industry for text-related classification tasks.

\begin{figure}[t]
\centering 
\includegraphics[width=0.95\linewidth]{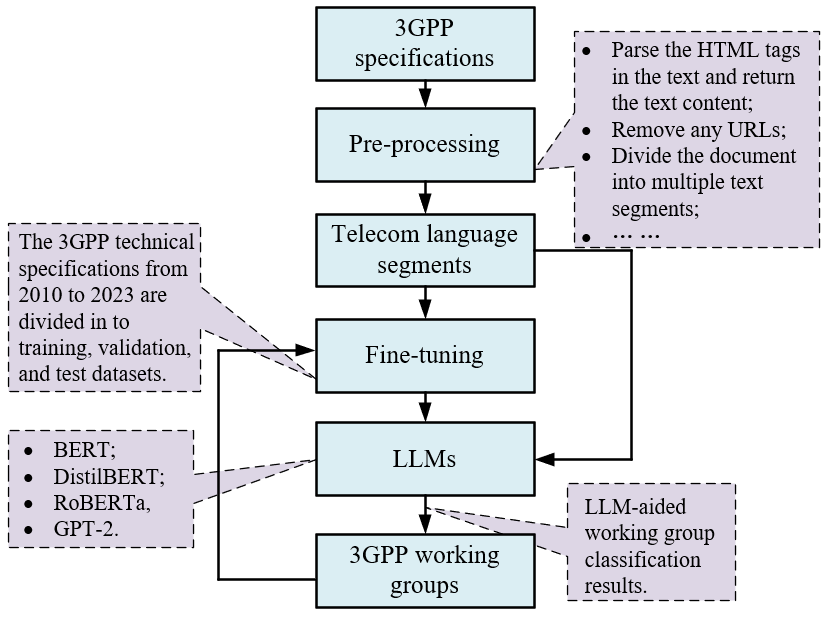}
\caption{Framework of LLM-aided 3GPP specification classification\cite{bariah2023understanding}.}
\label{fig-specification}
\end{figure}

\subsubsection{\rm \textbf{LLM-aided telecom standard classification}}
Telecom standards refer to agreed-upon specifications that ensure the interoperability, security, and reliability of telecom services. Standards play a critical role in global telecoms \cite{ibarrola2023evolution}, such as 3G, 4G, and emerging 5G for mobile communications, IEEE 802.11 for Wi-Fi, and ITU-T recommendations. 
For instance, 3GPP is the main organization for telecom standard development, which includes three technical specification groups, and each specification group consists of multiple working groups.
Given the large number of existing specifications with diverse topics, Lina \textit{et al.} proposed to use \blue{LLMs} for specification classification, classifying the text into an existing working group automatically\cite{bariah2023understanding}.  
Fig.\ref{fig-specification} summarized the key processes of using \blue{LLMs} to classify the 3GPP specifications. 
With proper pre-processing and fine-tuning, the experiment in \cite{bariah2023understanding} can achieve an  80\% accuracy even if only 20\% text segments are used. 
The experiment results also prove that increasing the length of technical text segments can significantly improve classification accuracy. 

Textual descriptions and documents are frequently involved in the telecom industry, e.g., user comments, standard specifications, technical and troubleshooting reports, etc. Incorporating \blue{LLMs} for text processing and classification will contribute to more intelligent and reliable telecom networks.

\subsection{Image Classification}
\label{sec-image}
Computer vision is an important approach for environment sensing, and there have been many existing studies toward vision-aided 6G networks. 
For instance, vision-aided blockage prediction and beamforming are investigated in \cite{charan2021vision} and \cite{ahn2022towards}. Specifically, the authors assume that the cameras attached to the BS can capture the environment image and then use deep learning to detect objects and 3D user locations. 
These studies have shown the importance of incorporating computer vision and image processing in telecom fields to better sense the wireless environment. 
Therefore, efficient image processing and object classification are the prerequisites for realizing vision-aided wireless networks.   
For example, Civelek \textit{et al.} \cite{civelek2016automated} proposed an automated moving object classification technique in wireless multimedia sensor networks, and such schemes can also be exploited in previous studies such as \cite{charan2021vision} and \cite{ahn2022towards} for efficient object detection. 
In addition, Kim \textit{et al.} propose an edge-network-assisted real-time object detection framework \cite{kim2021edge}. Specifically, the vehicles can compress the image based on the region of interest and transmit the compressed one to the edge cloud. Considering the limited computation resources at the BS, this can be a useful technique for BS-edge-cloud image processing and environment sensing.

The wireless environment can be very complicated with walking pedestrians, moving vehicles, building blockages, and other obstacles. 
Therefore, it requires dedicated model training and fine-tuning to extract useful information and identify specific objects.
\blue{LLMs} have been pre-trained on a huge amount of real-world datasets, and some LLMs, such as flamingo \cite{alayrac2022flamingo} and GPT-4V \cite{yang2023dawn}, have proved versatile capabilities on various vision-related tasks, e.g., using text to generate images, describing given images, and object detection. 
For instance, Matsuura \textit{et al.} investigate the zero-shot image classification capabilities of the LLaVA model \cite{matsuuravisual}, and they found that the performance can be significantly improved with a combination of carefully crafted prompts, hierarchical classification strategies, and adjusted model temperatures.
Meanwhile, Pratt \textit{et al.} \cite{pratt2023does} also demonstrate that 
using LLM's knowledge can immediately improve zero-shot accuracy on a variety of image classification tasks, saving considerable manual effort. 
In addition, LLMs can also describe and summarize the image content for further classification, documentation, and processing, and an example is given in \cite{xu2024large} by generating the accident report of a car crash.

\begin{figure}[t]
\centering 
\includegraphics[width=0.75\linewidth]{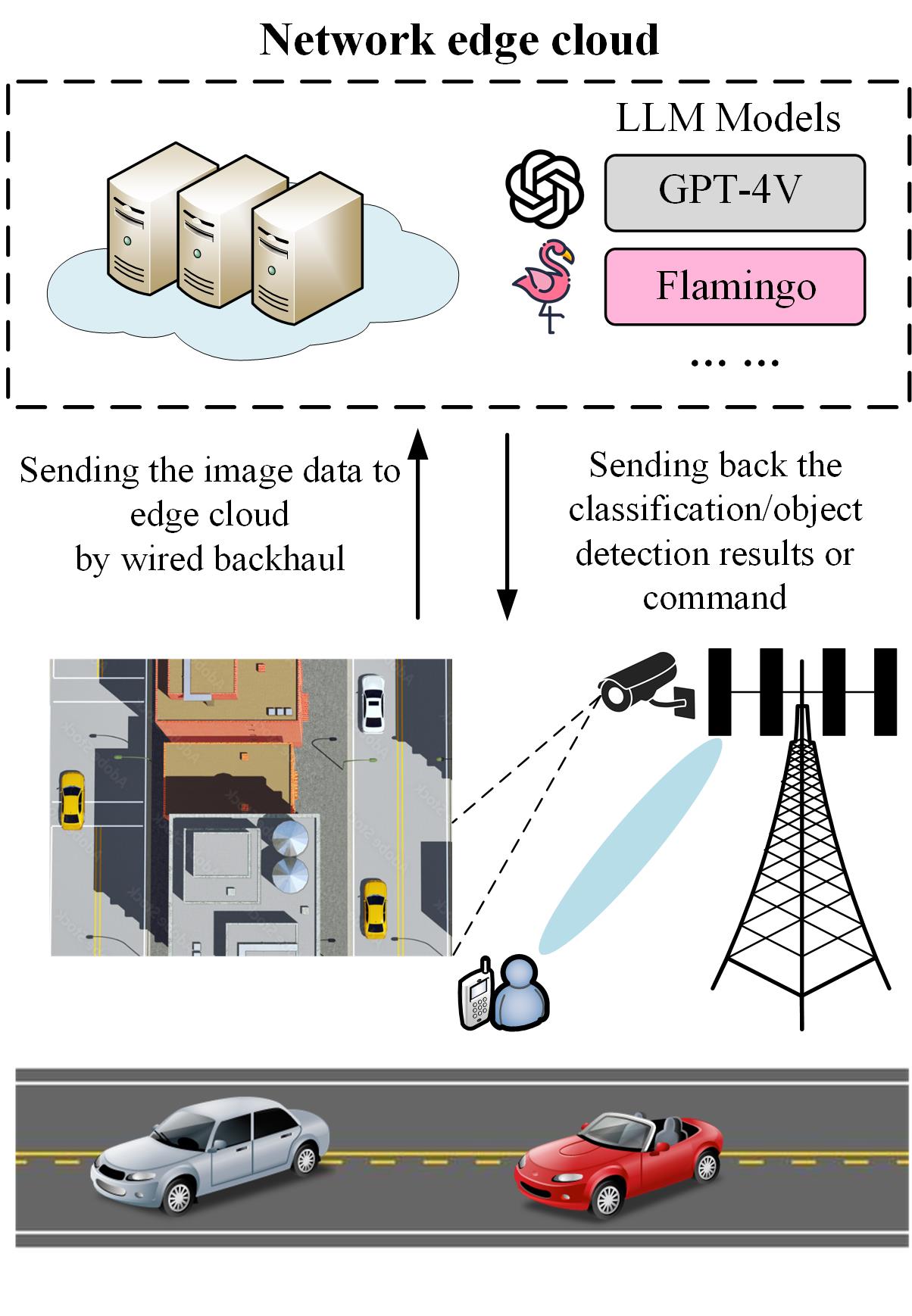}
\caption{Framework of LLM-aided computer vision in wireless networks.}
\label{fig-image}
\end{figure}

Finally, Fig.\ref{fig-image} presents an example of using \blue{LLMs} for image classification and object detection in radio access networks. 
In particular, the cameras attached to the BS can capture environmental images, and then the image data will be sent to the LLM at the network edge by wired backhaul.
The LLM can use computational resources at the edge cloud for image processing, classification, and detecting object locations such as vehicles, users, and blockage buildings. 
After that, the edge cloud can send back the classification and detection results to BSs, and then the BS can adjust the beamforming and hand-off decisions accordingly.

\subsection{Encrypted Traffic Classification}

Network traffic classification is an essential technique in telecom network management, which aims at identifying the category of traffic from various applications~\cite{bujlow2015independent}. 
Specifically, the widespread utilization of traffic encryption plays a significant role in protecting data and user privacy. However, it also presents challenges in capturing implicit and robust patterns within encrypted traffic, which is essential for network management. 
To tackle these challenges, conventional methods~\cite{van2020flowprint} usually extract features within encrypted traffic such as certificates to create fingerprints for classification through fingerprint matching, while these methods fall short with the advent of advanced encryption techniques. 
Additionally, existing ML-based studies~\cite{lin2021tscrnn} can automatically extract complex and abstract features to analyze encrypted traffic, resulting in notable performance improvement. However, these methods are heavily dependent on the amount and distribution of labelled training data, leading to limited generalization ability due to model bias.  

Recently, pre-training-based methods have achieved great breakthroughs across a wide range of application fields. In particular, pre-trained models are designed to learn data representations from unlabelled data, allowing these representations to be effectively applied to downstream tasks through fine-tuning models on labelled data.
In the context of encrypted traffic classification, Ma \textit{et al.}~\cite{shi2023bfcn} capture long-distance contextual relations within traffic sequence through BERT, and then integrate packet-level token semantic features at the forward and backward positions of BiLSTM, enhancing the BiLSTM attention to packet-level features.
BERT-BiLSTM is evaluated to identify the types of network communication application activities using the ISCX VPN dataset~\cite{sirinam2018deep}, which includes various \textit{pcap} files corresponding to different application activities. 
The dataset is comprised of 17 label categories, with each label representing a distinct type of application activity, including Email, Facebook, Gmail, Netflix, SCP, Skype, Youtube, and Spotify, among others.  
BERT-BiLSTM effectively distinguishes each application, achieving an overall accuracy of 99.70\%, precision of 99.34\%, recall of 99.51\%, and F1-score of 99.43\%, thereby surpassing the performance of traditional ML methods. 
The performance enhancement further indicates the advantages of BERT-BiLSTM in encrypted traffic classification: (1) BiLSTM captures the relevant feature of front and rear token sequences after BERT extracts general features of encrypted traffic, learning the long-distance relations within token sequences. (2) BiLSTM captures packet-level features and contextual relations by simultaneously integrating packet-level token semantic features at both forward and backward starting positions of BiLSTM.

\begin{figure}[t]
\centering
\setlength{\abovecaptionskip}{0pt} 
\includegraphics[width=1\linewidth]{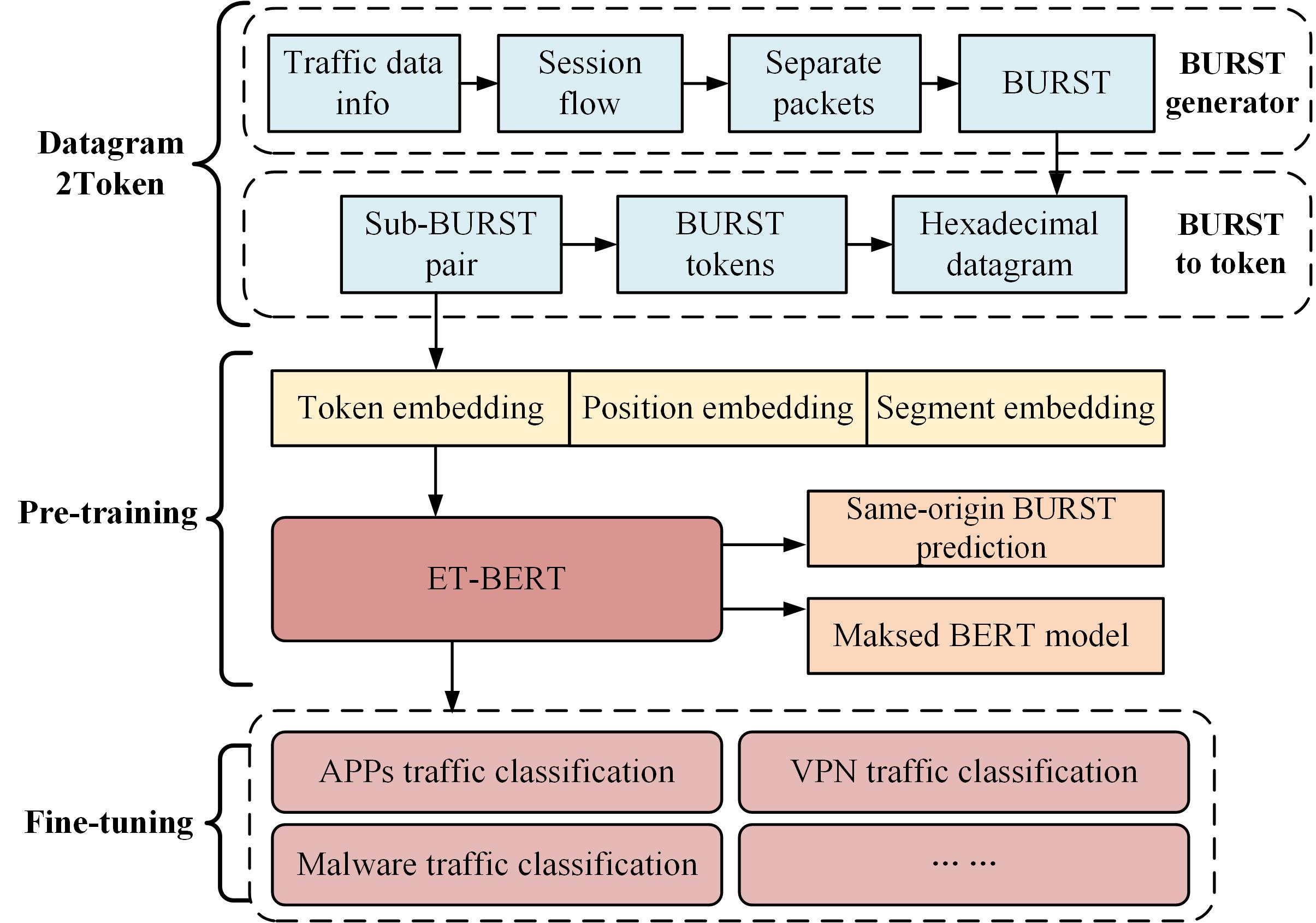}
\caption{Framework of LLM-based encrypted traffic classification\cite{lin2022bert}. "BURST" refers to a set of time-adjacent network packets originating from the request or the response in a single session flow, and therefore a group of BURSTs can characterize the network flow transmission patterns.}
\label{fig-trafficclassification}
\end{figure}

Moreover, Lin \textit{et al.}~\cite{lin2022bert} introduce Encrypted Traffic BERT (ET-BERT), as shown in Fig.~\ref{fig-trafficclassification}, which aims to learn generic traffic representations from large-scale unlabelled encrypted traffic. 
Concretely, Datagram2Token is first utilized to convert traffic flow into word-like tokens, through three steps: (1) BURST Generator extracts BURST time-adjacent network packets representing the session information. (2) BURST2Token applies a bi-gram model to convert the datagram of each BURST into token embeddings and divides a BURST into two segments for subsequent pre-training tasks. (3) Token2Embedding merges the token embeddings, position embeddings, and segmentation embeddings of each token as the input representations for pre-training.
To demonstrate the effectiveness and generalization of ET-BERT, the authors conduct experiments across several encrypted traffic classification tasks, e.g., general encrypted application classification \cite{van2020flowprint}, encrypted malware classification, encrypted traffic classification on VPN \cite{draper2016characterization}, with the remarkable improvements over existing state-of-the-art methods by 5.4\%, 0.2\%, and 5.2\%. 

Although ET-BERT exhibits a strong generalization capability across various tasks, the assumption of clean pre-training data presents challenges in secure traffic classification. 
This vulnerability is exposed particularly when attackers craft a poisoned model with backdoors by maliciously inserting low-frequency words as toxic embeddings. Such manipulation allows attackers to deceive the normally fine-tuned model during specific classification tasks.
Hence, how to construct toxic tokens within encrypted traffic can be potentially investigated as a promising direction in the field of LLM-based encrypted traffic classification.

\begin{table*}[!t]
\caption{Summary of LLM-enabled classification for telecom. }
\centering
\small
\setstretch{1.1}
\resizebox{1\textwidth}{!}{%
\begin{tabular}{|m{2cm}<{\centering}|m{4.5cm}<{\centering}|m{5cm}<{\centering}|m{4.5cm}<{\centering}|m{4cm}<{\centering}|}
\hline
LLM-based forecasting techniques & Main features & Prompt and fine-tuning requirements & Advantages compared with conventional approaches & Network classification application opportunities\\
\hline
Attack detection & Attack detection is vital for maintaining the security and reliability of telecom networks. LLMs have showcased a strong ability to capture discriminative information within multi-modal and heterogeneous network data. & Fine-tuning LLMs on security-specific language emerges as a promising approach for attack detection. This method allows fine-tuned \blue{LLMs} to maintain their proficiency in processing general English vocabulary while excelling at achieving specific security objectives.  &  LLM techniques have the strong advantages of processing both numerical traffic loads and descriptive security-related textual contents, e.g., ransomware and keylogger, achieving better performance than existing ML and DL algorithms.  &  \blue{LLMs} can be effectively employed for detecting cyber attacks~\cite{aghaei2022securebert, ferrag2024revolutionizing} and contributing to the mitigation and recovery strategies against such attacks~\cite{ferrag2024revolutionizing}. \\
\hline
Text classification & Text classification and processing are very useful for the telecom industry. \blue{LLMs} have shown great promise in understanding and processing text and languages. & Fine-tuning \blue{LLMs} on telecom language is required, e.g., network trouble report datasets and 3GPP technical specifications. There are no specific requirements for prompts.  & Automatic text classification and processing will greatly save human efforts on many document-related tasks, e.g., automated troubleshooting report generation and ranking.        &  The telecom applications include user enquiries and intent classification and analyses\cite{terra2021q}, automated trouble report classification\cite{yayah2021automated}, standard specification classification\cite{bariah2023large}.    \\
\hline
Image classification   &  Computer vision is a very useful technique for 6G networks, enabling 3D sensing for the environment. Some LLMs have shown impressive capabilities in image and vision-related tasks.  &  The study in \cite{matsuuravisual} shows that carefully crafted prompts are critical to improving the classification performance of \blue{LLMs}, e.g., "\textit{Fill in the blank, this is a picture of \{...\}  }". However, fine-tuning \blue{LLMs} on telecom-image datasets can improve classification accuracy. 
&  LLM's zero-shot learning capability can avoid the complexity of dedicated model training. For instance, \cite{matsuuravisual} achieved a satisfactory performance by pure prompt engineering without any fine-tuning.               
&  LLM-aided image classification can be used for blockage prediction\cite{ahn2022towards}, proactive beamforming and hand-off\cite{charan2021vision},  user localization \cite{yao2022joint}, etc.      \\
\hline
Traffic classification &  Network traffic classification is an essential technique in network management and security, which aims at identifying the category of traffic from various applications. LLM techniques have demonstrated remarkable performance in encrypted traffic classification.  & Fine-tuning \blue{LLMs} on labeled network data is crucial for ensuring their adaptability across various traffic classification scenarios, such as single packet and single flow classification.  &  LLMs are capable of learning generic traffic representations from extensive amounts of unlabeled, encrypted traffic without plaintext, resulting in extracting valuable insights from encrypted traffic for downstream traffic classification. & Traffic analyses and classification are very common tasks in telecom networks. LLMs can be effectively applied for encrypted traffic classification~\cite{lin2022bert, shi2023bfcn}.\\
\hline
\end{tabular}}
\label{tab-class-summary}
\end{table*}

\subsection{Discussions and Analyses}
Table \ref{tab-class-summary} summarized LLM-enabled classification techniques in terms of the main features, prompt and fine-tuning requirements, advantages, and network classification application opportunities. 
It shows LLM's versatile capabilities on different classification tasks, ranging from textual security logs and customer comments to images and network traffic files.

In particular, Section \ref{sec-security} demonstrates that LLM techniques have great potential for telecom network security.
Security is an important topic for telecom operations, and \blue{LLMs} can contribute through their classification and detection capabilities. 
In particular, the LLM can handle multi-modal and heterogeneous network data, e.g., CSI, traffic load level, network device logs and network configurations, and then extract useful network security information from these correlated inputs. 
Additionally, some LLMs can also recommend response and recovery strategies for network incidents\cite{ferrag2024revolutionizing}. This indicates the potential of building an end-to-end telecom security system, from status monitoring and attack detection to incident response and recovery.

Meanwhile, LLM can serve as a zero-shot classifier.
Telecom networks indicate a complicated dynamic environment, leading to various tasks.
Existing methods are usually task-specific, with dedicated designs for each incoming request.  
By contrast, some \blue{LLMs} have shown zero-shot classification capabilities\cite{matsuuravisual}. For instance, they can be directly used to classify images captured by the cameras on the BS, or analyze customer comments without prior training.   
Such a feature can be very useful in handling diverse tasks in complicated telecom systems such as object detection and user localization.
In addition, \blue{LLMs} have outstanding capabilities in processing text-related tasks, including both natural languages, such as customer comments and system language like network log files. 
These textual tasks are usually performed manually, but \blue{LLMs} can easily handle different classification and detection tasks with much less human intervention.

Finally, \blue{LLMs} can contribute to vision-aided telecom.
Sensing is increasingly important for wireless networks, and computer vision is an important approach to capturing wireless environment dynamics. 
With pre-trained real-world knowledge, \blue{LLMs} can be directly used for image and vision-related tasks, such as image description, image-text transformation, object detection, and image classification. 
In addition, LLM technologies also have advantages over conventional algorithms in terms of generalization capabilities. 
This means that LLMs can process various telecom tasks without extra training, e.g., blockage detection and prediction by using BS cameras\cite{ahn2022towards}, proactive beamforming and hand-off\cite{charan2021vision}, and user localization \cite{yao2022joint}.

\section{LLM-enabled Optimization Techniques for telecom }
\label{sec-optimize}

Optimization techniques are of paramount importance to telecom network management, and this section presents LLM-enabled optimization techniques. It first analyzes the motivations and optimization capabilities of LLMs, and then introduces LLM-aided reinforcement learning, black-box optimizer, convex optimization, and heuristic algorithms along with network optimization applications. Finally, we analyze and summarize the key findings.

\subsection{Motivations and Optimization Capabilities of LLM}

Optimization problems have been widely investigated in the communication field due to their critical importance. Existing optimization techniques can be categorized into several approaches\cite{zhou2023survey}: ML-based, convex optimization, heuristic algorithms, and black-box optimization. For instance, reinforcement learning is a widely considered ML algorithm to solve optimization problems\cite{zhou2021ran}. Meanwhile, fractional programming is a well-known convex optimization technique in wireless networks, e.g., decoupling signal strength with interference and noise to maximize the data rate\cite{shen2018fractional}. Heuristic algorithms are particularly useful for solving problems with integer control variables\cite{martins2006metaheuristics}, and black-box optimization is also a useful method to handle problems with unknown objective function structure \cite{alarie2021two}.

However, applying these techniques to telecom is not straightforward. For instance, the reward function is an important part of implementing reinforcement learning, but the corresponding design can be difficult without professional knowledge of telecom. Moreover, the reward function may be related to multiple network metrics such as delay, throughput, and packet drop rate, incorporating these metrics into the reward function usually follows a time-consuming trial-and-error manner\cite{devidze2021explicable}. Similarly, although there have been many commercial convex optimization solvers, e.g., CPLEX and LINDO\cite{anand2017comparative}, it is worth noting that optimization problems have to be formulated in standard form, i.e., relaxing specific constraints for convexity or continuity, which are considered as obstacles for the application of convex optimization.
To this end, existing studies have shown that LLM may offer new opportunities to overcome the theory-application gap between existing optimization techniques and real-world telecom applications. There are multiple advantages to exploiting LLM-enabled optimize techniques for telecom: 

Firstly, LLMs demonstrate a strong ability to follow human instructions. Specifically, the LLM agent has the potential to formulate problems, design algorithms, select models, and finally optimize the system performance based on human preferences and language instructions\cite{bubeck2023sparks}. With LLM-enabled intelligence, operators can easily manage the network operation using simple natural language input, and then LLM can automatically select proper ML algorithms to implement tasks with minimum human intervention.

\begin{table*}[!t]
\caption{Summary of LLM-aided Optimization Techniques Studies. }
\centering
\small
\setstretch{1.1}
\resizebox{1\textwidth}{!}{%
\begin{tabular}{|m{0.8cm}<{\centering}|m{8cm}<{\centering}|m{5.5cm}<{\centering}|m{5cm}<{\centering}|}
\hline
Refer- ences & Proposed LLM-aided optimization techniques & Key findings \& Conclusion & Telecom application opportunities\\
\hline
\cite{song2023self}  &  A LLM framework with a self-refinement mechanism for automated reward function design, where LLM can formulate an initial reward function based on natural language inputs.  & LLM-designed reward functions can rival or even surpass
manually designed reward functions in 9 robot control tasks.   &  \multirow{3}*{\makecell{Reinforcement learning is very\\ useful for network optimization, \\ and automatic reward design \\ \slash universal proxy reward function \\ is an appealing 
 approach to \\ lower the difficulty of applying \\ reinforcement leaning\\ techniques to various network \\ management scenarios.  }}\\
\cline{1-3}
\cite{kwon2023reward}  &   It considers a universal reward design by prompting the LLM as a proxy reward function, where the user provides a textual prompt with a few examples or a description of the desired behavior.  &  The generated rewards are well-aligned with the user’s objectives and outperform supervised learning approaches.  &  \\
\cline{1-3}
\cite{ma2023eureka}  & An LLM-aided reward design system with zero-shot
generation, code-writing, and in-context improvement capabilities. It performs evolutionary optimization over reward code.  &  It outperforms human experts on 83\% of the tasks, leading to an average normalized improvement of 52\%.  &     \\
\hline
\cite{shinn2023reflexion}  &  It proposed a novel framework to reinforce language agents through linguistic feedback. The agent verbally reflects on task feedback signals, maintaining the reflective text in an episodic memory buffer to induce better decision-making.  &  The proposed framework achieves a 91\% accuracy on the HumanEval coding benchmark, surpassing the previous state-of-the-art GPT-4 that achieves 80\%. &  \multirow{2}{*}{ \makecell{\blue{LLMs} have self-improvement \\ capability, which means they can \\ work as an agent to receive\\  network environment feedback and\\ improve the policies \\ based on textual input.  }  } \\
\cline{1-3}
\cite{yang2023large}  &  LLM generates new solutions from the prompt that
contains previously generated solutions with their values, then the new solutions are evaluated and added to the prompt for the next optimization step.      & The proposed prompt-design scheme outperforms human-designed prompts by up to 8\% on GSM8K\cite{cobbe2021training}, and by up to 50\% on Big-Bench Hard tasks\cite{suzgun2022challenging}.&     \\
\hline
\cite{guo2023towards} & Evaluating the optimization capabilities of \blue{LLMs} across diverse tasks and data sizes, including gradient descent, hill-climbing, grid-search, and black-box optimization. & 1) The LLM show strong optimization capabilities; 2) \blue{LLMs} perform well in small-size samples; 3) They exhibit strong performance in gradient-descent; 4) LLMs are black-box optimizers.  & Black-box optimizer is a promising approach to estimating the unknown loss function, which is especially useful since telecom networks become more complicated.  \\
\hline
\cite{chen2023diagnosing}   &  A natural language-based system that engages in interactive conversations about infeasible optimization models. It provides natural language descriptions of the optimization
model itself, identifies potential sources of infeasibility, and
offers suggestions to make the model feasible.  &  The proposed system can assist both expert and non-expert users in improving their understanding of the optimization models, enabling them to quickly identify the sources of infeasibility.  &  \multirow{2}*{\makecell{Convex optimization is a commonly \\ used technique for network \\ optimization, and integrating LLM \\ with convex optimization can bring \\ promising changes to network \\ optimization.}}  \\
\cline{1-3}
\cite{ahmaditeshnizi2023optimus}  &  An LLM-aided system that can develop
mathematical optimization models, write and debug solver code, develop tests, and check the validity of generated solutions. &  It achieves nearly 0.8 success rate in 41 linear programming and 11 mixed integer linear programming problems.  &     \\
\hline
 \cite{pluhacek2023leveraging}     &  Using LLMs such as GPT-4 to generate novel hybrid swarm intelligence optimization algorithms. &  Generated a novel meta-heuristic algorithm with pseudo-code by using 5 existing algorithms.  &\multirow{2}*{\makecell{Heuristic algorithms are naturally \\ compatible with LLM techniques, \\ since many heuristic rules can be \\ easily described by textual language.\\ It offers new opportunities\\ for selecting and design new \\ heuristic network optimization methods.  }   }\\
\cline{1-3}
 \cite{liu2023large}   &  Using general LLM serves as a black-box search operator for decomposition-based multi-objective evolutionary optimization in a zero-shot manner.  & The LLM operator only learned from a few instances can have robust generalization performance on unseen problems with quite different patterns and settings.  &        \\
\hline
\end{tabular}}
\label{tab-llm-optimization}
\end{table*}

Secondly, LLMs can lower the training and fine-tuning difficulties of ML-based network optimization.
Algorithm training is considered one of the main obstacles to realizing AI-enabled wireless networks, which is usually very time-consuming. By contrast, LLM has shown impressive few-shot or even zero-shot learning capabilities in many fields\cite{xian2018zero}. 
In particular, \blue{LLMs} can learn in context from few or zero network management and optimization examples and then generalize to incoming new tasks. By providing a handful of examples, the LLM agent can quickly learn the hidden patterns without any extra model training and fine-tuning, saving considerable time and effort for algorithm training in network management. 
In addition, such fast learning capability is critical to make rapid responses to network dynamics. This means that network optimization decisions can be efficiently adjusted based on traffic patterns, user types, and operator demands.

Finally, the rich real-world knowledge of LLM will contribute to network optimization algorithm modelling and design.
LLM is equipped with rich internalized knowledge about the world in the pre-training stage\cite{beltagy2019scibert}. Such diverse knowledge can contribute to comprehending user preferences, task requirements, and even optimization algorithm modelling and design. 
For instance, LLM can already understand the fundamental concepts of reinforcement learning and linear programming without any extra training, and both techniques are very useful in optimizing telecom networks. This real-world knowledge eliminates the gap between real-world network optimization demands and problem modelling and design.    
Table \ref{tab-llm-optimization} summarizes existing studies on LLM-aided optimization techniques, including proposed methods, key findings, and telecom application opportunities. 
Given these motivations and the benefits of applying LLM to telecom optimization, we will introduce state-of-the-art LLM-aided optimization techniques along with telecom network optimization applications. 


\subsection{LLM-aided Reinforcement Learning for Network Optimization}
\label{subsection-RL}
Reinforcement learning is one of the most important techniques for network optimization. It explores various sequential action combinations, e.g., network resource allocation strategies and signal transmission power level, to maximize the long-term reward, such as higher data rate or lower transmission delay\cite{zhou2022learning}. 
Many network optimization problems can be transformed into a unified Markov decision process (MDP), and then using reinforcement learning to improve network metrics dynamically. 
For instance, resource allocation is a very common problem in many telecom scenarios, in which allocation decisions, desired network performance metrics, and network dynamics are usually defined as actions, rewards, and states, respectively\cite{zhang2022federated}. 
However, it is worth noting that these definitions are usually intuitive and require expert knowledge of reinforcement learning techniques and telecom. Especially, most reward functions are manually designed using trial-and-error approaches, and the algorithm performance is affected by the hyperparameter selection, e.g., learning rate, batch size, and number of hidden layers.
Fortunately, LLM techniques provide new opportunities to overcome these bottlenecks. This section will introduce two LLM-aided reinforcement learning techniques: automatic reward function design and verbal reinforcement learning.

\subsubsection{\rm \textbf{Using LLM for reward function design}}
\label{subsubsection-reward}
A recent survey in \cite{booth2023perils} shows that 92\% reinforcement learning researchers use manual trial-and-error reward function design and 89\% indicate that the designed rewards lead to
unintended behaviour during algorithm training\cite{hadfield2017inverse}.
Such issues become more difficult in complicated telecom scenarios since various network elements are involved, e.g., users with diverse requirements, limited available resources, and dynamic network environments. 
To this end, LLM shows the capability of developing a universal approach for reward design, which will significantly lower the difficulty of using reinforcement learning.   
For instance, \blue{Song et al.} \cite{song2023self} proposed a self-refined \blue{LLM} for automated reward function design in robotics, achieving a comparable performance as manually designed functions. 
\blue{Kwon et al.} \cite{kwon2023reward} applied LLM as a proxy reward function, where the user provides a textual prompt with a few examples or a description of the desired behaviour. In the following, we will introduce how the automatic reward function is designed.

An MDP can be defined as a tuple $<S, A, R, T>$, where $S$ and $A$ are the set of environment states $s \in S$ and actions $a \in A$, respectively. 
$T$ is the transition probability with $T(s,s')=Pr(s'|s,a)$, indicating the probability of taking action $a$ under state $s$ and reaching the next state $s'$. $R$ is the reward with $R=\mathcal{F}(s,a)$, where $\mathcal{F}$ is the reward function that maps the states and action selection to an immediate reward\cite{sutton2018reinforcement}. 
The reward feedback $R$ will further affect the action selection policy $\pi$ with $a=\pi(s)$, which means the action selection is under the current state $s$. However, the definition of such a reward function $\mathcal{F}$ is not straightforward since mapping the state $s$ and action $a$ to a specific value requires considerable experience and trial-and-error tests. Therefore, the objective of reward design is to use LLM as a proxy reward function or generate a reward function automatically\cite{kwon2023reward,song2023self}. Given the above MDP fundamentals, extra prompt input is required as textual input for LLM. Consider a set of prompt string $l \in \mathcal{L}$ and a mapping function $\mathcal{M}$, we need to define: 
\begin{itemize}
    \item Task description $l_1$: The string or environment code to describe the target task\cite{ma2023eureka};
    \item Objective description $l_2$: The optimization objective or desired final states of the task;
    \item States and actions description $l_3$: It explains the definition of states and actions in the target task;
    \item Examples description $l_4$: It provides a trajectory or examples of the episode. Note that a trajectory usually serves as a demo, but it is not required in zero-shot learning.  
    \item A mapping function $\mathcal{M}$ that maps the textual output of LLM to a binary value, e.g., "\textit{good}" or "\textit{bad}". This binary value feedback indicates the quality criteria of the generation, and then the \blue{LLM} easily understands the overall feedback.  
\end{itemize}
Given these definitions, the LLM-aided MDP definition becomes $<S, A, R, T, \mathcal{L}, \mathcal{M}>$, where $\mathcal{L}$ is a set of prompts $l_1$ to $l_4$, and $\mathcal{M}$ is the mapping function. The reward function $\mathcal{F}$ in $R=\mathcal{F}(s,a)$ is defined by $\mathcal{F}:=\mathcal{G}(\mathcal{L}, \mathcal{M})$, where $\mathcal{G}$ is the inference of \blue{LLMs}. $\mathcal{F}:=\mathcal{G}(\mathcal{L}, \mathcal{M})$ shows that the design of the reward function $\mathcal{F}$ depends on prompt input $\mathcal{L}$ and the mapping function $\mathcal{M}$.      
Based on the LLM-aided MDP framework, using LLM for reward function design can be summarized as the following steps:
\begin{itemize}
    \item Step 1: Description input. Using language to describe the task, objective, states, and actions. If necessary, providing possible trajectory examples to \blue{LLMs}. Here an alternative approach is to feed the environment code to the LLM agent, and then using natural language to describe the task, which has been used in \cite{ma2023eureka}.
    \item Step 2: Initial reward function design, which will use Step 1 as input, and produce initial reward function designs.  
    \item Step 3: Reward function implementation. Using the reward function produced in Step 2 to train the reinforcement learning agent.
    \item Step 4: Evaluation and feedback. Evaluating the reinforcement learning training output and providing feedback to \blue{LLMs}.
    \item Step 5: Self-improvement. Sending the feedback and evaluation results to the LLM agent, and then LLM will produce a new reward function design. Repeating from Step 3 until the algorithm has the desired performance or reaches the maximum iteration number.  
\end{itemize}

\begin{figure*}[!t]
\centering
\setlength{\abovecaptionskip}{0pt} 
\includegraphics[width=1\linewidth]{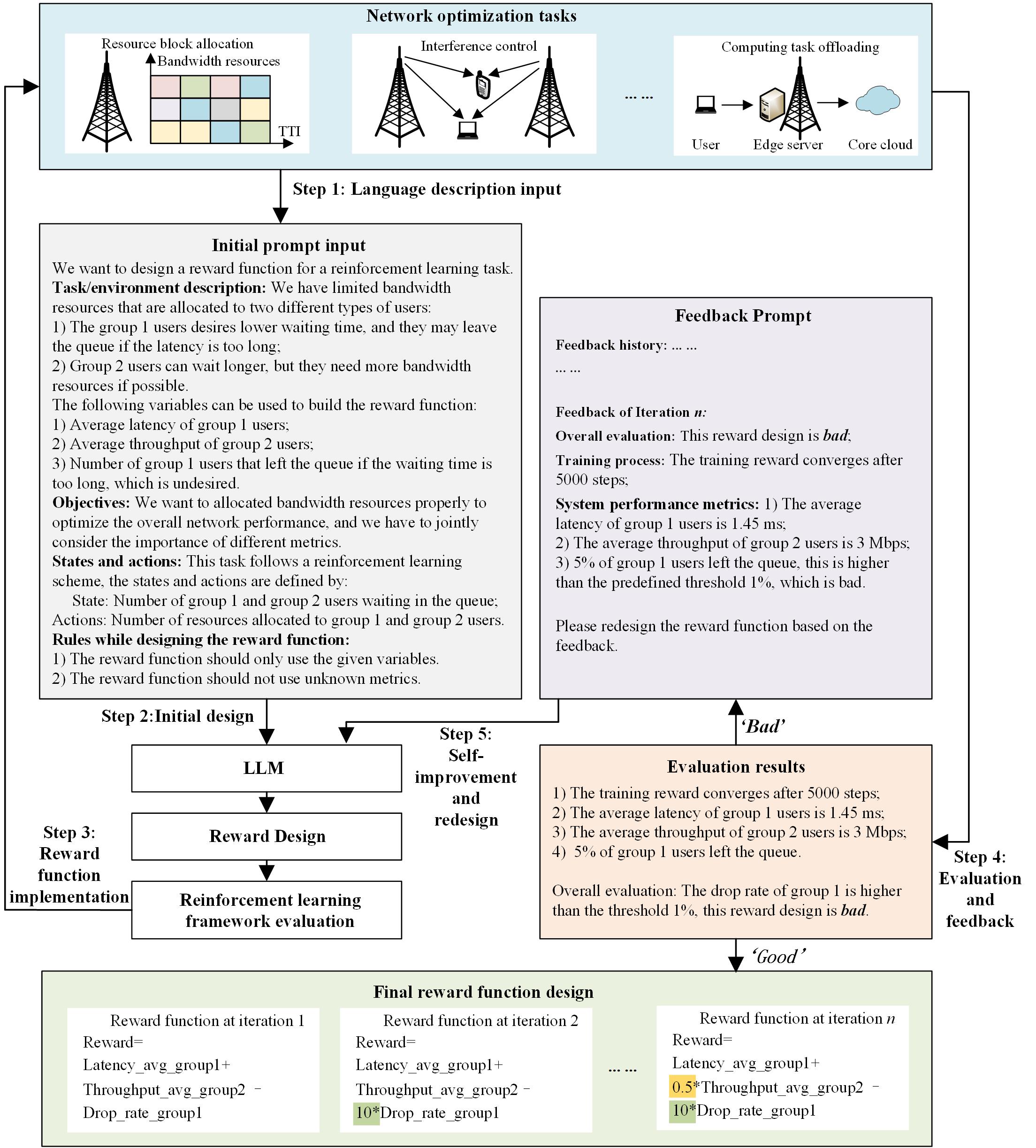}
\caption{LLM for reward design in network optimization.}
\label{fig-reward}
\end{figure*}

To better explain how LLM-aided reward design can be used for network optimization. Fig.\ref{fig-reward} shows the procedure of solving a simple resource slicing problem\cite{zhou2021ran}. In particular, it considers resource allocation as an example with two types of users. 
Group 1 indicates URLLC users that desire lower latency and higher reliability, and group 2 represents enhanced Mobile Broad Band (eMBB) users with high throughput demands. 
As shown in Fig.\ref{fig-reward}, the resource allocation task is described by natural language as input for LLM, including task description and user features, objectives, states and actions, and reward design rules. 
Note that we use "group 1" and "group 2" instead of "eMBB" and "URLLC" to lower LLM understanding difficulty. In addition, the features of the two groups have been clearly defined. 
Then, LLM will generate an initial reward function design and send the initial design to the reinforcement learning framework for evaluation. After that, we will receive and analyze the evaluation results, e.g., convergence and system metrics. 
For instance, the evaluation results in Fig.\ref{fig-reward} show that the 5\% drop rate of group 1 users is much higher than the predefined threshold 1\%, and therefore the overall evaluation for this design is "bad". It is worth noting that the final evaluation of "good" or "bad" depends on the user's predefined criteria, which varies between different scenarios.

In Fig.\ref{fig-reward}, if the evaluation result is "bad", then a detailed feedback summary is provided with possible suggestions, e.g., \textit{the drop rate of group 1 users is too high}. Given this feedback, the LLM agent will redesign the reward function and repeat the process from Step 3. On the other hand, if the evaluation result is "good", the system will output the final reward function design. The bottom module also shows an example that the reward function is improved by iterations. e.g., the coefficient of $Drop\_rate\_group1$ is increased from 1 to 10, preventing dropping group 1 users. The coefficient of $Throughput\_avg\_group2$ is also improved to balance the latency and through metrics of two groups.

Reward function design is a prerequisite for applying reinforcement learning to telecom, and LLM-aided automatic reward function design significantly lowers the difficulty. 
However, it is worth noting that some reward functions can be very complicated in the telecom field, which may include transformation functions like $arctan$ or $sigmoid$ and diverse network metrics. These design problems can be more complicated if multiple network elements are simultaneously involved, such as vehicle networks and RISs \cite{zhou2023survey}. The simulations in \cite{song2023self, kwon2023reward, ma2023eureka} have demonstrated LLM's capabilities in reward design for robotics and logic games, but the application in the telecom field is still an open issue.

\begin{figure}[!t]
\centering
\includegraphics[width=0.95\linewidth]{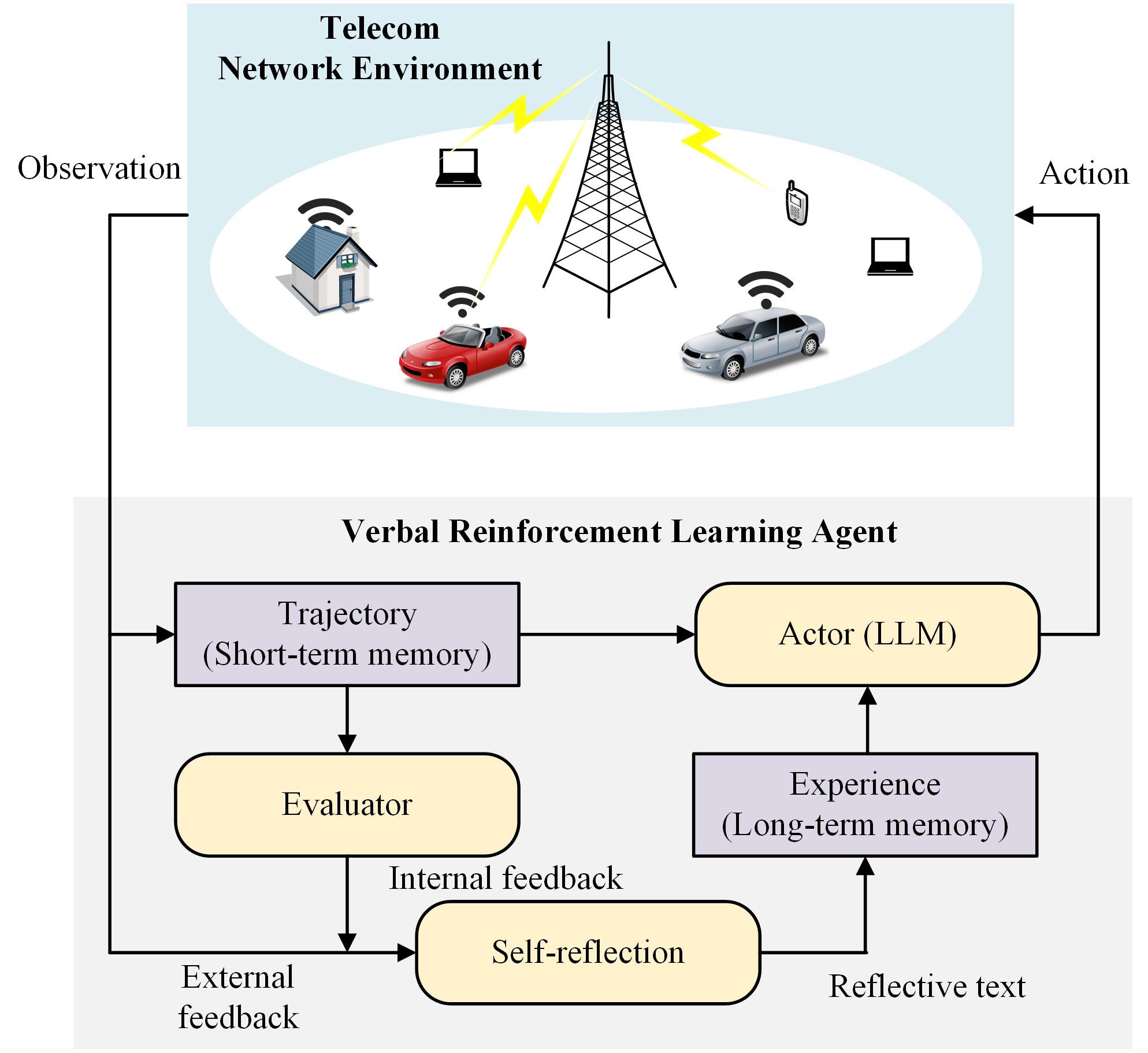}
\caption{LLM-aided verbal reinforcement learning for network optimization.}
\label{fig-verbal}
\end{figure}

\subsubsection{ \rm \textbf{Verbal reinforcement learning via LLM}}
Section \ref{subsubsection-reward} proves that LLM can use the feedback to improve previous solutions. Given this self-improvement capability, a promising optimization technique is to consider LLM as an agent, interacting with the environment to explore optimal policy. Verbal reinforcement learning is proposed in \cite{shinn2023reflexion}, and achieved satisfied performance across diverse tasks, including sequential decision-making, coding, and language reasoning. 
Fig.\ref{fig-verbal} shows an example of using verbal reinforcement learning for radio access network optimization, and the agent consists of the following modules:
\begin{itemize}
    \item Actor: The actor is built upon an \blue{LLM}, which is specifically prompted to generate actions, e.g., network control and optimization strategies. Based on short-term and long-term memories, the actor can apply various methods to produce actions, such as CoT \cite{wei2022chain} and ReAct \cite{yao2022react}. These advanced prompt techniques can improve the actor's capability of reasoning and planning, which can better adapt to the complicated decision-making of network optimization problems.   
    \item Evaluator: The evaluator is a critical module to assess the performance of the actor. In particular, it takes the short-term trajectories as input and produces a reward score that shows the action quality. The evaluator can be defined in various approaches, e.g., a specified reward function or heuristic criteria. For instance, in resource allocation problems, the evaluator can be defined by a reward function with network metrics, or a heuristic like "\textit{all the users' requirements have been fulfilled}". We still consider the radio access network as an example. The evaluator's internal feedback could be "\textit{The average latency of network edge users is too high, and 10\% edge users' communication demand is dropped. The overall performance of this trajectory is bad.}"    
    \item Self-reflection: The self-reflection module is the most important part of the verbal reinforcement learning scheme, providing useful feedback instructions to the actor. Specifically, with external feedback from the environment and internal feedback from the evaluator, the self-reflection module can generate more detailed feedback to the actor, which is far more informative than a pure reward value in conventional reinforcement learning. A feedback example could be "\textit{Cell edge users should have more resources if available, and cell edge means users that are far away from the BS than other users.}"       
    \item Short-term and long-term memories: The memory mechanism consists of short-term and long-term memories. The long-term memory indicates important lessons learned from previous experience, while the short-term memory shows recent decisions and performance. This is an intuitive approach that is similar to the human brain with fine-grain recent details and important lessons from long-term memory. With the self-reflection mechanism, the long-term memory will automatically learn important rules, e.g., "\textit{Cell edge users should have more available resources; Type 1 users are delay sensitive, they should have higher priority.}"   
\end{itemize}

Compared with conventional reinforcement learning, the LLM-aided verbal learning technique has multiple advantages for telecom optimization: 
1) Lowering the difficulty of implementing network optimization. Verbal reinforcement learning avoids the difficulty of tuning hyperparameters like learning rate, batch size, and training frequency. This will significantly lower the difficulty of applying artificial intelligence to network optimization. 
2) Allowing for language instructions to guide network optimization policies. Specifically, the LLM-aided system allows for language instructions to guide the agent exploration, which is much more efficient than existing strategies such as $\epsilon$-greedy policy. 
Experienced network operators can provide language instructions to \blue{LLMs} directly, and no ML knowledge is required.
3) Reasoning and interpretable explanations for algorithm performance. One crucial advantage of LLM-aided systems is that they provide interpretable explanations of the algorithm and telecom system performance, and these experiences can further help understand network management policies. 

Despite the advantages, LLM-aided reinforcement learning is still at a very early stage, and there are very few studies that apply this technique to the telecom field. In addition, specific telecom domain knowledge may be required to let the LLM better understand user demand. Therefore, professional wireless knowledge datasets such as TeleQnA in \cite{maatouk2023teleqna} may be required to fine-tune the \blue{LLM}.

\subsection{LLM as a Black-box Optimizer}
\begin{figure}[!t]
\centering
\includegraphics[width=0.85\linewidth]{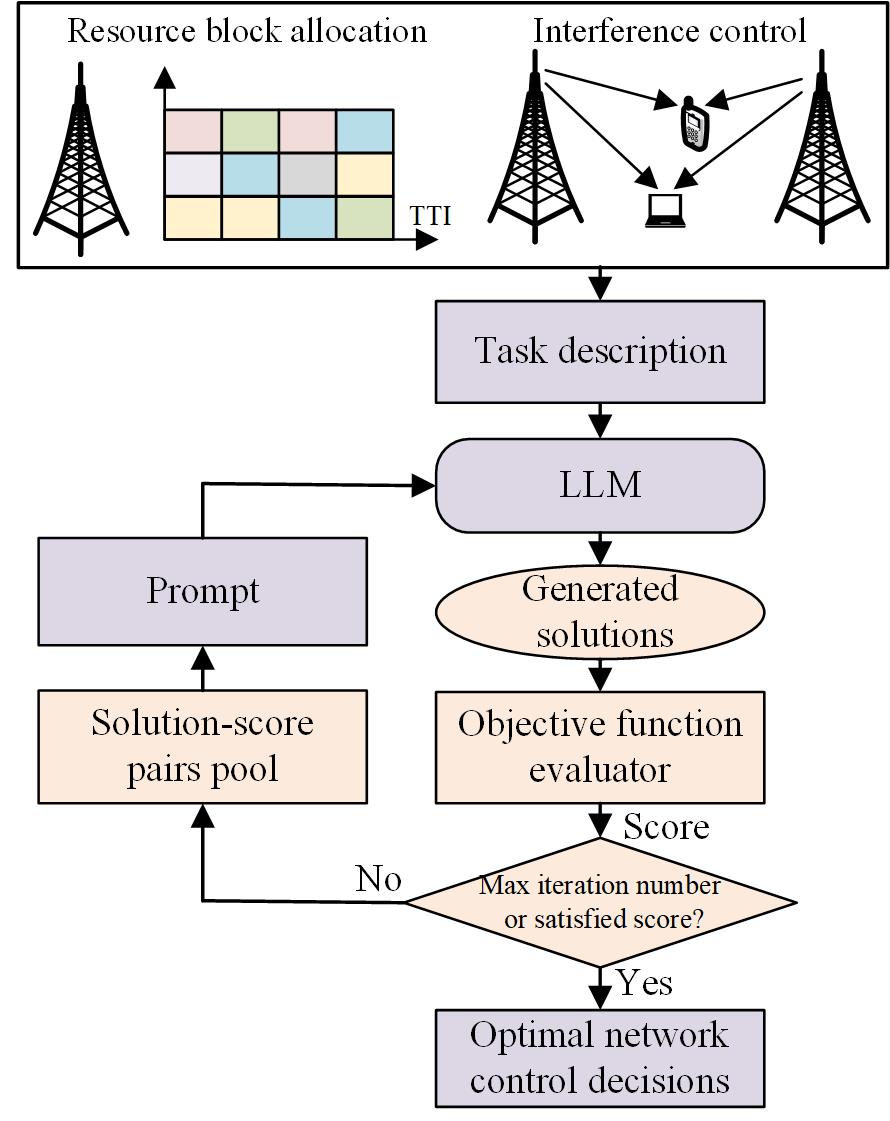}
\caption{LLM-as a black-box optimizer for telecom.}
\label{fig-black}
\end{figure}

Black-box optimizer is also an appealing approach for network optimization problems. It refers to the task of optimizing an objective function $f: X \rightarrow R$ without access to any other information about $f$, e.g., gradients or the Hessian\cite{golovin2017google}.
Telecom networks will become more and more complicated in the 6G era, and black optimization can avoid the complexity of building dedicated optimization models.
Existing studies have shown that LLM has the black-box optimization capability to fit an unknown loss function\cite{guo2023towards}. 
Fig.\ref{fig-black} shows an example of using LLM in a black-box manner. It starts by describing the optimization task, and then LLM will generate an initial solution. The generated solution will be evaluated by the objective function evaluator, e.g., average or sum data rate, average latency, etc. If the evaluated score is satisfied or it is the maximum iteration number, then the system will output the final solution. Otherwise, the current solution is sent to a solution-score pairs pool, and a new prompt will be generated accordingly for LLM as input. Here the solution-score pair pool includes past experience and corresponding scores. By comparing the similarities of high-score solutions, the LLM can generate better solutions iteratively with few-shot learning capabilities.
To better understand how LLM can be used as a black optimizer for network optimization, we provide an example of BS power control\cite{ha2013distributed}:
\begin{itemize}
    \item Initial task description module: 
     \begin{tcolorbox}[title={BS power control task description} ]
     \textit{“We have an interference control task related to wireless network management. We need to control the power level of two BSs to maximize the average data rate. We need you to provide the transmission power of these two BSs, and adjust the power based on provided feedback"}.
    \end{tcolorbox}
  
    \item Prompt inputs module for black-box optimization: 
    \begin{tcolorbox}[title={Prompt input for black-box optimization} ]
    \textit{“Below are some previous power levels and the corresponding data rate, which are arranged in descending order}. \vspace{5pt}
    
    Input: $P\_level\_1$: 14 dBm, $P\_level\_2$: 17 dBm; Output $Avg\_rate$: 1.1 Mbps;\\
    ... ...  \qquad \qquad ... ... \qquad \qquad ... ... \qquad \\
    Input: $P\_level\_1$: 22 dBm, $P\_level\_2$: 15 dBm; Output: Average data rate is 1.8 Mbps;\\
    Input: $P\_level\_1$: 25 dBm, $P\_level\_2$: 22 dBm; Output: Average data rate is 2.5 Mbps. \vspace{5pt}

    \textit{Give me a new power level input that is different from all the traces above and has a higher average data rate than any of the above"}.
    \end{tcolorbox}
\end{itemize}

After the above prompts input, one can use the output to update the candidate solutions and then repeat this process as shown in Fig.\ref{fig-black} until obtaining a satisfactory solution.
The main advantage of black-box optimization is that it avoids the complexity of defining dedicated optimization models, which have been used to automatically construct the wireless network optimization model in \cite{zhang2019toward}, and to optimize cellular network coverage and capacity in \cite{dreifuerst2021optimizing}. 
For the LLM-aided black-box optimizer, the existing example quality may affect the output results, and the algorithm performance cannot be guaranteed.  
However, telecom management usually has stringent requirements on solution qualities to guarantee the service level, which can be an obstacle to using LLM techniques.

\begin{figure*}[!t]
\centering
\includegraphics[width=1\linewidth]{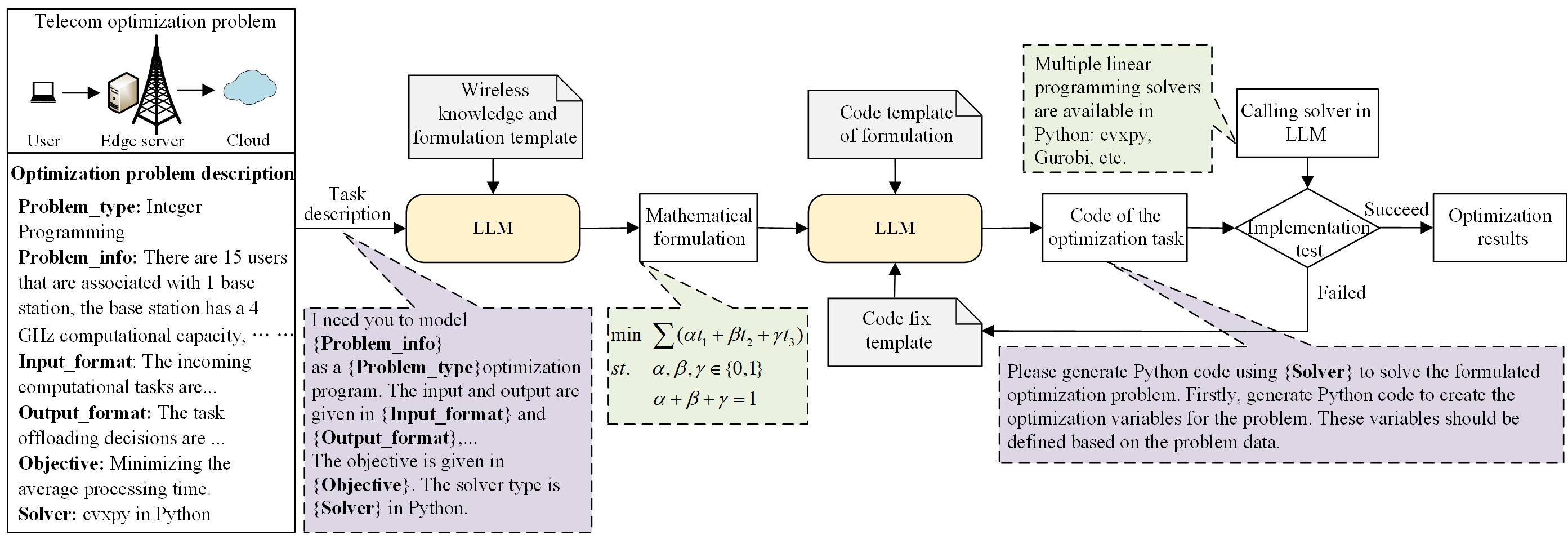}
\caption{LLM-aided Convex Optimization Problems. }
\label{fig-convex}
\end{figure*}

\subsection{LLM-aided Convex Optimization for Telecom}
\label{sec-convex}
Convex optimization is a crucial technique for telecom networks, and it is commonly used in many scenarios\cite{zhou2023survey}. For instance, fractional programming is especially useful for wireless network optimization due to the fractional terms in communication systems such as signal-to-interference-plus-noise ratio (SINR) and energy efficiency, which is applied to wireless power control and beamforming\cite{shen2018fractional}. Convex optimization can provide stable and efficient solutions, especially when closed-form solutions are achieved. 
However, deploying convex optimization techniques usually requires dedicated problem modelling, transformation, and relaxation since the original problems may be non-convex. 
Therefore, the requirement for expert knowledge may prevent the application of convex optimization techniques.  
To improve the accessibility of convex optimization, the authors in \cite{chen2023diagnosing} propose to use the LLM to diagnose the infeasibility of optimization problems, aiming to relax or remove some infeasible constraints, and LLM is used for convex optimization problem modelling, code generation and solving in \cite{ahmaditeshnizi2023optimus}. The experiments in \cite{chen2023diagnosing} and \cite{ahmaditeshnizi2023optimus} have demonstrated that LLM has the potential to improve convex optimization techniques.

Fig.\ref{fig-reward} shows the key steps of using LLM to solve network convex optimization problems with the following modules:
\begin{itemize}
    \item Problem modelling and description: Transforming the network optimization problem into a standard form is the first step of automatic problem modelling. Fig.\ref{fig-reward} presents some key elements of defining the problem, including problem type, problem information, input and output format, objective and solvers. Specifically, problem type specifies the type of this problem, e.g., linear programming, mixed-integer linear programming, quadratic programming, etc. 
    Problem information includes the core description of the problem, which defines the relationship between input and output variables. Then, input and output variables show the expected input and output variables along with definitions, i.e., network decision variables and output metrics. 
    Objectives and solvers give the optimization objective and applied solvers. Such a standard form and description will lower the difficulty of LLM understanding.    
    \item Telecom knowledge and formulation templates: Telecom optimization requires professional network knowledge. LLM has learned fundamental knowledge in the pre-training period such as calculating information capacity using Shannon's formula. However, using state-of-the-art telecom knowledge and formulation templates to fine-tune the \blue{LLM} can better improve the modelling accuracy. For instance, a dataset named TeleQnA is defined in \cite{maatouk2023teleqna}, and it includes nearly 10000 communication field questions from both standards and research articles.    
    \item LLM and Solvers: Existing studies have shown that LLM can use the advanced features of existing solvers such as Gurobi and cvxpy to solve the problems\cite{diamond2016cvxpy}. For instance, \cite{ahmaditeshnizi2023optimus} observed that LLM can use the function $gurobi.abs$ to model $\mathcal{L}$1-norm objective instead of adding auxiliary constraints and variables. It demonstrates that LLM has the potential to take advantage of existing solvers to address complicated optimization problems. In addition, if the implementation fails, code-fix templates can also be included to address the issues automatically and rerun the test. 
\end{itemize}
In summary, Fig.\ref{fig-convex} shows an example of solving network optimization problems in an end-to-end manner. Given a proper problem description, the LLM-aided system can automatically model the problem, generate code, and call the server to solve and debug the problem. Such a scheme has been used in \cite{ahmaditeshnizi2023optimus} to solve 41 linear programming and 11 mixed-integer linear programming problems and achieved a nearly 0.8 success rate for small-scale problems using GPT-4. 
The study in \cite{ahmaditeshnizi2023optimus} also observed that the success rate could be further improved by adding supervised tests and data augmentation. 

Despite the great potential, it is worth noting that telecom networks have become more and more complicated, and there are many complicated large-scale and non-convex optimization tasks. For example, RIS-related optimization problems usually include multiple control variables, e.g., RIS phase-shift control and BS passive beamforming, which are usually optimized in an alternating approach. It still requires dedicated human effort to transform the problems into standard forms\cite{zhou2023cooperative}. However, LLM-aided automatic convex optimization is still a promising approach that will save human time and effort on network optimization problem modelling and solving.

\begin{figure*}[!t]
\centering
\setlength{\abovecaptionskip}{0pt} 
\includegraphics[width=1\linewidth]{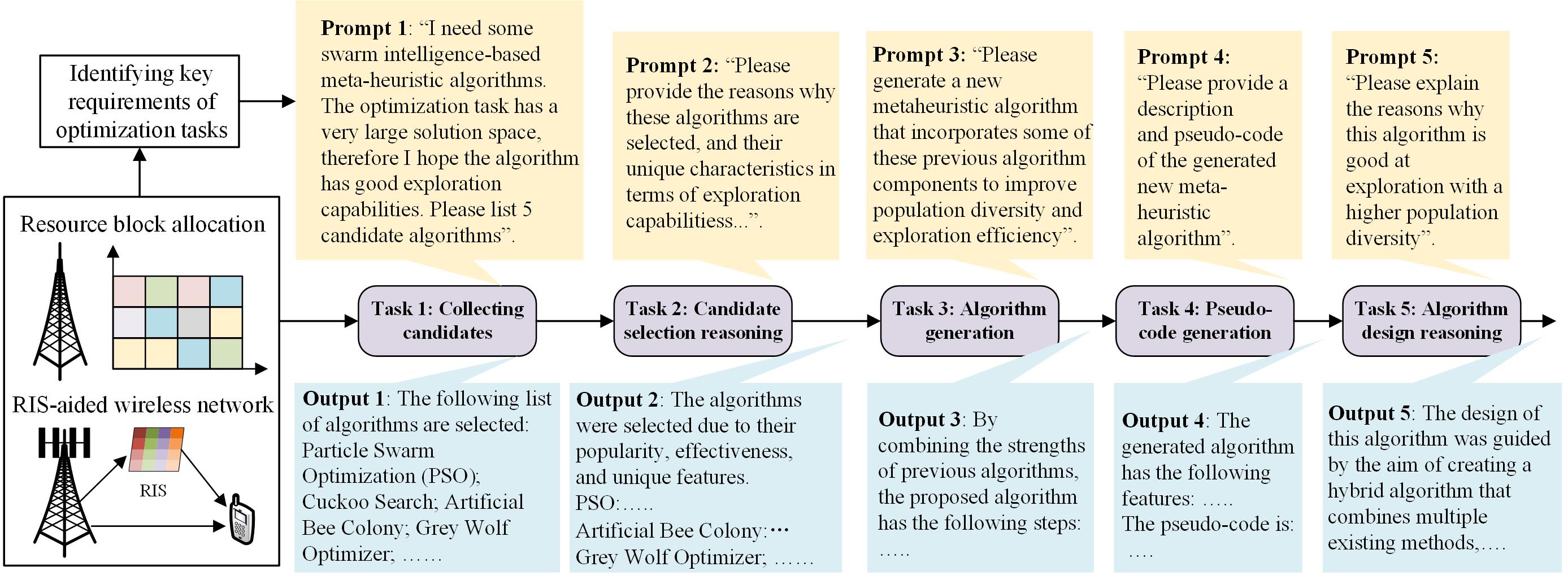}
\caption{LLM-aided meta-heuristic algorithm generation.}
\label{fig-heuristic}
\end{figure*}

\subsection{LLM-based Heuristic Algorithm Design }
\label{subsection-heu}

Heuristic algorithms are very useful techniques for network management and optimization. Specifically, they apply diverse heuristic rules to select near-optimal solutions with low design and computational complexity\cite{zhou2023heuristic}. 
Heuristic algorithms are particularly useful for solving optimization problems with integer control variables, which are very frequently formulated in telecom. 
For instance, the phase-shift optimization of RISs is considered as a very difficult problem with integer control variables and large solution space, and genetic algorithms and particle swarm optimization are used in \cite{dai2021reconfigurable} and \cite{zhi2022power} to solve this problem. 
In addition, heuristic algorithms are intuitively compatible with \blue{LLMs}, since heuristic rules can be easily described by natural language and instructions. 
For example, swarm-based methods are very widely used heuristic algorithms, e.g., genetic algorithm, particle swarm optimization, and grey wolf optimizer, providing near-optimal solutions by iteratively searching for better solutions. However, the number of these algorithms has grown significantly in the past decade, and selecting the proper algorithm to solve specific network optimization problems has become more difficult. 
Given the reasoning and understanding capabilities, LLM offers promising changes for selecting and designing novel meta-heuristic algorithms. 

Fig.\ref{fig-heuristic} presents an example of using LLM to design novel swarm-based meta-heuristic algorithms for network optimization, which consists of 5 tasks. Such a decomposition and CoT approach can considerably lower the prompt difficulty and improve output performance\cite{pluhacek2023leveraging}. 
The first step is to identify the key requirements of optimization tasks. For example, RISs consist of hundreds of small units, and each requires dedicated phase-shift control, leading to a large solution space. 
Therefore, Prompt 1 in Fig.\ref{fig-heuristic} requires the candidate algorithm to have “\textit{good exploration capabilities}”, and the first instruction is to “\textit{Please list 5 candidate algorithms}”. Then, the next task is to identify the key components of these algorithms. For instance, inertia weight and local and global best mechanisms are two key components in particle swarm optimization, and then LLM can better understand the functionality of each unique heuristic rule. After that, Tasks 3 and 4 will generate the step-by-step design and pseudo-code of a new swarm-based meta-heuristic optimization algorithm. Most importantly, Task 5 will take full advantage of LLM's reasoning capability, and explain how this novel algorithm is designed with step-by-step motivations. 

In summary, Fig.\ref{fig-heuristic} presents an automatic approach for novel meta-heuristic algorithm design, which can be very useful for telecom network control and optimization. 
For instance, many network control scenarios require rapid responses for environment dynamics such as traffic load level and user demand changes, and LLM-aided systems in Fig.\ref{fig-heuristic} have the potential to generate novel heuristic algorithms with fast convergence and low computational complexity. Additionally, such a scheme can also be used to generate new heuristic network protocols or management policies, significantly saving human efforts in terms of creation and design\cite{bariah2023large}.

\begin{table*}[!t]
\caption{Summary of LLM-based Optimization  Techniques for Telecom. }
\centering
\small
\setstretch{1.05}
\resizebox{1\textwidth}{!}{%
\begin{tabular}{|m{2cm}<{\centering}|m{3.3cm}<{\centering}|m{3cm}<{\centering}|m{3.5cm}<{\centering}|m{3.5cm}<{\centering}|m{3.5cm}<{\centering}|}
\hline
LLM-based optimization applications & Main features & Prompt/ Input requirements & Advantages compared with existing approaches & Potential issues  & Network optimization application opportunities\\
\hline
 LLM-aided reward function design &  Reward function is a crucial part of reinforcement learning-enabled network optimization, and LLM provides automatic reward function design by using its self-improvement and understanding capabilities.  & Task/environment description; Objective description; States and actions; Examples or demos. A mapping function/criteria to evaluate the design to "good"/"bad". & Automatic reward function design can significantly save human effort in applying reinforcement learning to network optimization tasks. Automatic reward function design has produced comparable performance as human manual design. &  1) Automatic reward design is still at a very early stage, and there are few applications that explore such a novel technique in the telecom field; 2) The prompt has to be carefully designed to describe the target task, which is known as prompt engineering.   &  Reinforcement learning is a very useful technique for telecom network management, and automatic reward design is a promising technique to enable artificial general intelligence, which is particularly useful for small-scale optimization problems to save human effort.       \\
\hline
 Verbal reinforcement learning & It considers LLM as an agent, exploring the environment and accumulating experiences. Using the self-improvement capability to improve previous solutions and obtain a higher reward.  &  1) Self-evaluator will provide critical feedback to the actor for improved performance; 2) Short-term and long-term memories are crucial for the actor to distinguish between good and bad actions.    &  1) Avoiding the difficulty of tuning hyperparameters like learning rate, batch size, and training frequency; 2)  Allowing for language instructions to guide network optimization policies; 3) Providing  Reasoning and interpretable explanations for algorithm performance.   & 1) The evaluator and self-reflection modules have to be carefully designed to generate useful experience; 2) It may have exploration-exploitation difficulty, since the agent relies on previous experience to produce new solutions.   &  Verbal reinforcement learning can be very useful for solving problems that have been well-defined with small action spaces and immediate rewards, which is very common in telecom networks, e.g., resource allocation and association. \\
\hline
LLM as a black-box optimizer  &  Black-box optimization is a useful approach for network optimization, and LLM has been demonstrated to have the black-box optimization capability to fit an unknown loss function.  &  1) Task description; 2) Previous input and output examples, and then asking for a better solution based on previous input and output.   &  Black-box optimization avoids the complexity of building dedicated optimization models and transformations, which can be very time-consuming in complicated telecom network environments.  &  The performance of using an LLM black-box optimizer cannot be guaranteed, which relies on the quality of the provided input and output examples.   &  Black-box optimization is a promising technique for telecom network, but it may have difficulty providing stable and reliable results. The reasoning capability of LLM may shed light on solving this problem.          \\
\hline
LLM-enabled convex optimization  &  LLM provides end-to-end automatic solutions for convex optimization techniques, including problem modelling, code generation, and solver implementation.  & 1) The problem has to be defined in standard form, so then the LLM can understand and model it; 2) Telecom knowledge and formulation template are required; 3) Existing solvers have to be specified for the LLM to solve the problem.  
&  1) Automatic problem modelling is an especially promising technique, significantly saving human effort; 2) It enables automatic problem-solving in an end-to-end manner, requiring minimum human intervention.    &  Some convex optimization problems in the telecom field are extremely complicated with coupled control variables and highly non-convex objectives and constraints. These problems can be very difficult to solve automatically.   &  Many network control problems can be formulated as convex optimization problems, and LLM-aided convex optimization has great potential to solve these problems efficiently with much less human effort.  \\
\hline
LLM for heuristic algorithms &   Heuristic algorithms are inherently compatible with LLM, since many heuristic rules can be easily described by natural language. LLM offers opportunities for heuristic algorithm selection and design for specific network optimization tasks.  &  It may require a series of prompts in a CoT manner, including candidate algorithm selection, analyses, new algorithm and pseudo-code code generation, and reasoning.  &  Automatic heuristic algorithm selection and design will considerably save human time on algorithm analyses and design. It can also provide reasoning and analyses of the generated results.     &  The generated heuristic algorithms still need to be tested and verified. Such an automatic design cannot guarantee the performance of the algorithm that was produced.  &  Heuristic algorithms are widely used for telecom network optimization and management, and LLM has promising potential for heuristic algorithm selection and design, producing novel algorithms that can better serve telecom networks.         \\
\hline
\end{tabular}}
\label{tab-summary-opt}
\end{table*}

\subsection{Discussions and Analyses}
\label{subsection-opti-diss}
Subsections \ref{subsection-RL} to \ref{subsection-heu} have introduced various LLM-aided optimization techniques along with telecom applications. Table.\ref{tab-summary-opt} summarizes various LLM-aided optimization techniques, including main features, prompt and fine-tuning requirements, advantages compared with existing approaches, and network optimization application opportunities. 
In the following, we summarize our key findings and analyses.

Firstly, task description is crucial for network optimization. Task description is the first step of using LLMs, which requires accurate and standard input, e.g., input and desired output format, objective and specific rules. In addition, these tasks are usually closely related to telecom domain knowledge, and LLM may have difficulty understanding some professional concepts. For example, in Section \ref{subsection-heu}, the LLM may already have some general knowledge of RIS technology, but they are unable to directly understand the difficulty of RIS phase-shift control, which is a very professional domain-specific knowledge. Therefore, the task description has to be carefully designed, which will directly affect the LLM output. 
Meanwhile, prompt design is the key to network optimization problems. Previous sections have demonstrated that prompt is one of the most important approaches to take advantage of LLM's capabilities, and there have been various prompt engineering techniques, e.g., CoT\cite{wei2022chain}, ReAct \cite{yao2022react}, zero-shot instruction \cite{kojima2022large}, etc. Therefore, understanding the function of prompt engineering is crucial for applying LLM to solve optimization problems. For instance, in reward design problems, the feedback prompt is critical to improve the reward design step by step. In the heuristic algorithm design problem in Section \ref{subsection-heu}, the output completely depends on the user prompt input to the LLM agent.

\blue{In addition, several of the above optimization approaches rely on the feedback mechanism, in which the solutions are iteratively improved based on previous answers and environment feedback. 
For instance, the reward function design is iteratively improved by involving the evaluation results and feedback prompts. 
Similarly, in verbal reinforcement learning, the LLM agent can adjust the action selections to obtain a higher reward based on environmental feedback. 
Therefore, the design of these prompts is crucial for improving LLM's performance, e.g., dedicated feedback and evaluator prompt designs.}
On the other hand, balancing exploration-exploitation is a common obstacle for many optimization problems. This problem becomes severe when the action space is larger, which is very common in telecom networks. 
Therefore, how to use the LLM's self-improvement capability and meanwhile balance the exploration-exploitation is very important.

\blue{Finally, note that many optimization problems can be very complicated in wireless networks by involving multiple control variables, network elements and layers. It may require step-by-step problem decomposition, formulating multiple objectives, and alternating optimization, e.g., jointly optimizing the transmission data rate and signal coverage. 
Handing these optimization problems needs strong planning capabilities, which is still a challenge for current LLM research fields.}

\section{Time Series LLM for Prediction Problems }
\label{sec-predict}

Prediction tasks are crucial in telecom networks that involve predicting future trends, demands, and behaviours based on historical data, e.g., predicting network traffic, customer demand, equipment failures, and service usage. 
%
This section will introduce time series models for prediction problems in wireless networks, including pre-training foundation models, frozen pre-trained, fine-tuning, and multi-modality LLMs.

\subsection{Motivations}

Conventional prediction algorithms in the telecom domain rely on statistical and time-series analysis to estimate the output. 
However, telecom data is usually non-linear, non-stationary, and influenced by various external factors, leading to challenges in capturing complex patterns and relationships.
While these traditional methods have been effective to some extent, they may struggle with the complexity and dynamic nature of telecom data.
Recently, LLM technologies have shown promise in addressing the challenges of time-series prediction due to their ability to handle complex data structures and adapt to changing patterns. 

Firstly, \blue{LLMs} provide a universal and generalizable model for telecom network prediction. 
Given historical data, conventional prediction approaches must train a new model to adapt to incoming target tasks. 
These methods usually require extensive feature engineering and manual tuning, which can be time-consuming and may not generalize well across different scenarios.
By contrast, the versatility of LLMs makes them suitable for processing diverse forms of time-series data, and such adaptability is crucial given the vast volumes of data generated in telecom. 
%
%
Moreover, LLM's capability to continuously learn and adapt to new data patterns helps mitigate the concept drift problem, ensuring that the models remain relevant and effective over time. 
As a result, the integration of LLM techniques in time-series prediction offers a promising avenue for developing more robust and generalizable models that can better handle the complexities of data in telecom.

Meanwhile, \blue{LLMs} have excellent ICL capabilities, which means that they can perform new tasks by leveraging contextual information in demonstrations. 
In particular, it means that the LLM can directly learn from the provided examples, and map the input-output relationships without extra model training.
Such a prediction method is much more efficient than conventional prediction methods. Meanwhile, it is also more accessible since no professional knowledge of model training/fine-tuning is required.  
In addition, multi-modal LLM-enabled prediction can also be combined with sensing in telecom networks.
In particular, multi-modal \blue{LLMs} can process and integrate information from various data types, such as text, images, audio, and time-series data. 
In addition, sensing is an important part of envisioned 6G networks, aiming to integrate environmental information into communication networks, e.g., the image captured by street cameras or satellites, 3D LiDAR maps and WiFi sensing. 
In the context of telecom prediction, multi-modal LLMs can combine sensing data with numerical time-series data to generate more accurate context-aware prediction, which can be particularly useful in 6G.

Given the great potential, it is important to investigate time series LLM techniques and applications in telecom networks. 
In the following, we will introduce various LLM-based prediction methods and applications to telecom networks.

\begin{figure*}[!t]
\centering
\setlength{\abovecaptionskip}{0pt} 
\includegraphics[width=0.95\linewidth]{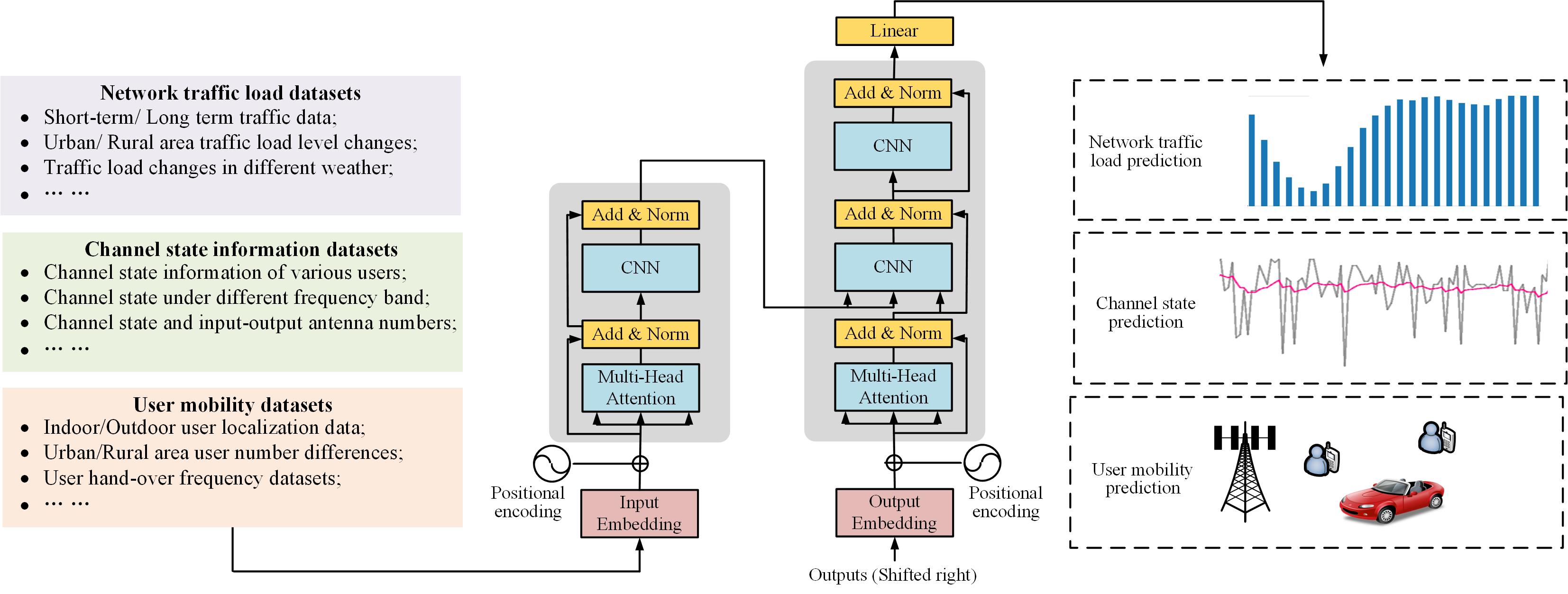}
\caption{Encoder-decoder-based TimeGPT for prediction problems in telecom networks\cite{garza2023timegpt1}.}
\label{fig-prediction}
\end{figure*}

\subsection{Pre-training Foundation Models for Zero-shot Prediction}
\label{sec-foundation-predict}
The pursuit of training a general-purpose foundation model for time-series data is driven by the desire to address the inherent challenges associated with diverse and dynamic data.
%
Traditional time-series methods may struggle to adapt to the non-stationary properties of real-world data, where the statistical characteristics of the series can change over time due to evolving patterns and trends.
For instance, the network traffic load level can be affected by many factors, including time, area, environment buildings, service types, etc. It usually requires dedicated model design and training for each target task, and then extracts the underlying patterns from history datasets\cite{razghandi2023smart}.
By contrast, a general-purpose foundation model aims to overcome these challenges by leveraging the advancements in LLM technologies. 
The following will first formulate the problem of training a foundational \blue{LLM} for time-series prediction, and then discuss different tokenization mechanisms and model architectures.

\subsubsection{\rm \textbf{Problem formulation}}
The primary goal of a foundation model is to design a zero-shot forecasting scheme that utilizes the past $t$ time points of a time series as input to predict the future $h$ time points. 
Let the input context be $y_{1:L} := \{y_1, \ldots, y_L\}$ and the prediction horizon be $y_{L+1:L+H}$. 
The model, denoted as $f_\theta$ (parameterized by $\theta$), aims to map the context to the horizon, i.e., $f_\theta : (y_{1:L}) \rightarrow \hat{y}_{L+1:L+H}$.
In this setting, the prediction model $f_\theta$ maps the feature space $\mathcal{X}$ to the dependent variable space $\mathcal{Y}$. 
The spaces are defined as $\mathcal{X} = \{y_{[0:t]}, x_{[0:t+h]}\}$ and $\mathcal{Y} = \{y_{[t+1:t+h]}\}$, where $h$ is the prediction horizon, $y$ is the target time series, and $x$ are exogenous covariates. 
The prediction task is to estimate the conditional distribution:
\begin{equation}
\mathbb{P}(y_{[t+1:t+h]}| y_{[0:t]}, x_{[0:t+h]}) = f_\theta(y_{[0:t]}, x_{[0:t+h]})
\end{equation}



\subsubsection{\rm \textbf{Tokenization mechanisms}}
\label{sec-token}
Motivated by ViT~\cite{dosovitskiy2021image}, many existing works use patching to convert the raw input sequences to tokens.
In particular, each time series $x_{[0:t]}$ is segmented into a series of patches, which may overlap or be distinctly separate. 
The patch length is denoted as $P$, and the stride, representing the non-overlapping interval between consecutive patches, is denoted as $S$. 
Consequently, this patching technique produces a sequence of patches $x_p \in \mathbb{R}^{P \times N}$, where $N$ denotes the number of patches, calculated by $N = \left\lfloor \frac{L-P}{S} \right\rfloor + 2$. 
Before patching, $S$ repetitions of the final value $x_t$ are appended to the sequence's end.
This tokenization mechanism effectively reduces the number of input tokens from $L$ to roughly $L/S$, which significantly diminishes the memory space consumption and computational intensity.

\subsubsection{\rm \textbf{Model architecture}}

Most existing works employ either encoder-decoder or decoder-only architecture as the backbone model to train a time-series foundation model.

\textbf{Encoder-decoder:} 
The encoder-decoder transformer architecture stands out for its remarkable efficiency and efficacy, primarily attributed to its self-attention mechanism~\cite{vaswani2017attention}. 
%
Fig.\ref{fig-prediction} shows an example named TimeGPT that exemplifies the application of the encoder-decoder transformer for prediction problems\cite{garza2023timegpt1}. 
In particular, TimeGPT inputs a historical sequence of data points to predict future values.
The inputs are added relative positional embedding, which demonstrates higher capability to handle long sequences than the original absolution positional embedding of the transformer~\cite{vaswani2017attention}.
Its encoder captures temporal dependencies within the historical context, encoding it into a latent space, while the decoder utilizes this encoded information to predict future values.
As shown in Fig.\ref{fig-prediction}, with its specialized architecture, TimeGPT can address the intricacies of time-series data, such as trends and seasonality, which makes it an ideal model for telecom time-series predicting such as network traffic load, channel state, user mobility, etc. Once pre-trained, such a universal model can be used for various prediction tasks without extra training. 
By contrast, conventional methods such as RNN and DNN are usually task-specific, and training a new model from scratch for each incoming new task is time-consuming.

\begin{figure}[!t]
\centering
\setlength{\abovecaptionskip}{0pt} 
\includegraphics[width=0.94\linewidth]{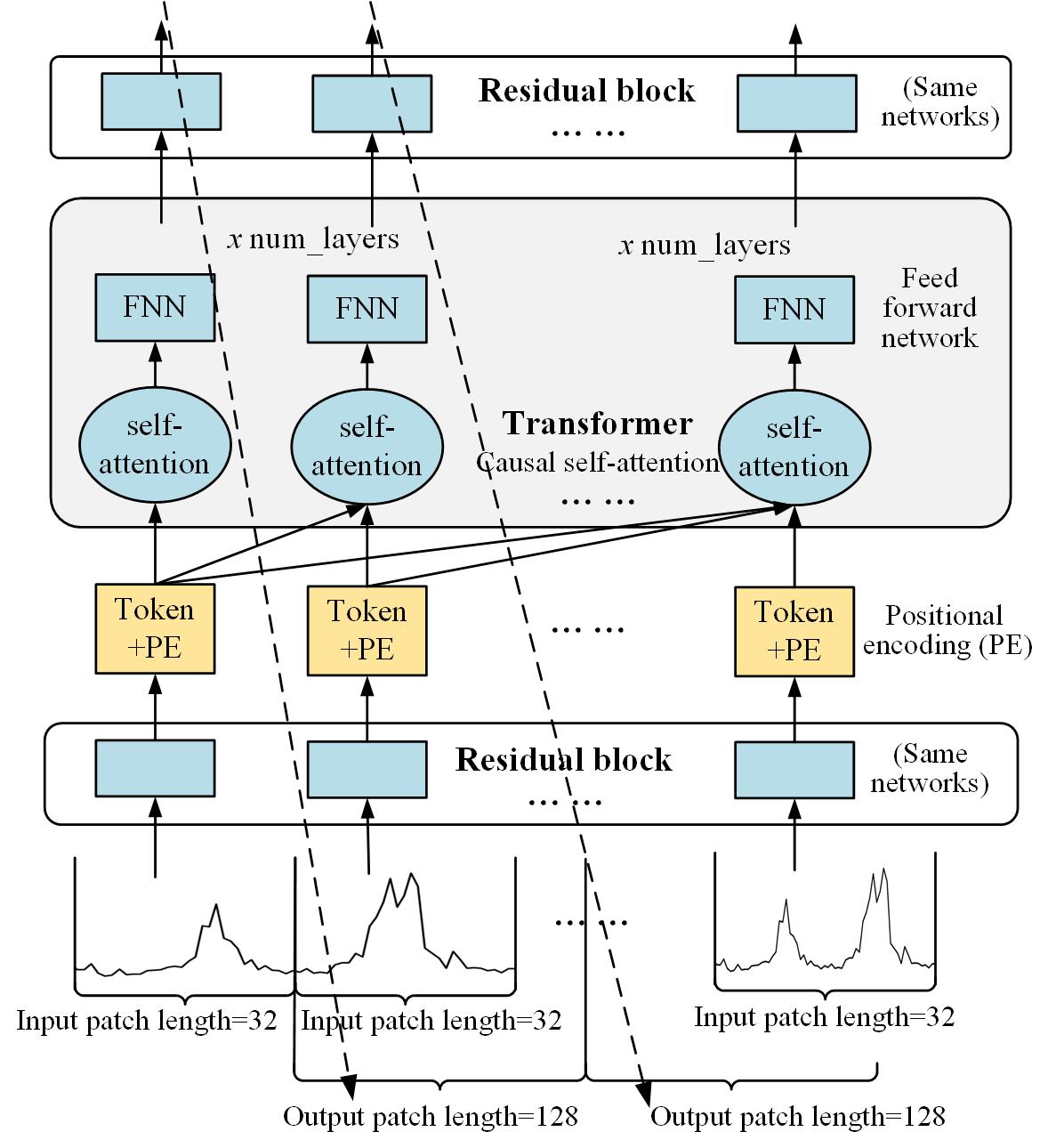}
\caption{A decoder-only model named TimesFM for time-series prediction proposed in \cite{das2024decoderonly}.}
\label{fig-prediction-decoder}
\end{figure}

\textbf{Decoder-only:} Even though encoder-decoder models exhibit impressive effectiveness for handling sequences, decoder-only models become more popular in recent years.
\blue{Like the encoder-decoder architecture, the decoder-only model must first tokenize the raw inputs and then incorporate positional embeddings.}
The essential difference between encoder-decoder and decoder-only models is that the bidirectional attention is used by encoder~\cite{vaswani2017attention}, which means each token is attending to all other tokens.
In contrast, the decoder-only model employs casual attention, where each token cannot attend to tokens after it, but can only look at tokens before it.
\blue{The causal attention mechanism enhances prediction models because it is well-suited for time-series forecasting tasks. In time-series forecasting, we typically predict future values based on historical data. Causal attention allows each token to consider all preceding tokens, meaning it attends to all events that occurred before the current time frame. Moreover, with these small modifications, the attention score matrices in decoder-only models are triangle matrices, which always have full-column rank, resulting in better expressibility~\cite{dong2023attention}.}
%
As shown by Fig.\ref{fig-prediction-decoder}, TimesFM~\cite{das2024decoderonly} employs decoder-only architecture to train a time-series prediction model.
Unlike traditional LLM techniques, which predict one element at a time, TimesFM is designed to predict extended future sequences in a single step, enhancing accuracy for long-term predictions. 
%
This flexibility also extends to inference; given a series, the model can predict its immediate future in fewer steps than a model with equal-length input and output segments would require. 
Such fast inference could be an appealing feature for telecom applications, because many prediction tasks require rapid response to network dynamics, such as channel state, short-term traffic changes, and indoor user locations. 
Conventional prediction methods usually take a long training time to adapt to such network environment changes. By contrast, TimesFM has the potential to capture short-term patterns instantly, which aligns with the fast-response requirements of telecom networks. 



%
In summary, the key differences between encoder-decoder and decoder-only architecture can be found by comparing Fig. \ref{fig-prediction} and Fig. \ref{fig-prediction-decoder}. 
In particular, the encoder-decoder design in Fig. \ref{fig-prediction} includes an encoder to encode the raw features into latent representations by using bidirectional attention. 
In contrast, the decoder-only scheme in Fig. \ref{fig-prediction-decoder} illustrates causal attention, e.g., the first token is attended by all other tokens, and the second token is attended by all except the first token.

With a backbone model and an effective tokenization mechanism, one can train a time-series prediction foundation model for telecom with a mixture of different datasets. 
Meanwhile, understanding the tokenization and model architecture differences is crucial for designing and pre-training a time series LLM for telecom applications. 
For instance, the tokenization mechanism introduced in the previous Section \ref{sec-token} is very useful for telecom applications, since telecom networks are associated with a large number of network devices and end users, generating a huge number of datasets, such as historical CSI, traffic load level\cite{kousias2023large}, and network performance metrics\cite{raca2020beyond}.
Therefore, reducing the number of input tokens can lower the pre-training difficulty of \blue{LLMs}, especially considering that network edge devices usually have limited computational resources. 

\subsection{Frozen Pre-trained LLM for Prediction}
\label{sec-frozen}

Rather than developing a specific \blue{LLM} for prediction, frozen pre-trained LLM refers to approaches that directly adapt a general-domain LLM to prediction tasks.
This section delves into using a pre-trained LLM for prediction tasks without the necessity for further fine-tuning. 
There are two primary approaches: prompting-based and preprocessing-based methods.
Specifically, the prompting-based methods include hard and soft prompts. 
Hard prompts employ rigid and predefined textual structures to present time-series information in a format that is intuitive for the language model.
Conversely, soft prompts adopt a more nuanced strategy by integrating trainable embeddings within the input that subtly guide the language model's predictions.
Meanwhile, preprocessing-based methods aim to reformat the time series numerical values into a representation that aligns more seamlessly with LLM's tokenization process, rather than introducing extra template tokens or trainable embeddings.

\begin{table*}[ht]
\centering
\caption{Three hard prompt examples for prediction tasks in telecom.}
\label{tab:template-examples}
\small
\setstretch{1.1}
\resizebox{0.9\textwidth}{!}{
\begin{tabular}{|l|l|l|}
\hline
\multirow{3}*{\makecell{ Network \\traffic load \\ prediction}} & Input prompt (source) & From $\{t_1\}$ to $\{t_{obs}\}$, network $\{U_m\}$ experienced $\{x_{{t_1}:t_{obs}}\}$ GB of traffic each hour. \\
\cline{2-2}
& Question & What will the data traffic be on $\{t_{obs+1}\}$? \\
\cline{2-2}
& Output (target) & The network will experience $\{x_{t_{obs}+1}\}$ GB of traffic. \\
\hline
\multirow{3}*{\makecell{ Network \\ users\\ prediction}} &  Input prompt (source) & From $\{t_1\}$ to $\{t_{obs}\}$, the BS had $\{x_{{t_1}:t_{obs}}\}$ active connections each day. \\
\cline{2-2}
& Question & What will the BS utilization be on $\{t_{obs+1}\}$? \\
\cline{2-2}
& Output (target) & The BS will have $\{x_{t_{obs}+1}\}$ active connections. \\
\hline
\multirow{3}*{\makecell{ Customer \\service\\ prediction}} & Input prompt (source) & From $\{t_1\}$ to $\{t_{obs}\}$, customer service received $\{x_{{t_1}:t_{obs}}\}$ calls each week. \\
\cline{2-2}
& Question & How many service calls will be received in the week of $\{t_{obs+1}\}$? \\
\cline{2-2}
& Output (target) & There will be $\{x_{t_{obs}+1}\}$ service calls received. \\
\hline
\end{tabular}
}
\end{table*}

\textbf{1) Prompting-based methods}
In leveraging prompt engineering, two predominant prompting strategies are utilized: hard prompts~\cite{xue2023promptcast} and soft prompts~\cite{lester2021power}. 

\textbf{Hard prompts} (\(P_{\text{hard}}\)) involve pre-pending a fixed textual instruction \blue{and} query to the input data sequence\blue{, or fit the raw input data into a carefully designed template. 
In this way, we can transform numerical values into textual contexts that can be processed by pre-trained large language models.
By leveraging the impressive generalizability of pre-trained LLMs, hard prompting techniques can yield high prediction accuracy in zero-shot settings.} 
Specifically, the model input for a time-series \(x_{[1:T]}\) with a hard prompt is thus formalized as the concatenation \([P_{\text{hard}}; x_{[1:T]}]\), which directs the model to generate a prediction in response to the prompt. 
When designing hard prompts for time-series prediction with language models, the general guideline is to transform numerical data into a format that mimics natural language constructs~\cite{xue2023promptcast}. 
This involves two main components: input prompts and output prompts. 
Input prompts provide historical context and highlight the target time step for prediction, while output prompts focus on the desired prediction value, serving as the ground truth label for training or evaluation. 
Table~\ref{tab:template-examples} presents several telecom examples of designing hard prompts for specific tasks, such as network traffic load prediction, network user number prediction, and customer service prediction.
The process mirrors the source/target structure common in machine translation tasks or can be likened to a question-answering setting, with the context as background information and the question seeking future insights. 
Then the output prompt becomes the answer to this query, such as "\textit{the number of active users, traffic load level at specific times, and predicted customer calls next week}". 
\blue{According to PromptCast~\cite{xue2023promptcast}, this simple approach achieves comparable or superior prediction accuracy across various datasets, demonstrating its effectiveness on bridging the gap between raw numerical sequences and language-based data representations, further facilitating the use of language models for prediction tasks traditionally handled by numerical methods.}

\begin{figure}[!t]
\centering
\includegraphics[width=0.87\linewidth]{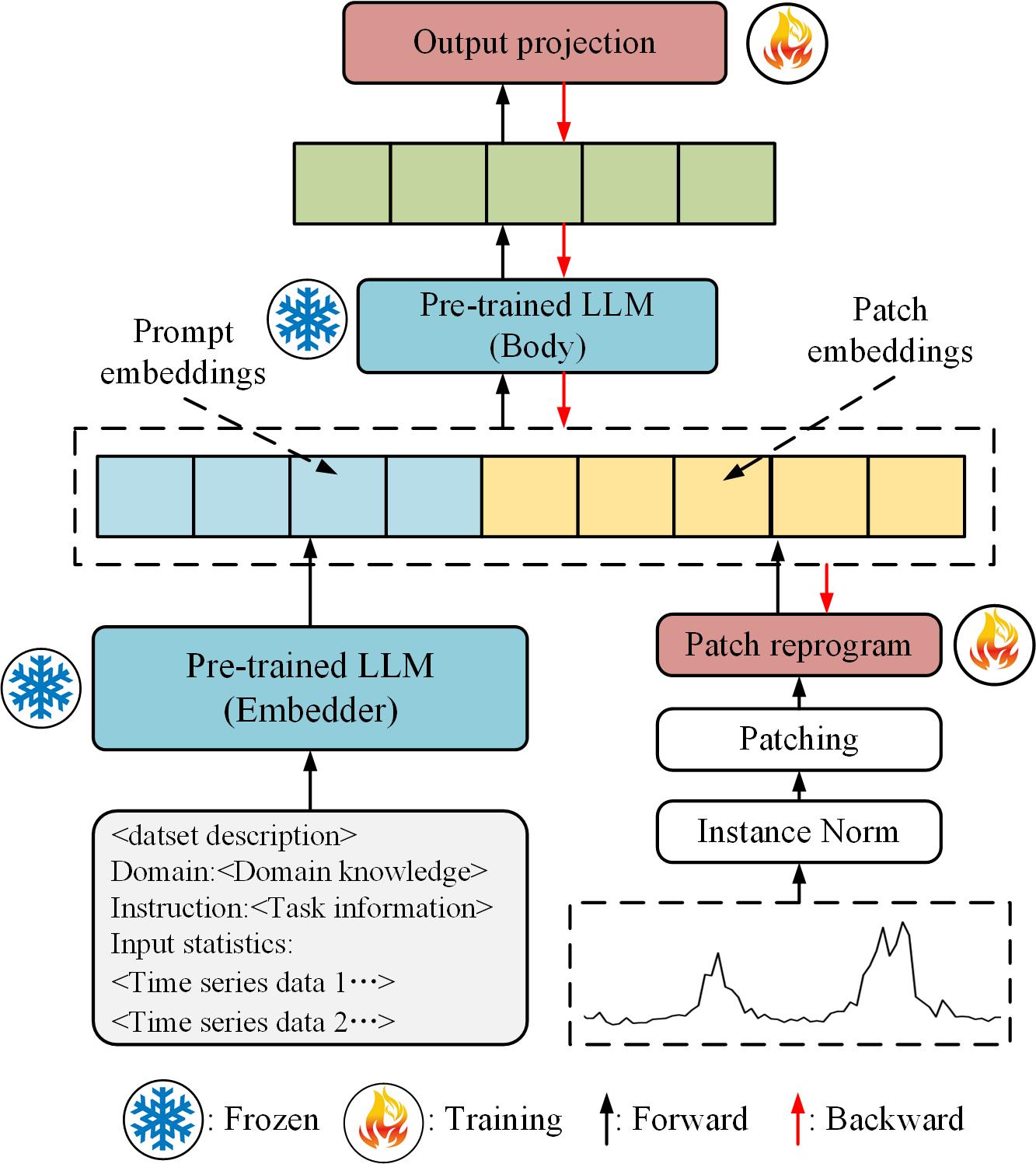}
\caption{ The model framework of TIME-LLM with soft prompt\cite{jin2024timellm}.}
\label{fig-timellm}
\end{figure}

Conversely, \textbf{soft prompts} (\(P_{\text{soft}}\)) introduce trainable embeddings that are optimized during training to influence the model's prediction subtly~\cite{jin2024timellm}. 
The input for a soft prompt is represented as \([P_{\text{soft}}; x_{[1:T]}]\), where \(P_{\text{soft}}\) constitutes a series of parameters that are fine-tuned to enhance the predictive capability of the model. 
This adjustable approach allows the model to internalize and apply nuanced guiding signals without the rigidity of fixed textual cues.
Fig. \ref{fig-timellm} shows an example of using soft prompts in \cite{jin2024timellm}.
\blue{TIME-LLM utilizes two types of inputs: a textual description of domain knowledge with some in-context examples, and a time-series input. The textual description is tokenized and processed through the embedding layers of the pre-trained LLM to generate latent representations, termed as prompt embeddings. When a time-series input is received, it is tokenized and embedded through a method called patching, which includes a specialized embedding layer (patch reprogram), resulting in patch embeddings. The pre-trained LLM then takes the concatenated prompt and patch embeddings as inputs and produces outputs via an output projection layer. Throughout this process, all parameters of the pre-trained LLM remain frozen, requiring training only for the custom embedding layer to connect the time series and textual data.} 
When designing soft prompts for time-series prediction using a pre-trained LLM, there are a couple of guiding principles and design choices. Unlike hard prompts, soft prompts require no explicit textual additions to the input data. 
The general approach involves encoding time series data into a format that the LLM can process, harnessing its underlying capabilities to discern patterns and generate predictions.
To utilize soft prompts prediction for telecom efficiently, one might consider the specific characteristics of the telecom time-series data, such as traffic patterns or usage trends, to design the transformation and reprogramming steps that align the time-series data with the model's language understanding capabilities.

\textbf{2) Preprocessing-based Methods}
%
The preprocessing-based method leans on the LLM's inherent ability to detect and follow patterns within generalized sequences, devoid of reliance on any specific language structure. 
In particular, when numerical values are adeptly transformed into textual strings, prediction with the model adheres to standard language model sampling methods.
%
Therefore, tokenization plays a pivotal role because it shapes the model's perception of numerical patterns. 
LLMTIME~\cite{gruver2023large} proposes two ways to preprocess the raw data:
\begin{itemize}
    \item \textbf{Introducing extra space:}
For example, GPT-$3$'s tokenizer might dissect the number $42235630$ into $[422, 35, 630]$, which complicates arithmetic operations. 
To address this, a preprocessing step is introduced where digits are separated by spaces, and time steps by commas, ensuring uniform tokenization of each digit: "4 2 2 3 5 6 3 0". 
With this small change, the tokenizations are completely different.
Each digit now is processed by the model individually.

\item  \textbf{Eliminating decimal points and rescaling}
Given a fixed precision, the decimal points are redundant and unnecessary.
Decimal points are excluded under fixed precision to optimize context length, transforming a series $"0.123, 1.23, 12.3, 123.0"$ into $"1 2, 1 2 3, 1 2 3 0, 1 2 3 0 0"$. It provides a straightforward approach to processing the inputs. 
\end{itemize}
In terms of telecom application potentials, these two preprocessing techniques provide a simple but efficient approach to using LLM techniques for prediction.
They eliminate the need for careful designs of prompts, which can better adapt to various prediction tasks in telecom. Preprocessing-based methods have the potential to generate prediction results instantly based on given raw network data input.

\subsection{Fine-tuned LLM Prediction} \label{sec:fine-tune4predict}
\label{sec-fine-tune}
\begin{figure}[!t]
\centering
\includegraphics[width=0.7\linewidth]{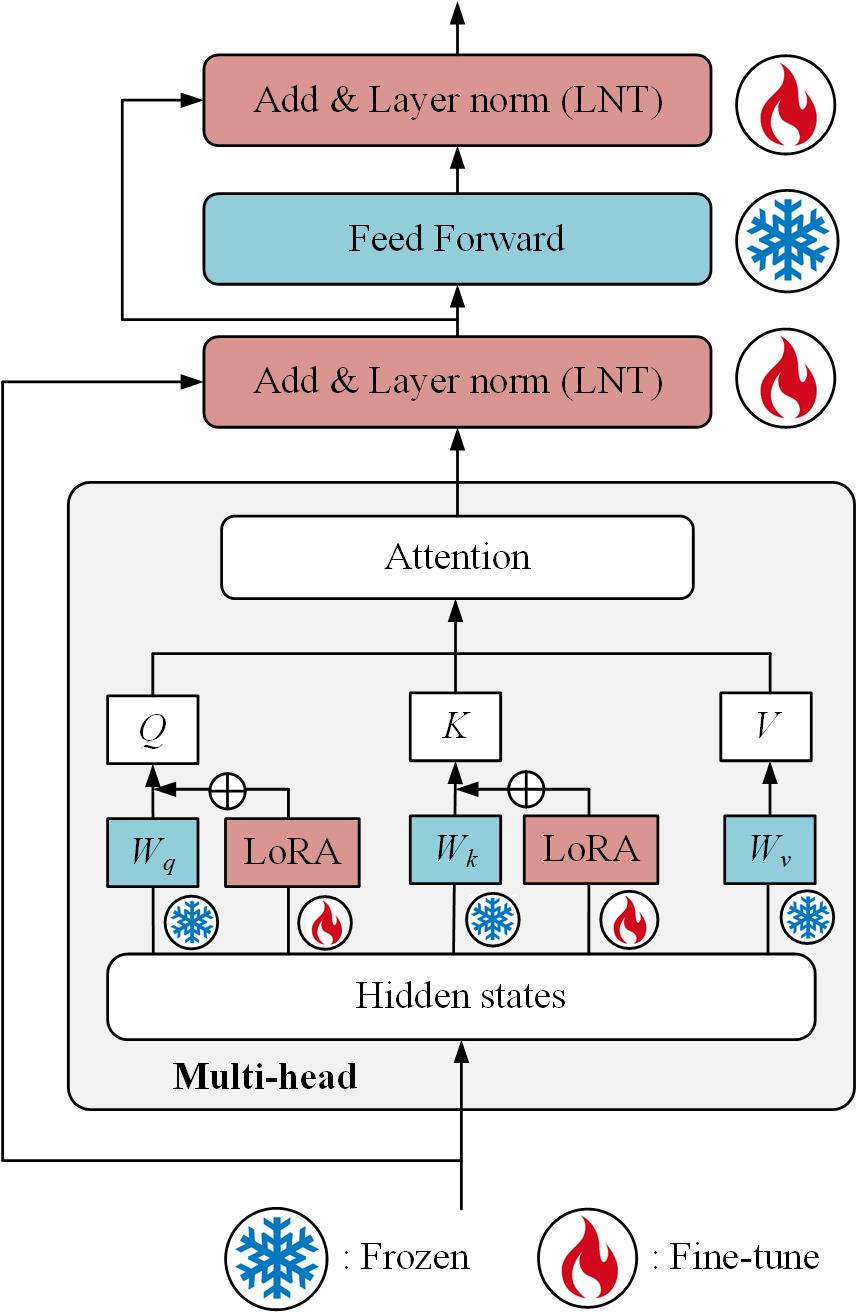}
\caption{ The model framework of LLM4TS framework\cite{chang2024llm4ts}. $Q$, $K$, $V$ are the query, key, value vectors respectively. $W_q$, $W_k$, $W_v$ are the matrices used for generating query, key and value vector.}
\label{fig-llm4ts}
\end{figure}

Fine-tuning pre-trained \blue{LLMs} presents a significant advancement for time-series prediction, offering a powerful alternative to traditional prediction approaches~\cite{chang2024llm4ts, zhou2023fits}. 
General-domain \blue{LLMs}, initially pre-trained on extensive linguistic data, can be fine-tuned to capture the unique temporal patterns inherent in time-series data. 
This process equips LLMs with the ability to effectively prediction in domains where data scarcity or specificity presents challenges to conventional deep learning models. 
%
In the pursuit of efficiency and practicality, most recent works have shifted towards parameter-efficient fine-tuning methods like Low-Rank Adaptation (LORA)~\cite{hu2021lora} and Layer Normalization Tuning (LNT)~\cite{lu2021pretrained}. 
In particular, LORA adapts pre-trained models to new tasks by modifying the weight matrices of the model's layers.
Given a weight matrix $W \in \mathbb{R}^{d \times m}$ in a pre-trained model, LORA fine-tuning introduces two low-rank matrices $A \in \mathbb{R}^{d \times r}$ and $B \in \mathbb{R}^{r \times m}$, where $r$ is the rank and $r \ll \min(d, m)$. 
The weight matrix $W$ is updated as:
\begin{equation}
W' = W + \Delta W, \quad \text{where} \quad \Delta W = AB.
\end{equation}
The matrices $A$ and $B$ are the parameters learned during fine-tuning while the original weights $W$ are kept frozen. 
This results in a model that is fine-tuned for the task at hand with only a small increase in the number of parameters.
Many works of fine-tuning LLMs proposed applying the technique to the query (Q) and value (V) matrices in attention layers, showing notable results without extending it to all parameters within the attention or feed-forward layers. 
In the context of time-series prediction, however, this selective fine-tuning may require adjustment. 
As shown in Fig. \ref{fig-llm4ts}, LLM4TS~\cite{chang2024llm4ts} applies LORA fine-tuning to the query (Q) and key (K), achieving state-of-the-art performance.
\blue{It augments the pre-trained model with additional trainable components and thus incorporates changes to the model's weights, unlike the soft prompting to modify the inputs.}
Using LORA allows for retaining the general capabilities of the LLM while imbuing it with domain-specific knowledge, ensuring that the time-series prediction model is both specialized and robust.

On the other hand, LNT offers a focused approach to adapt pre-existing parameters in transformer blocks to specific tasks. 
LNT specifically targets the affine transformation parameters within the layer normalization components of a transformer model. 
These parameters, such as scale and shift, originally set to ensure standardized input distribution across network layers, become trainable to allow the model to retain its learned representations while fine-tuning the time-series prediction.
%
%
As shown in Fig. \ref{fig-llm4ts}, LLM4TS~\cite{chang2024llm4ts} employs both LNT and LORA fine-tuning for the query and key.
A similar strategy can be found in \cite{zhou2023fits}, which freezes all attention and feed-forward layers, and only fully fine-tunes the embedding layers and applies the LNT.
Incorporating LNT in the fine-tuning process, in the context of adapting pre-trained \blue{LLMs} for time-series prediction, provides a mechanism for the model to adjust its internal normalization to better fit the dynamics and scale of the time-series data. 

These parameter-efficient fine-tuning methods, such as LoRA and LNT, are crucial for the practical deployment of \blue{LLMs} in telecom. 
Lin \textit{et al.} claims that applying LoRA to GPT-3 can reduce the number of trainable parameters from 175.2 billion to 37.7 million\cite{lin2023pushing}, and combining LoRA with federated split learning can significantly reduce computing and communication latency at the mobile edge.

\begin{figure*}[!t]
\centering
\setlength{\abovecaptionskip}{0pt} 
\includegraphics[width=1\linewidth]{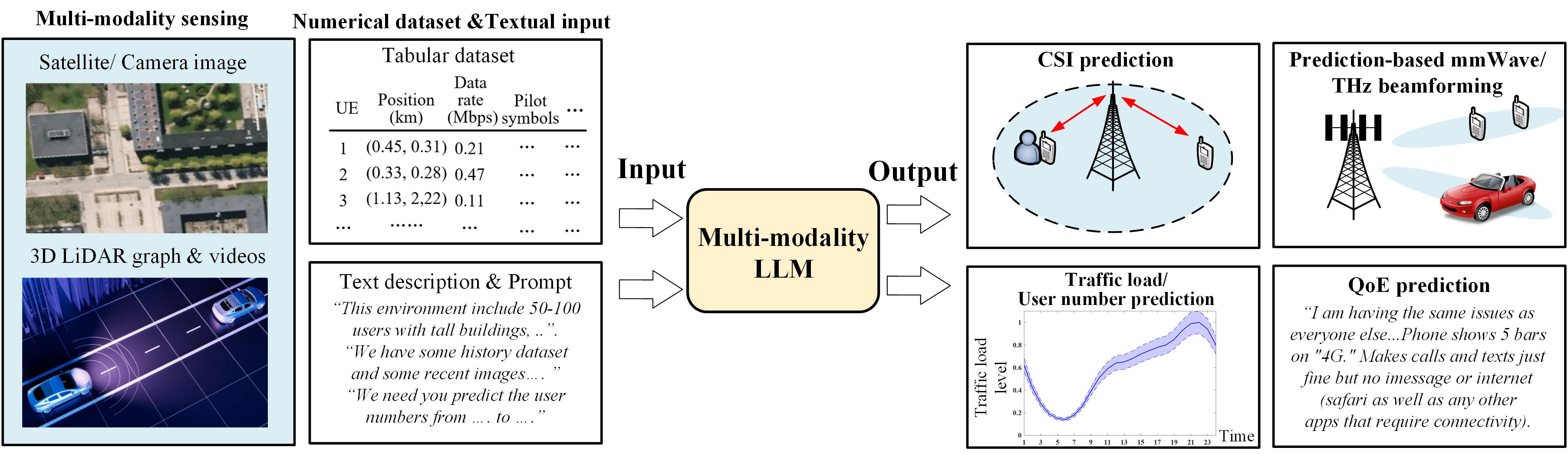}
\caption{Multi-modality LLM for prediction problems in wireless networks.}
\label{fig-multi-moda}
\end{figure*}

\subsection{Multi-modal LLM for Telecom Prediction}
\label{sec-multimodal}

Multi-modal learning is a promising feature of LLM techniques, aiming to process related information from multiple modalities, such as text, audio, image, video, 3D maps, graphs, etc~\cite{zhang2023meta}. A multi-modal LLM can use diverse encoders to extract features from different modalities into desired outputs, indicating a more comprehensive and flexible approach to process information. Such multi-modal capabilities can be particularly useful for integrating sensing and communication, which is a crucial technique in 6G networks.
In particular, as shown in Fig. \ref{fig-multi-moda}, LLMs can include multiple inputs with diverse modalities, e.g., the image captured by satellite or street cameras, 3D LiDAR maps and videos collected by vehicles. Sensing has become a critical part of envisioned 6G networks, and multi-modal LLMs are capable of making the most of the collected sensing information. On the other hand, LLMs can also include conventional tabular-based numerical input, and further consider textual input and prompt instructions. 
With multi-modal inputs, LLM agents can better understand the surrounding environment and then make more accurate predictions for network dynamics.

\textbf{1) Channel state information (CSI) prediction:} CSI plays an increasingly vital role in wireless networks, enabling the transmitter to adjust the transmission parameters based on current channel conditions and, therefore, achieve better performance. Prediction-based methods are appealing methods to obtain instantaneous CSI. For instance, Jiang et al.~\cite{jiang2020deep} applied deep learning for CSI prediction using generated or historical data. 
Most existing studies consider single modality input, which usually consists of tabular-based numerical data such as historical CSI. However, the real-world environment can be more complicated, and CSI may be affected by many other factors such as weather conditions and dense buildings \cite{luo2018channel, sen2022terahertz}. 
These multi-modal inputs, such as weather maps and building distributions, can provide a more comprehensive understanding of the signal transmission environment, but jointly processing these inputs is beyond the capabilities of existing techniques.       
Multi-modal LLMs offer promising solutions by jointly considering diverse modalities and data sources, producing more accurate CSI prediction results. In addition, users can provide textual prompt instructions, which are easy-accessible and user-friendly for non-researcher users.

\textbf{2) Prediction-based mmWave\slash THz beamforming:} 
The increasing traffic demand and limited bandwidth resources make mmWave and THz communications promising techniques. However, these high-frequency transmissions are highly directional and vulnerable to signal blockages. 
Consequently, efficient beamforming and alignment are required to achieve reliable mmWave and THz networks. For instance, Ke et al.~\cite{ke2019position} applied a Gaussian process-based ML scheme for UAV position prediction and UAV-mmWave beam-tracking, and Shah et al.~\cite{shah2022multi} deployed LSTM networks to predict multiple mmWave beams from multiple cells. 
These predictions usually consider numerical input, especially historical data \cite{ke2019position}. 
Charan et al.~\cite{charan2021vision} introduced computer vision-aided techniques for signal blockage prediction, using cameras on BSs to capture possible blockages and then initiate user hand-off beforehand. 
However, it is still limited to a single image modality with limited environmental information.
By contrast, multi-modal LLMs can take holographic input from the environment, and jointly consider historical tracks, instant images, and text instructions, etc. These comprehensive inputs can produce more accurate and reliable prediction results, contributing to efficient mmWave and THz beamforming.

\textbf{3) Traffic load prediction:}
Accurate traffic load prediction is the prerequisite of efficient network management, which is related to user numbers, service types, time periods, and so on. Similar to CSI prediction, most existing studies take numerical datasets as single modal input \cite{alekseeva2021comparison, 9685424}. 
For instance, Alekseeva et al.~\cite{alekseeva2021comparison} compared the performance of seven ML algorithms (including Bagging, Random Forest, Gradient Boosting, Linear Regression, Bayesian Regression, Huber Regression, and SVM Regression) on the task of traffic load prediction. 
Their findings indicate that Boosting-based methods demonstrate superior performance when handling large volumes of load data, yet incurring significant training costs.
Hu et al.~\cite{9685424} integrated a sequence of AutoEncoders to extract multiple sets of latent temporal features from historical load data for load prediction, which ensures that extracted feature sets are representative of the entire load data.
Most existing studies mainly consider two factors: the spatial correlation between nearby BSs and the temporal dynamics captured in historical data. 
However, Abdullah et al.~\cite{abdullah2022weather} suggested that various meteorological factors, such as rain, wind, and temperature, can also significantly influence the volume of traffic loads. 
Consequently, multi-modal LLMs may be used to harness diverse information streams, including spatial BS corrections, temporal historical traffic loads, and environmental factors, facilitating accurate load prediction and providing effective network management and service delivery.

\begin{table*}[!t]
\caption{Summary of LLM-based Prediction for Telecom. }
\centering
\small
\setstretch{1.1}
\resizebox{1\textwidth}{!}{%
\begin{tabular}{|m{1.8cm}<{\centering}|m{4cm}<{\centering}|m{4cm}<{\centering}|m{3.5cm}<{\centering}|m{5.5cm}<{\centering}|}
\hline
LLM-based prediction techniques & Main features & Input and fine-tunning requirements & Advantages compared with conventional approaches & Telecom prediction application opportunities\\
\hline
\makecell{ Pre-training \\foundation \\ models}    & Leverages pre-trained models on diverse time series datasets to capture general temporal patterns. It means training \blue{LLMs} from scratch specifically for prediction purposes. & Requires a large corpus of time-series data for initial pre-training, but collecting these datasets may be difficult in telecom; fine-tuning may be needed for specific telecom tasks.  & The model has zero-shot prediction capabilities, and quickly adapts to new tasks with minimal fine-tuning; captures a wide range of temporal dynamics. & A prediction foundation model for telecom can handle various short-term or long-term prediction tasks, such as traffic load prediction, CSI prediction, user number estimation, and so on.\\
\hline
\makecell{ Frozen \\ pre-trained \\ LLM \\ for prediction}    & Using pre-trained \blue{LLMs} without fine-tuning their parameters, including prompting-based and preprocessing-based methods.   & Tokenization and embedding of time series data; For prompt-based methods, the prompt format must be carefully designed; No fine-tuning for \blue{LLMs} is required. & Low computational cost and design complexity; Leveraging the generalization capabilities of a pre-trained LLM directly. & 
This technique is particularly useful for short-term prediction, such as short-term traffic load and network performance prediction. The low computational cost can also adapt to network edge, even mobile applications.\\

\hline
\makecell{Fine-tuned \\ LLM \\ prediction}    & Adapts a pre-trained LLM to telecom-specific prediction tasks through fine-tuning techniques such as LoRA and LNT. & It requires time series data for fine-tuning; may require parameter-efficient fine-tuning like LoRA and LNT to improve the efficiency. & Fine-tuning can incorporate telecom domain knowledge into \blue{LLMs}, improving the accuracy and specificity of telecom tasks. &  Fine-tuning LLMs can better adapt to specific tasks in telecom, e.g., collecting specific datasets to fine-tune an LLM for user localization. It is more flexible than pre-trained models from scratch, and more reliable than pure prompting-based methods. \\

\hline
\multirow{2}*{\makecell{ Multi-modality\\ prediction}}  & Using \blue{LLMs} to jointly consider multi-modal environment information, e.g., tabular data, text, and image, aiming to provide more accurate prediction results.   &  It requires multi-modal input and proper prompt to predict desired output; \blue{LLMs} can be specifically pre-trained/fine-tuned to further improve the performance and generalization capabilities. &  Multi-modal can take advantage of inputs with various modalities. With these comprehensive inputs, LLMs can better predict the network dynamics than existing methods.  &  Sensing is an important part of 6G networks, and multi-modal sensing can provide more comprehensive input for \blue{LLMs}, producing more accurate prediction results by jointly considering various inputs, e.g., more accurate traffic load prediction and beam steering.   \\
\hline
\end{tabular}}
\label{tab-mlsummary}
\end{table*}

\textbf{4) Quality of Experience (QoE) prediction:}
QoE is a measure of the customer's experiences of specific services, which is a useful metric in diverse mobile scenarios, such as mobile edge computing\cite{mitra2013context}, edge caching \cite{liang2017enhancing}, and resource allocation \cite{sousa2020survey}. 
QoE is closely related to the user's natural language comments. An example is given by \cite{mitra2013context}: “\textit{I am having the same issues as everyone else...Phone shows 5 bars on 4G." Makes calls and texts just fine but no imessage or internet (safari as well as any other
apps that require connectivity). Right now the two things I
have noticed are that I’m more likely to have it work late at
night (11pm-2am) and more likely to have it work when I’m
outdoors...}"    
Most existing studies predict QoE by extracting the key attributes of users, devices, applications, and networks for modeling and measuring. 
However, this comment indicates a specific network issue "\textit{no imessage or internet at midnight}".  
Extracting such an informative and specific user experience to several attributes could lead to considerable information loss, and therefore the service provider cannot fully understand the user's demand. 
With multi-modal LLMs, user's textual comments and network numerical metrics can be jointly evaluated, providing a comprehensive evaluation of the network performance and user experiences. In addition, multi-modal LLMs can also be used to generate and predict user experience using LLM's comprehension and reasoning capabilities.    



\subsection{Discussions and Analyses}

Table \ref{tab-mlsummary} summarized the LLM-enabled prediction techniques in terms of main features, input and fine-tuning requirements, advantages, and telecom prediction application opportunities. We summarize the key findings as follows.

Firstly, large-scale time-series datasets are important for building Time-LLM for telecom.
Previous sections have demonstrated LLM's potential for solving time-series prediction problems.
However, it is worth noting that time-series datasets are prerequisites of pre-training Time-LLM, and then the LLM can understand and capture the hidden patterns of the input data.
Despite the importance, collecting such datasets can be difficult in telecom due to various data formats and sources, different network operators, customer privacy, etc.

Secondly, prompting and preprocessing-based methods are the most efficient approaches to using LLM for prediction tasks.
Compared with pre-training and fine-tuning, the discussions in Section \ref{sec-frozen} demonstrate that prompting is one of the most straightforward methods of using an LLM for prediction tasks. 
Such an advantage can still be explained by LLM's impressive zero-shot learning capabilities. 
In addition, preprocessing input data is another simple method. Transforming numerical values into textual strings can make the most of LLM's capabilities in processing standard language tasks.
These two methods are particularly useful for short-term prediction problems in telecom with instant responses.

In addition, previous sections also show that parameter-efficient fine-tuning methods are critical for LLM deployment in telecom.
Section \ref{sec:fine-tune4predict} introduced two parameter-efficient methods, LoRA and LNT, to fine-tune LLM for prediction tasks.
Efficient fine-tuning methods can improve overall computing efficiency,  lowering the demand for computational resources.
These features are very useful for processing various tasks ranging from generation and classification to prediction problems in the telecom field such as in Section \ref{sec-fine-tune}, especially considering limited computational and storage resources at the network edge.

Finally, multi-modal LLM has great potential for telecom applications.
Incorporating multi-modality LLM into telecom has been discussed in multiple existing studies\cite{bariah2023large}, and Section \ref{sec-multimodal} investigates the potential for telecom prediction problems such as CSI prediction, prediction-based beamforming, and QoE prediction.
The key motivation is that multi-modal information from multiple sources can contribute to prediction accuracy, and such enhancement can further improve the network operator's decision-making.

\section{Challenges and Future Directions of LLM-empowered telecom}
\label{sec-challenge}

This section will introduce the challenges of realizing LLM-empowered telecom, including telecom-domain LLM training, practical LLM deployment in telecom, and prompt engineering for telecom applications. 
Then, we identify several future directions, e.g., LLM-enabled planning, model compression and fast inference, overcoming hallucination problems, retrieval augmented-LLM, and economic and affordable LLMs.

\begin{table*}[!t]
\caption{\blue{\textbf{Summary of Telecom datasets for LLM}}}
\centering
\small
\setstretch{1.3}
\resizebox{1\textwidth}{!}{%
\begin{tabular}{|m{4.8cm}<{\centering}|m{4cm}<{\centering}|m{4cm}<{\centering}|m{3.5cm}<{\centering}|m{2.5cm}<{\centering}|}\hline
Dataset &Task  &Document size  &Question size  &Open-source  \\\hline
5GSum\cite{karim2023spec5g} &Summarization  &713 articles  &N/A  &Yes  \\\hline
Tspec-LLM\cite{nikbakht2024tspec} &Summarization  &30,137 documents  &100 questions  &Yes  \\\hline
TeleQnA\cite{maatouk2023teleqna} &Question answering  &N/A  &10,000 questions  &Yes  \\\hline
NetEval\cite{miao2023empirical} &Question answering  &N/A  &5,732 questions  &Yes  \\\hline
TeleQuAD\cite{holm2021bidirectional} &Question answering  &N/A  &2,021 questions  &No  \\\hline
StandardsQA\cite{roychowdhury2024unlocking} &Question answering  &N/A  &2,400 questions  &No \\\hline
ORAN-Bench-13K\cite{gajjar2024oran} &Question answering  &116 documents  &13,952 questions  &Yes  \\\hline
5GSC\cite{karim2023spec5g} &Sentence classification  &2,401 sentences  &N/A &Yes\\\hline
\end{tabular}}
\label{tab-datasets}
\end{table*}

\subsection{Challenges of Applying LLM Techniques to Telecom}

\subsubsection{\rm \textbf{Telecom-domain LLM training}}

Previous sections have shown the importance of building telecom-specific LLMs, e.g., telecom-domain question answering\cite{holm2021bidirectional}, telecom troubleshooting\cite{bosch2022integrating}, and standard specification classification\cite{bariah2023understanding}. 
Despite the great potential, training an \blue{LLM} specifically for telecom presents unique challenges due to the complex nature of \blue{telecom} networks. In the following, we will analyze this challenge in detail.

Sufficient \blue{telecom-related datasets} are prerequisites for training a telecom LLM. 
Unlike general-domain \blue{LLMs}, which can leverage large-scale text corpora on the internet, obtaining a sizable dataset exclusively focused on communication networks can be challenging. 
Existing studies usually focus on one specific task and then build the corresponding dataset, e.g., the trouble report dataset \cite{bosch2022integrating}, 3GPP specification dataset \cite{bariah2023understanding}, and telecom question answering dataset \cite{holm2021bidirectional}. 
However, these datasets are usually small-scale and task-specific, and a comprehensive large-scale dataset should include network-related documents, standard specifications, protocols, textbooks, research papers, and other relevant sources.
Maatouk \textit{et al.} started the exploration in \cite{maatouk2023teleqna} by building a dataset with 10000 telecom-related questions and answers, including around 25000 pages and 6 million words. 
More efforts are needed to provide more comprehensive and diverse datasets for telecom LLM training.

Meanwhile, it is worth noting that telecom networks involve a large number of various concepts such as network protocols, routing algorithms, network topologies, network security, etc.
Therefore, teaching an LLM to comprehend and reason about these complex concepts requires a robust training strategy. 
An effective approach is to pre-train the LLM on a large-scale general language corpus and then fine-tune it on specific communication network datasets, e.g., datasets for the BS services, historical datasets for prediction, or datasets for edge computing-related tasks. 
In addition, balancing model size and performance is crucial. 
\blue{LLMs} trained on large-scale datasets tend to be computationally expensive and memory-intensive. Appropriate model size can reduce the burden on energy and computation resources during pre-training and fine-tuning phases. 
In addition, balancing model size and performance is crucial to ensure practical usability, especially considering scenarios with limited computational capacity such as vehicles and mobile phones.  
Therefore, techniques like model compression, knowledge distillation, or utilizing specialized hardware accelerators can be explored to reduce the model's size and enhance its efficiency without compromising its understanding of communication networks.

\blue{The above analyses show that obtaining domain-specific datasets is one of the main bottlenecks of training telecom LLMs. Several datasets have been released recently to address this challenge. As shown in Table \ref{tab-datasets}, there exist different types of datasets, focusing on tasks including text summarization, question answering, and sentence classification. Those datasets are mainly extracted from telecom-related documents. By integrating these datasets, LLMs can be trained to understand and respond to customer queries more effectively, predict and mitigate network issues, and identify fraudulent activities with higher accuracy, ultimately leading to improved operational efficiency and customer satisfaction in the telecom sector.}

\subsubsection{\rm \textbf{Practical LLM deployment in telecom} }

To leverage the benefits of LLM techniques, the models should be properly deployed in telecom networks. 
Specifically, \blue{LLMs} can be deployed at different levels, including central cloud, network edge, or user devices. 
We have introduced the features of each approach in Section \ref{sec-deploy}. However, the related studies are still in very early stages, and the proposed schemes mainly focus on system-level design and definitions. 
The following will discuss the key challenges and difficulties for practical LLM deployment in telecom networks.

Firstly, many real-world wireless applications have stringent requirements for service delay, e.g., autonomous driving and robotic control.
With such time constraints, using a central cloud-based LLM to process these latency-critical tasks can be inappropriate, since the task uploading and solution downloading may increase the service delay.
Additionally, if the task involves image and video processing, the uploading and downloading process will significantly increase the latency, especially considering the limited backhaul capacity.
For instance, the image classification tasks introduced in Section \ref{sec-image} require rapid responses for signal blockage prediction and autonomous driving\cite{charan2021vision} \cite{ahn2022towards}, and processing these requirements on cloud can be impractical due to high service latency.
In addition, the LLM inference time will also contribute to system latency, ranging from 0.58 to 90 seconds. 
Therefore, the service time should be very carefully evaluated before using a LLM for latency-critical applications. 
Network edge provides an efficient approach for computational task processing, and edge intelligence has become an appealing direction to deploy ML algorithms in telecom networks.
However, network edge servers have limited computational or storage capacity, and LLMs are usually computationally intensive with large model sizes, which may prevent edge-LLM deployment.

\blue{To this end, hybrid deployment can be an ideal solution by combining central cloud, edge, and user device deployments, providing a balance between scalability, low latency, and privacy.}
Deploying \blue{LLMs} at different levels, including the central cloud, network edge, or user devices, offers unique opportunities and challenges in telecom applications. 
\blue{For example, large-scale LLMs such as GPT-4 and LLama3-70b are deployed at the central cloud to handle tasks with high-quality requirements on the generated content. Meanwhile, small-scale LLMs are implemented at the network edge or even on user devices for latency-sensitive tasks.}
However, coordinating LLMs at different levels can be challenging and requires dedicated designs. \blue{For example, how to select appropriate LLMs for diverse user tasks such as lower latency and price, higher generation quality, or multi-modal tasks.}

\subsubsection{\rm \textbf{Prompt engineering for telecom applications }}

Prompt engineering is a crucial aspect of utilizing LLM techniques effectively, as it plays a significant role in guiding the model's behaviour and generating desired outputs. 
\blue{Specifically, prompt engineering refers to the process of structuring an instruction that can be interpreted and understood by generative AI models, and then produce the desired output.
For instance, few-shot learning can be considered one of the prompt engineering approaches, in which LLMs can learn from examples and demonstrations to improve their performance on target tasks. 
Compared with pre-training or fine-tuning an LLM, prompting has a much lower cost on computational resources. 
In particular, it relies on LLM's inference capabilities, indicating the most straightforward and efficient approach to use LLMs.
The high efficiency of prompting techniques aligns well with many telecom applications, which usually require fast responses to network dynamics, e.g., changing channel conditions, user numbers, network traffic level, etc.}
However, designing prompts for telecom applications presents unique challenges due to the domain complexity.

\blue{Firstly}, telecom networks encompass a wide range of concepts, protocols, and technologies, making it challenging to distill the necessary information into a concise prompt. 
The diverse nature of the domain requires a deep understanding of networking principles and the ability to capture specific nuances related to network architectures, protocols, performance optimization, and security. 
To design effective prompts, researchers must identify the most relevant components and provide concise yet comprehensive instructions to LLMs.
Meanwhile, prompt designs should strike a balance between being specific enough to guide the LLM in generating accurate and contextually appropriate responses, while also remaining general enough to handle a wide range of network-related queries or tasks. 
Achieving this balance is crucial as overly specific prompts may limit the model's ability to generalize, while a general prompt may lead to irrelevant responses. 

Moreover, telecom tasks often require the LLM to consider contextual information and situational variables. For example, network troubleshooting may involve analyzing network logs, diagnosing performance issues, or identifying security vulnerabilities. Designing prompts that take into account the relevant context and guide the model to consider appropriate factors can significantly enhance the accuracy and relevance of the generated responses. Techniques like providing explicit context cues or utilizing conditional generation can be explored.

To summarize, prompt design of LLMs for telecom applications poses a significant challenge due to the intricate and constantly evolving nature of the domain. 
Crafting effective prompts necessitates a profound comprehension of networking principles, the capacity to strike a balance between specificity and generality, and an awareness of contextual factors. 
\blue{To this end, a practical solution is to publicize some standard prompting templates. They will provide fundamental suggestions for prompt designs of each kind of task. For instance, network optimization-related tasks should specify the optimization objective and control variables, providing feedback or examples for previous selections.
In addition, researchers and industry experts play a crucial role in developing these standards, since the template design requires professional knowledge and understanding of telecom networks. Then these templates will be able to deliver precise, pertinent, and impartial responses in telecom applications.}

\subsection{Future Directions}

\subsubsection{\rm \textbf{\blue{Multi-modal LLMs for telecom}}}
\blue{Multi-modality is a crucial direction for LLM development, indicating seamless integration of information with various modalities such as text, image, audio, video, etc.
Such a capability may serve many applications in future telecom networks.
For instance, sensing has become a critical pillar for envisioned 6G networks, and multi-modal LLMs can utilize 3D multi-modal data, e.g., text, satellite or street camera images, 3D LiDAR maps and videos, to provide a holographic understanding for wireless signal transmission environment \cite{bariah2023large}. 
A specific example of mmWave/THz beamforming has been discussed in Section VII-E, and it aims to predict signal transmission blockage by using multi-modal input such as image and video. 
Meanwhile, with multi-modal information of the 3D environment, we can also have better CSI estimation results, which is critical for signal transmission and network management \cite{bariah2023large}. 
In addition, Xu \textit{et. al} also introduced an example in \cite{xu2024large}, which utilizes LLMs to generate a traffic accident report by using videos collected by vehicles. This video-to-text generation can also be used to analyze the videos collected by UAVs to describe the wireless signal transmission environments. 
}

\subsubsection{\rm \textbf{LLM-enabled planning in telecom}}
\blue{Multi-step planning and scheduling are critical for handling many tasks in the telecom field. 
For instance, Section \ref{sec-code-multi} has introduced an example of coding wireless projects with step-by-step prompting.
Meanwhile, many optimization problems with multiple network elements and control variables have to be solved by dedicated planning \cite{zhou2023survey}.   
However, recent benchmarks have shown that \blue{LLMs} struggle with tasks requiring complex planning and sequential decision-making, which may prevent the direct application to many telecom tasks.} 
Some existing studies such as \cite{du2023power} and \cite{xiang2023toward} propose to improve the multi-step planning capabilities by step-by-step and CoT prompting. 
Despite the satisfactory performance in \cite{du2023power} and \cite{xiang2023toward}, they require dedicated analyses to manually decompose a complicated task into multiple sub-tasks.  
Therefore, future studies should aim at developing better algorithms for planning that can be integrated into \blue{LLMs}, and such multi-step planning capability is crucial for solving telecom-domain tasks. 
This might involve incorporating structured reasoning and problem-solving frameworks into the models, enabling them to break down tasks into smaller and more manageable sub-tasks. 
Therefore, automated task decomposition can be an attractive solution to improve the planning performance of \blue{LLMs}. However, automatically decoupling one complicated task into multiple sub-tasks is still very challenging in the telecom field.
Additionally, another solution could be integrating simulation environments directly within the training process, allowing models to practice and refine their planning skills in a controlled setting before applying them to real-world tasks. It allows the \blue{LLMs} to improve the planning performance by trial-and-error before applying it to telecom tasks.

\subsubsection{\rm \textbf{\blue{LLM for resource allocation and network optimization}}}

\blue{Resource management is a fundamental and crucial problem for network operation, e.g., transmission power and bandwidth resources allocation. 
The above Section VI introduced various LLM-enabled optimization techniques for telecom applications, including LLM-aided reinforcement learning, black-box optimizer, LLM-aided convex optimization and heuristic algorithm design. 
These analyses of existing studies have revealed the potential of using LLM to optimize network performance. For instance, verbal reinforcement learning can take the network operator's human language instructions as input to improve task performance, and LLMs can design novel heuristic algorithms based on specific task demands for network resource allocations. 
LLM-based optimization has two crucial advantages: Firstly, LLM can integrate human languages into the optimization procedure, which makes network management more accessible with much lower complexity; Secondly, LLM can provide reasons and explanations for their decisions, and this capability is crucial for understanding complicated systems such as telecom networks.
Despite the advantages, it is worth noting that some network optimization problems can be extremely complicated with coupled control variables and correlated network elements. Solving these optimization tasks may require dedicated design and multi-step scheduling, which is still a challenge in the LLM field.  
}

\subsubsection{\rm \textbf{\blue{LLM-enhanced machine learning for telecom}}}
\blue{ML algorithms have been widely applied to wireless networks and achieve satisfactory performance. For example, reinforcement learning is one of the most widely used ML techniques for network optimization, and deep neural networks have been extensively explored to predict CSI.
Section VI-B introduced LLM-enhanced reinforcement learning by automating the reward function design, indicating a promising direction to explore LLM-enhanced ML algorithms.
For instance, Sahu \textit{et al.} investigated LLM-aided semi-supervised learning for the task of extractive text summarization, in which they proposed a prompt-based pseudo-labelling strategy with LLMs\cite{sahu2023enchancing}.
Multi-agent learning also has many crucial applications in wireless networks, and existing studies show that LLM-based multi-agents have many promising features \cite{guo2024large}. 
In summary, LLM brings new opportunities to make conventional ML algorithms more accessible and explainable when applied to telecom networks.
}

\subsubsection{\rm \textbf{\blue{Real-world implementations of LLM for telecom industry}}}
\blue{
The study in \cite{alizadeh2023llm} by Apple introduced a method for efficiently running LLMs on devices with limited DRAM capacity. This advancement is particularly beneficial for telecom-specific applications that rely on on-device LLM. By bringing LLM capabilities onto the device, Qualcomm also integrates on-device AI into smartphones to provide faster and more personalized services without relying on cloud-based solutions\cite{qualcomm}. 
Qualcomm aims to improve user privacy and reduce latency, enabling applications such as real-time language translation, advanced camera features, and direct contextual assistance on the user's device. 
On-device AI also allows for continuous operation without internet dependency, which is crucial for maintaining service quality in areas with poor connectivity. 
Additionally, solutions such as Kinetica's SQL-GPT enable telecom professionals to interact with data using natural language queries, converting these queries into SQL for quick and effective analysis\cite{smith-2023}. 
This approach democratizes access to data insights, empowering employees to make faster and more informed decisions. 
These applications demonstrate the transformative potential of LLMs in the telecom industry, enhancing security, operational efficiency, and customer experience. By continuing to innovate with LLMs, telecom companies can stay ahead in an increasingly AI-driven landscape, providing superior services and maintaining a competitive advantage.}

\subsubsection{\rm \textbf{Model compression and fast inference for network edge and mobile applications}}
The model size is one of the key bottlenecks of applying LLMs to the telecom domain, leading to stringent requirements for computational and storage capacities. 
Therefore, compressing the model size to adapt to network edge and mobile applications becomes a promising direction. 
In addition, it will also contribute to the fast inference of \blue{LLMs}, since many wireless applications require rapid response time and low latency.
For instance, Xu \textit{et al.} proposed an on-device inference model specifically designed for efficient generative natural language processing tasks\cite{xu2023llmcad}, achieving a $9.3 \times$ faster generation speed. Such a technique can be very promising for LLM-enabled mobile applications in telecom, enabling faster response time for user inquiries. 
Meanwhile, it is worth noting that compressing the model size may degrade the LLM performance, and how to balance the model size and performance requires more research efforts. 
It calls for novel model compression and pruning techniques to reduce the storage and computation burdens at the network edge; on the other hand, standard metrics must be defined to evaluate the performance of \blue{LLMs} in the telecom domain, e.g., accuracy and hallucination probability as we introduced in previous Section \ref{sec-evaluation}.

\subsubsection{\rm \textbf{Overcoming hallucination problems in telecom applications}}
Hallucination, or the generation of factually incorrect or nonsensical information, remains a significant issue for LLM applications. 
Specifically, it means that the LLM may generate some nonsensical answers or solutions for the given telecom task. 
Hallucination can severely undermine the reliability and credibility of LLM-generated content, degrading the performance on many downstream tasks. For instance, a nonsensical answer may be generated when using LLM for telecom question answering. Overcoming these issues is critical for telecom applications to guarantee network service quality and reliability.
To this end, future research should focus on developing methods to reduce hallucination and improve the factual accuracy of model outputs. 
This could include enhancing the training datasets with more verified and reliable sources, implementing post-generation verification steps, or incorporating cross-referencing mechanisms within the model.
Additionally, exploring the use of external knowledge bases and real-time fact-checking during the generation process could help mitigate this issue.
Recently, it has been demonstrated that under specific evaluation conditions, LLMs exhibit exceptional zero-shot capabilities in assessing factual consistency~\cite{luo2023chatgpt}. 
This underscores their potential to become leading evaluators of hallucination in various contexts.
Moreover, techniques such as adversarial testing can also help in assessing their susceptibility to hallucination, where models are deliberately presented with complex or misleading inputs.

\subsubsection{\rm \textbf{Retrieval augmented-LLM for telecom}}
Retrieval augmentation is an important direction for LLM development, which retrieves facts from an external knowledge base to ground \blue{LLMs} on the most up-to-date information.
Telecom networks are constantly evolving and updating, and retrieval augmentation has great potential for telecom applications. In particular, retrieval-augmented LLM can improve the quality and relevance of the generated responses since the LLM has access to more accurate and relevant information.
However, current retrieval-augmented generation models increase the context length, which in turn decreases the efficiency of the model due to the added computational cost, which may lead to severe slow-response issues. 
Such slow response may increase the overall network latency and degrade the service quality. It may prevent the application of some scenarios with tight delay budget, which is very common in wireless networks.
Therefore, future research could focus on improving the efficiency of retrieval-augmented generation by optimizing retrieval mechanisms to balance context relevance and length. This could involve developing more advanced indexing and search algorithms that require less memory. 
Additionally, dynamically adjusting the amount of retrieved information based on the query's complexity could help maintain or improve efficiency without sacrificing the quality of the output.

\subsubsection{\rm \textbf{Economic and affordable \blue{LLMs}}}
Despite the great potential and advantages, training an \blue{LLM} can be financially expensive. For instance, the training expenses for GPT-4 exceeded \$ 100 million, and the LLaMa2 70B model was trained on 2048 GPUs A100 for 23 days with \$ 1.7 million estimated cost \cite{gpt-cost}. 
Although training some smaller models such as LLaMa2 7B can be much cheaper, 
the affordability of LLM techniques is still one of the main concerns. 
For instance, the study in \cite{gpt-cost222} shows that using GPT-4 to support customer service can cost more than \$ 21,000 per month for a small business.
Meanwhile, there are many LLM APIs with various prices, including the prompt cost proportional to the prompt length, generation cost related to the generation length, and a possible fixed cost per query. 
For example, it costs \$30 for 10M tokens using OpenAI’s GPT-4, while only \$ 0.2 for GPT-J hosted by Textsyth\cite{chen2023frugalgpt}.
The financing cost of training, fine-tuning, and deploying \blue{LLMs} will significantly affect the application in telecom networks.
\blue{However, advancements like OpenAI's GPT-4o mini, a cost-efficient small model, offer promising solutions. Priced at just \$ 0.15 per million input tokens and \$ 0.60 per million output tokens, GPT-4o mini is more affordable than previous frontier models and over 60\% cheaper than GPT-3.5 Turbo. This affordability, combined with its superior performance in reasoning tasks, mathematical reasoning, and coding proficiency, enables a broad range of cost-effective applications. 
Similarly, Llama3-8b is also an affordable small-scale model with fast inference speed. These small-scale models may alleviate the economic cost of LLMs with fewer parameters, lower training and fine-tuning costs, and faster inference time. 
For instance, telecom companies can use GPT-4o mini for customer support chatbots that handle vast conversation histories or for network management systems that analyze extensive performance metrics in real time.}
Given the heterogeneous prices and service quality, it is of great importance to evaluate the financing cost of deploying \blue{LLMs} in telecom networks, e.g., balancing the possible performance improvement and the LLM deployment cost, and using \blue{LLMs} in an economic manner for telecom applications. However, this direction has limited existing studies, and it still requires more research efforts.


\section{Conclusion}
\label{sec-conc}


Recently, \blue{large language models (LLMs)} have shown great promise in many fields, specifically for language-related tasks such as summarization and question and answering. LLM-based solutions have also been primarily investigated in the telecom field. In this work, we aim to present a comprehensive survey on \blue{LLMs} for Telecom.  
In particular, we first introduced the LLM fundamentals. We present a comprehensive overview of the model architecture, pre-training, fine-tuning, inference and utilization, evaluation, and deployment of LLM-based solutions. 
Then, a comprehensive survey of existing works on the key techniques and applications in terms of generation, classification, optimization, and prediction problems is presented.
These investigations and analyses have proven that LLMs have outstanding potential to bring artificial general intelligence to the telecom field using in-context and zero-shot learning capabilities. 
Finally, we discussed the key challenges, such as data sets and cost, as well as future research opportunities of LLM-empowered telecom. We hope this work can serve as a good reference for researchers and engineers to better understand the existing works, potentials, challenges, and opportunities of applying \blue{LLMs} for the telecom field.

\normalem
\bibliographystyle{IEEEtran}
\bibliography{Reference}

\end{document}